\documentclass[a4paper,11pt]{article}
\pdfoutput=1 

\usepackage{jheppub} 

\usepackage[T1]{fontenc} 

\usepackage{IEEEtrantools}

\usepackage{subcaption}
\usepackage{acronym} 
\usepackage{graphicx}
\usepackage{float}
\usepackage{lineno}

\newcommand\cmt[1]{}

\acrodef{SM}{Standard Model}
\acrodef{BSM}{Beyond the Standard Model}
\acrodef{LHC}{Large Hadron Collider}
\acrodef{AI}{Artificial Intelligence}
\acrodef{ML}{Machine Learning}
\acrodef{NN}{Neural Network}
\acrodef{DDP}{Data Directed Paradigm}
\acrodef{RPV}{R-parity violating}
\acrodef{DM}{Dark Machine}
\acrodef{FCN}{fully connected network}
\acrodef{PDF}{Probability Density Function}
\acrodef{MSE}{Mean Squared Error}
\acrodef{DR}{dynamic range}
\acrodef{PLR}{profile likelihood ratio}
\acrodef{LR}{ likelihood ratio}
\acrodef{MLP}{multi-layer perceptron}
\acrodef{LQ}{Lepto Quark}

\title{\boldmath Automatizing the search for mass resonances using BumpNet}

\author[a,1]{Jean-Fran\c{c}ois Arguin,\note{Corresponding author.}}
\author[a,b]{Georges Azuelos,}
\author[a]{Émile Baril,}
\author[c]{Ilan Bessudo}
\author[a]{Fannie Bilodeau,}
\author[c]{Maryna Borysova,}
\author[c,2]{Shikma Bressler,\note{Corresponding author.}}
\author[d]{Samuel Calvet,}
\author[d]{Julien Donini,}
\author[c]{Etienne Dreyer,}
\author[c]{Michael Kwok Lam Chu,}
\author[d]{Eva Mayer,}
\author[a]{Ethan Meszaros,}
\author[c]{Nilotpal Kakati,}
\author[d]{Bruna Pascual Dias,}
\author[a,e]{Joséphine Potdevin,}
\author[c]{Amit Shkuri,}
\author[a]{Muhammad Usman,}

\affiliation[a]{Group of Particle Physics, Universit\'e de Montr\'eal, Montr\'eal QC; Canada.}
\affiliation[b]{TRIUMF, Vancouver BC; Canada.}
\affiliation[c]{Department of Particle Physics and Astrophysics, Weizmann Institute of Science, Rehovot; Israel.}
\affiliation[d]{LPCA, Universit\'e Clermont Auvergne, CNRS/IN2P3, Clermont-Ferrand; France.}
\affiliation[e]{Institute of Physics, Ecole Polytechnique F\'ed\'erale de Lausanne (EPFL), Lausanne; Switzerland}

\emailAdd{jean-francois.arguin@umontreal.ca}
\emailAdd{georges.azuelos@umontreal.ca}
\emailAdd{emile.baril.1@umontreal.ca}
\emailAdd{ilan.bessudo@cern.ch}
\emailAdd{fannie.bilodeau@umontreal.ca}
\emailAdd{maryna.borysova@weizmann.ac.il}
\emailAdd{shikma.bressler@weizmann.ac.il}
\emailAdd{scalvet@clermont.in2p3.fr}
\emailAdd{julien.donini@clermont.in2p3.fr}
\emailAdd{etienne.dreyer@weizmann.ac.il}
\emailAdd{eva.mayer@cern.ch}
\emailAdd{ethan.meszaros@umontreal.ca}
\emailAdd{nilotpal.kakati@cern.ch}
\emailAdd{bruna.pascual.dias@cern.ch}
\emailAdd{josephine.potdevin@cern.ch}
\emailAdd{amit.shkuri@weizmann.ac.il}
\emailAdd{muhammad.usman.1@umontreal.ca}

\abstract{
The search for resonant mass bumps in invariant-mass distributions remains a cornerstone strategy for uncovering Beyond the Standard Model (BSM) physics at the Large Hadron Collider (LHC). Traditional methods often rely on predefined functional forms and exhaustive computational and human resources, limiting the scope of tested final states and selections. This work presents BumpNet, a machine learning-based approach leveraging advanced neural network architectures to generalize and enhance the Data-Directed Paradigm (DDP) for resonance searches. Trained on a diverse dataset of smoothly-falling analytical functions and realistic simulated data, BumpNet efficiently predicts statistical significance distributions across varying histogram configurations, including those derived from LHC-like conditions. The network's performance is validated against idealized likelihood ratio-based tests, showing minimal bias and strong sensitivity in detecting mass bumps across a range of scenarios. Additionally, BumpNet's application to realistic BSM scenarios highlights its capability to identify subtle signals while managing the look-elsewhere effect. These results underscore BumpNet's potential to expand the reach of resonance searches, paving the way for more comprehensive explorations of LHC data in future analyses.}

\begin{document} 
\maketitle
\flushbottom


\newcommand{\HT}{$H_\mathrm{T}$}
\newcommand{\pT}{$p_\mathrm{T}$}
\newcommand{\MET}{$E_\mathrm{T}^\mathrm{miss}$}
\newcommand{\ZLR}{$Z_\mathrm{max}^\mathrm{LR}$}
\newcommand{\Zpred}{$Z_\mathrm{max}^\mathrm{pred}$}
\newcommand{\DZ}{$\Delta Z_\mathrm{max}$}

\newcommand{\SB}[1]{\textbf{\textcolor{blue}{[SB: #1]}}}
\newcommand{\SC}[1]{\textbf{\textcolor{red}{[SC: #1]}}}
\newcommand{\BP}[1]{\textbf{\textcolor{purple}{[BP: #1]}}}
\newcommand{\JD}[1]{\textbf{\textcolor{magenta}{[JD: #1]}}}
\newcommand{\EB}[1]{\textbf{\textcolor{cyan}{[EB: #1]}}}
\newcommand{\JFA}[1]{\textbf{\textcolor{cyan}{[JFA: #1]}}}
\section{Introduction}

Despite its success in describing the elementary particles and their interactions, the \ac{SM} is still incomplete \cite{Weinberg:2018apv}. Many models extending \ac{BSM} have been developed over the years predicting the existence of new resonances. Thus, the search for such resonances, either theoretically-predicted or model-agnostic, is a core strategy for discovery in experimental high-energy physics (e.g., recently \cite{ATLAS:2024tzc,ATLAS:2024gyc,ATLAS:2023ssk,CMS:2024ulc,CMS:2023hwl,CMS:2022eud} and many more). 

To date, several dozens of searches for BSM resonances were carried out by the ATLAS and CMS  collaborations at the \ac{LHC} at CERN. No significant deviation from the \ac{SM} predictions was observed. Despite this huge effort, the \ac{LHC} data is far from being fully exhausted \cite{Kim:2019rhy}. In particular, the majority of searches conducted so far focused on two-body decays in exclusive selections, e.g., exclusive di-lepton \cite{CMS:2021ctt,ATLAS:2019erb}, di-jet \cite{ATLAS:2019fgd,CMS:2019gwf} and di-photon \cite{ATLAS:2017ayi,ATLAS:2021uiz,CMS:2018dqv} searches. 
Among the hundreds of invariant-mass distributions that can be constructed from final states with four or fewer objects (e.g., leptons, jets, photons, etc.), we estimate that only a small fraction (approximately 5\%) has been investigated for resonant mass bumps at the LHC to date \cite{atlas_searches_results}.

Novel artificial intelligence and machine learning techniques provide new opportunities to improve the reach of resonant searches. In Ref.~\cite{ATLAS:2020iwa}, the ATLAS collaboration employed semi-supervised \ac{ML} technique to enhance the signal of massive resonances decaying into two large-cone jets over the background exploiting the mass of the two jet product. Attempts to enhance resonant signals using autoencoders were made in Ref.~\cite{ATLAS:2023ixc}. 

In both searches, once the signal was enhanced, traditional background modeling methods were employed, specifically side-band fitting to a predefined functional form.
While it was shown that there are tens of thousand of different potential signal regions at the LHC \cite{Chekanov:2023dby}, these time-consuming resources and methods limit the number of final states and selections that could be tested. Other \ac{ML}-based method for resonant searches were proposed but, to date, none were employed to real experimental data. Details can be found in, e.g., Ref.~\cite{Belis:2023mqs} and references therein.

The \ac{DDP} approach introduced in Refs.~\cite{Volkovich:2021txe,Birman:2022xzu,Bressler:2024wzc} leverages a theoretically well-established property of the \ac{SM} combined with a tool designed for efficiently identifying deviations from this property. This enables rapid and systematic exploration of numerous final states and selection criteria in the search for \ac{BSM} physics. Specifically, the bump-hunt \ac{DDP} \cite{Volkovich:2021txe} uses a \ac{NN} to rapidly map invariant-mass distributions into statistical inference, thus reducing significantly the time it takes to identify bumps in the data and allow to rapidly scan a large number of final states and selections. To prove the concept, it was demonstrated that a simple fully connected \ac{NN} can accurately predict the bin-by-bin significance of mass bumps in the smoothly falling invariant-mass distributions expected from \ac{SM} backgrounds. 
However, the method was demonstrated only using synthetic (non-realistic) data and under several caveats: a fixed number of mass bins; a narrow \ac{DR} with 100–10,000 entries per bin; background shapes given by analytical, smoothly falling functions (all concave); and Gaussian-shaped signals with a fixed width of three bins.

In this work, we present BumpNet—a generalization of the bump-hunt \ac{DDP}. Utilizing a more advanced \ac{NN} architecture and a richer training dataset, we show that a single network can accurately predict significance for histograms with varying numbers of bins, mass widths, and \ac{DR}, for backgrounds generated from smoothly falling analytical functions as well as simulations of realistic physics processes and real data.
The BumpNet architecture and training dataset are detailed in Section \ref{sec:methodology}, followed by a detailed performance study on Gaussian and data-like signals in Sections \ref{sec:performanceGaussian} and \ref{sec:performanceDataLike}, respectively. Also in
Section~\ref{sec:performanceDataLike}, a realistic analysis is mimicked by searching for \ac{BSM} signals in a total of $\mathcal{O}(10^4)$
histograms, introducing initial strategies to address the significant look-elsewhere effect (LEE) inherent in such comprehensive analyses.
We conclude the work in Section \ref{sec:conclusions}.

\section{Methodology}
\label{sec:methodology}

BumpNet is a generalization of the bump-hunt \ac{DDP} \cite{Volkovich:2021txe}, featuring significant enhancements in both the network architecture and the training dataset. These improvements enable BumpNet to more effectively capture the intricacies of realistic histogram shapes and predict the statistical significance $z$ of potential resonant signals. Specifically, BumpNet predicts the statistical significance for excesses of events (``bumps'') based on the \ac{LR} test for positive and negative signals~\cite{Cowan:2010js}. When given an invariant-mass distribution, the \ac{NN} outputs a $z$ distribution indicating where and how likely it is that the data contains a bump. BumpNet is trained in a supervised manner using generated training and testing datasets. The enhanced network architecture and training procedure are detailed in Section~\ref{sec:Znet}, while the data preparation is described in Section~\ref{sec:datasets}.

\subsection{BumpNet architecture and training procedure} 
\label{sec:Znet}

\begin{figure*}[ht!]
    \centering
    \includegraphics[width=0.99\textwidth]{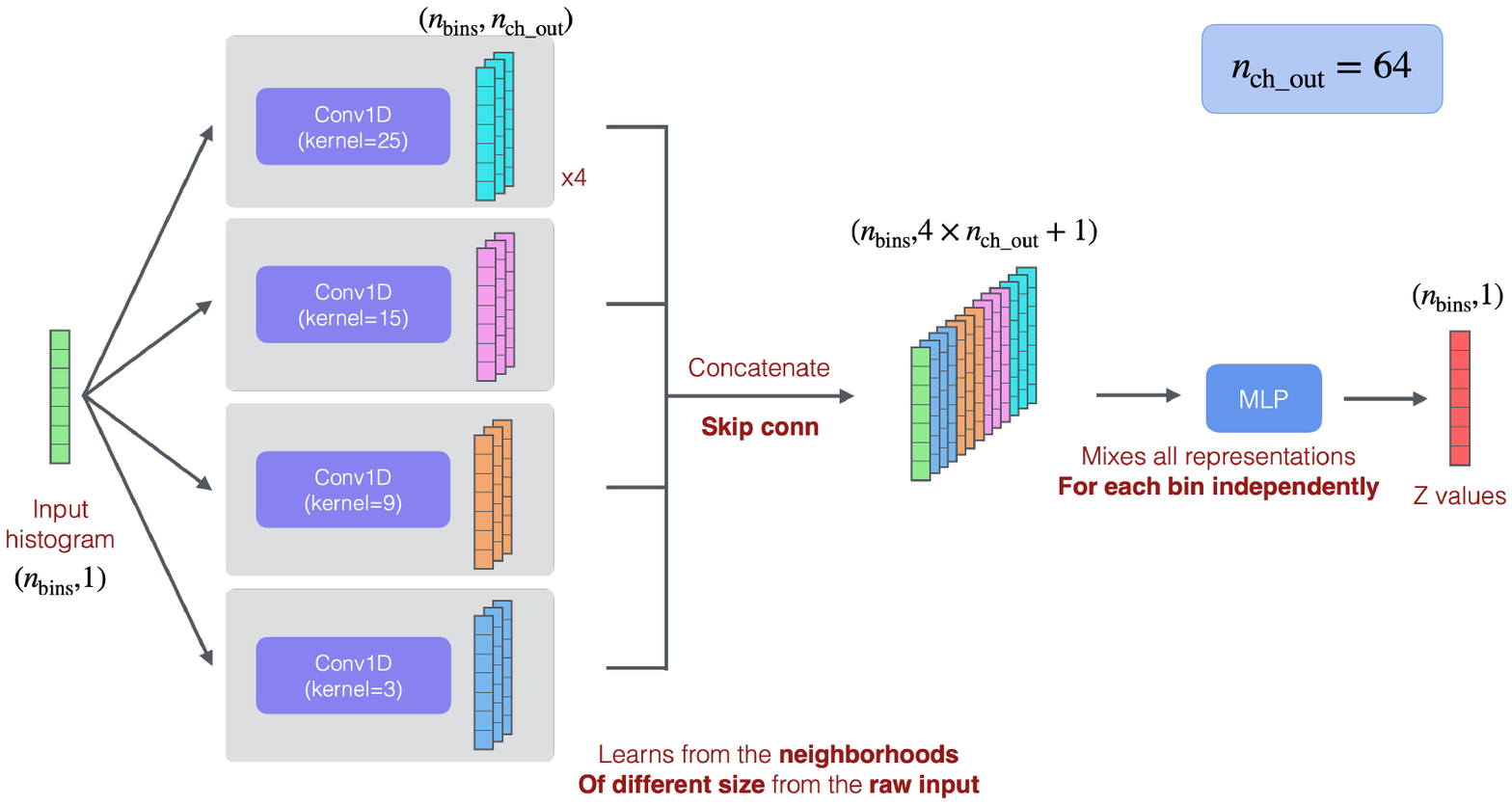}
    \caption{The BumpNet architecture}
    \label{fig:znet3}
\end{figure*}

BumpNet is designed to process smoothly-falling invariant-mass histograms and predict statistical significance of resonant signals. The architecture is illustrated in Figure~\ref{fig:znet3}. The network accepts an input tensor of dimensions $(n_{\text{bins}}, 1)$, where $n_{\text{bins}}$ is the variable number of bins in the invariant-mass histogram. This ability to handle variable-sized inputs represents a refinement over the bump-hunt \ac{DDP}~\cite{Volkovich:2021txe}. This input tensor is processed in parallel by four convolutional stacks. Each stack consists of four convolutional layers with 64 channels and ReLU activations~\cite{nair2010rectified}, all using the same kernel sizes within a stack: 3, 9, 15, or 25. The varying kernel sizes enable the model to capture features at different scales: smaller kernels learn local relationships, while larger kernels capture broader smoothly- falling background patterns. Zero padding ensures consistent input and output lengths for each convolutional layer.

The outputs from the convolutional stacks are concatenated. To preserve the raw input values, a skip connection combines the original input with the concatenated output, resulting in a representation of shapes $(n_{\text{bins}}, 4 \times 64 + 1)$. This representation is then fed into a multilayer perceptron (MLP), which processes each bin independently while allowing their representations to interact. The MLP consists of four layers: the first three are linear layers with ReLU activations and 128, 64, and 32 neurons, respectively; the final layer is linear  with a single output neuron providing the predicted significance for each bin.

To train the network, datasets are generated from known underlying signal and background shapes, enabling the computation of the true statistical significance of any signal using the LR method~\cite{Cowan:2010js}. Each input histogram is constructed by adding a localized gaussian signal, $S$, to a smoothly-decaying background curve, $B$. The injected gaussian signals are narrow, corresponding to a width of one histogram bin (the histogram bin size is discussed further in Section~\ref{sec:DMsamples}). To simulate realistic statistical fluctuations, both the signal and background histograms are Poisson-fluctuated bin by bin. Since the underlying signal and background shapes are known, the true statistical significance per histogram bin can be calculated using the LR method; this per-bin true significance serves as the target for training. More details on the data preparation is provided in the next section. Each input-target pair, referred to as a "sample," consists of an input histogram and the corresponding target significance distribution. Before being utilized by the \ac{NN}, all samples are globally scaled to the interval $[0,1]$ using a linear transformation.

The network is trained using the Adam optimizer~\cite{kingma2017adammethodstochasticoptimization} to minimize the mean squared error (MSE) loss function over 300 epochs. A scheduler learning rate "CosineAnnealingLR" that decays from 0.001 to 0 in a cosine way is used with a batch size of 5000. The training sample comprises approximately 3 million samples, one third of the background shapes obtained from analytical functions and two thirds derived from realistic simulated data, with 10\% reserved for validation. Both types of background shapes are described in the next section. Once trained, the \ac{NN}'s predictions are validated for consistency and convergence of the loss value.

\subsection{Dataset preparation}
\label{sec:datasets}

The training and evaluation datasets are produced by adding synthetic Gaussian signals on top of background invariant-mass distributions. The background distributions are obtained either from analytical functions (Section~\ref{sec:functions}) or from realistic simulated data using the public Dark Machines dataset~\citep{Aarrestad:2021oeb} (Section~\ref{sec:DMsamples}).

\subsubsection{Smoothly-falling functions}
\label{sec:functions}

Similarly to Ref.~\cite{Volkovich:2021txe}, smoothly-falling backgrounds are modelled by randomly selecting one of the following eleven functional forms for each sample:
%
\begin{equation} \label{eq1}
\begin{split}
&be^{-ax},\ \ ax+b,\ \ \frac{1}{ax}+b,\ \ \frac{1}{ax^2}+b,\ \ \frac{1}{ax^3}+b,\\
&\frac{1}{ax^4}+b,\ \ a\left(x-x_2\right)^2+y_2,\ \ -a\cdot\ln\left(x\right)+b,\\
&\left(y_1-y_2\right) \cos \left(a\left(x-b\right)\right)+y_2,\ \ \cosh\left(a\left(x-x_2\right)\right)+b\\
&\frac{1}{ax^n}+b, n\in[0.01,10].
\end{split}
\end{equation}
\noindent
The parameters \( a \) and \( b \) are defined such that each curve decays between two randomly selected points, \((x_1, y_1)\) and \((x_2, y_2)\). Here, \( x_1 \) and \( x_2 \) are the centers of the extreme mass bins within the range \([10,\ 5,000]\)\,GeV, with \( x_1 < x_2 \), and $y_1$ and $y_2$ are the number of events in the first and last bin. They are randomly drawn within the range \([0,\ 10^5]\) and sorted such that \( y_1 > y_2 \). To prevent the range from being too small, a minimum separation of 100\,GeV is required between \( x_1 \) and \( x_2 \).

\begin{figure}[htbp]
\centering
    \begin{tabular}{cc}
\subfloat[]{\includegraphics[width=.45\textwidth]{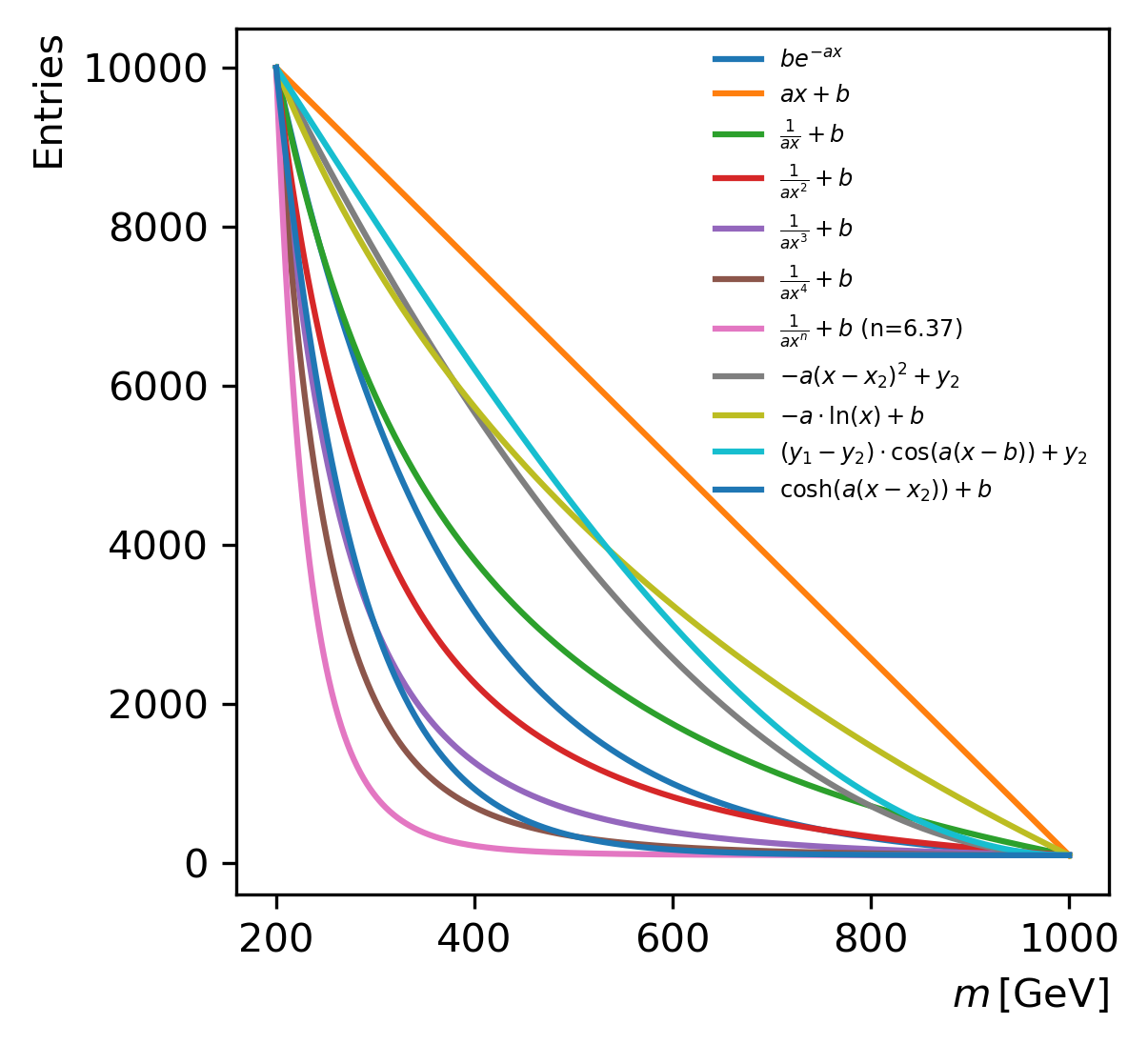}\label{fig:functions-a}}
\subfloat[]{\includegraphics[width=.45\textwidth]{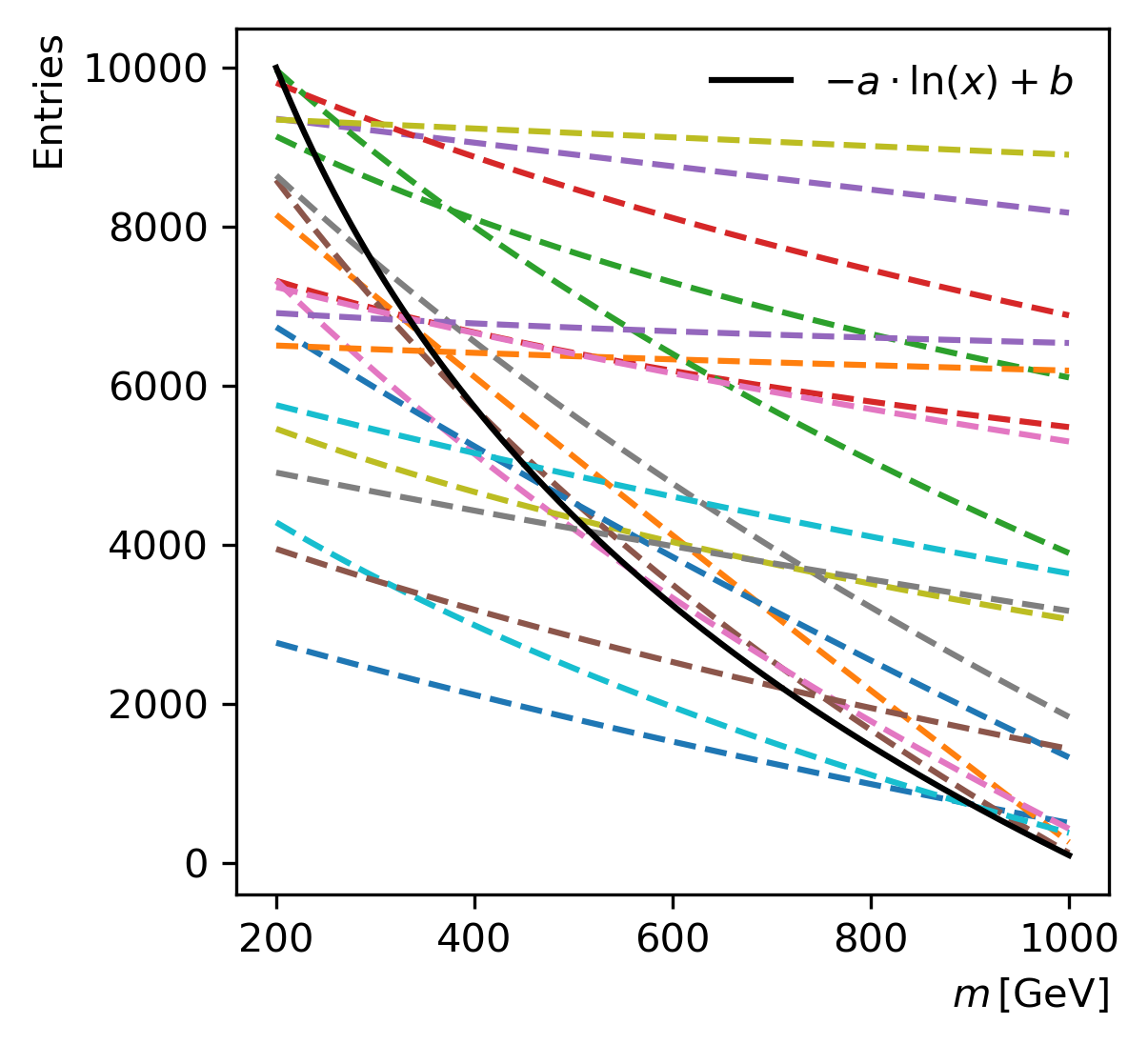}\label{fig:functions-b}}
    \end{tabular}
\caption{\subref{fig:functions-a}: Example of the 11 functions defined by two fixed points. \subref{fig:functions-b}: The effect of stretching the function $-a\cdot\ln\left(x\right)+b$.} \label{fig:functions}
\end{figure}

As seen in Figure~\ref{fig:functions-a}, using this scheme, each pair \((x_1, y_1)\) and \((x_2, y_2)\) defines a limited set of curves, covering a relatively narrow range of possible background shapes. To increase variability, an additional step was added. Two additional points \((x_i, y_i)\) and \((x_j, y_j)\) are randomly drawn such that \( x_1 < x_i < x_j < x_2 \), and \( y_1 > y_i > y_j > y_2\). The values of \( a \) and \( b \) are evaluated relative to \((x_i, y_i)\) and \((x_j, y_j)\) and the resulting function is stretched to match the full \([x_1,x_2]\) range. The process is demonstrated in Figure~\ref{fig:functions-b} for the function $-a\cdot\ln\left(x\right)+b$.


\subsubsection{Monte-Carlo Dark Machines samples}
\label{sec:DMsamples}

Histograms derived from smoothly falling functions cannot fully capture the features of histograms encountered in real data. In particular, kinematic cuts can sculpt the invariant-mass shapes, flattening the lower mass range of the histograms. To emulate this behavior, and to test the performance of BumpNet in a realistic setup, simulated data produced for the Dark Machines initiative\footnote{\href{https://www.darkmachines.org/}{https://www.darkmachines.org/}} is used.

\paragraph{Description of the \ac{DM} samples}
The \ac{DM} samples include a background dataset that emulates 10 fb$^{-1}$ of $pp$ collisions at $\sqrt{s}$ = 13 TeV. This dataset contains the 26 highest cross-section SM processes expected at the LHC,  ranging from dijets down to $t\bar{t}WW$. Each process is weighted according to its cross-section, resulting in a realistic dataset that simulates data coming out of the LHC. This dataset is described in detail in Ref.~\citep{Aarrestad:2021oeb}. The \ac{DM} samples also include  15 signal benchmarks, some of which are used and described in Section~\ref{sec:resBSM} to test the performance of BumpNet in a realistic setup. All of the \ac{DM} samples have been generated with \textsc{MadGraph} \cite{Alwall:2014hca} interfaced with \textsc{Pythia} \cite{Sjostrand:2014zea} and then passed through a detector simulation using DELPHES3 \cite{deFavereau:2013fsa}.


We adopt the same object definitions as the \ac{DM} project, specifically:
\begin{itemize}
    \item Electrons with transverse momentum \pT $>$ 15 GeV and pseudorapidity $|\eta|<2.5$.
    \item Muons with \pT $>$ 15 GeV and $|\eta|<2.7$. 
    \item Photons with \pT $>$ 20 GeV and $|\eta|<2.37$. 
    \item Jets with \pT $>$ 20 GeV and $|\eta|<2.8$. 
\end{itemize}
The jets can also be $b$-tagged, indicating  a high probability of originating from a $b$-quark.

To limit the potentially huge number of \ac{SM} events needed to emulate 10 fb$^{-1}$ of LHC data, in particular for the highest cross-section processes such as dijet events, the \ac{DM} project has split its data into four channels that include additional requirements. The two \ac{DM} channels that yield the largest number of events are used to train and to test BumpNet:
\begin{itemize}
\item \textit{Channel 2b}: \HT $>50$ GeV, \MET $>$ 50 GeV, $\geq$ 2 leptons (electron or muon), yielding 340,000 events;
\item \textit{Channel 3}: \HT  $>600$ GeV, \MET $>$ 100 GeV, yielding 8,500,000 events.
\end{itemize}
Here \HT\ is the scalar sum of the \pT\ of all jets in an events,  and \MET\ is the missing transverse energy.


\paragraph{Additional object definitions}
To more closely align with an analysis performed on real data that attempts to maximize the use of available information, we define additional objects beyond those provided by \ac{DM} project. Pairs of same-flavor, opposite charge leptons with a mass within 15~GeV of $m_Z=91.18~\mathrm{GeV}$ are identified as $Z$-boson candidates\footnote{There can be multiple $Z$ candidates in one event, as long as they do not share a lepton.  If a lepton could be assigned to two $Z$-boson candidates, the $Z$ candidate with the mass closest to $m_Z$ is kept.}. Leptons associated with a 
$Z$-boson candidate are removed from the list of leptons used in further analysis.

Jets with a mass between 60 and 110 GeV is assumed to originate from the hadronic decay of boosted vector bosons ($W$ or $Z$) and are labelled as "$V_h$". Those with a mass between 110 and  200 GeV are tagged as hadronically-decaying top-quark candidates (labelled "top", or "T" in the figures) and jets with a mass greater than 200 GeV are labeled by "HM" (High Mass), as they may originate from boosted, high-mass \ac{BSM} particles. 

\paragraph{Signal regions}


A large number of exclusive signal regions, or categories, are created by considering all possible combinations of electrons, muons, photons, \( Z \) bosons, \( V_h \) candidates, top quark candidates, and HM candidates. Standard jets are considered separately later. This is performed up to the maximum number of these objects observed in the \ac{DM} samples, which corresponds to five for electrons, muons, photons, top quarks, and HM candidates; three for \( Z \) bosons; and six for \( V_h \) candidates. Categories containing exactly two charged leptons that do not form a \( Z \) candidate are further divided into two exclusive categories based on the charge of the lepton pairs: opposite-sign (OS) and same-sign (SS). Approximately 63,000 such categories are defined from the \ac{DM} dataset.

The event categories defined above are used as the basis for further analysis and are additionally subdivided using kinematic selections on missing transverse energy (\( E_{\mathrm{T}}^{\text{miss}} \)), transverse momentum (\( p_{\mathrm{T}} \)) of the leading object in the event (such as a muon or top quark), and the number of \( b \)-tagged jets. These selections involve only a lower threshold on the kinematic variable and are therefore not statistically exclusive of other regions derived from the same parent category.


After applying kinematic cuts, each category is further subdivided into exclusive subcategories based on the number of standard jets—that is, jets (\( b \)-tagged or not) that have a mass less than 60\,GeV. The maximum jet multiplicity considered is determined by requiring that the subcategory with the highest jet multiplicity contains at least 100 events, ensuring sufficient statistics. This subdivision by jet multiplicity is utilized in Section~\ref{sec:resBSM}, where correlations of true \ac{BSM} signals occurring at the same mass value and object combination across neighboring jet bins are exploited to identify true positive signals.


\paragraph{Histogram production}

Once the subcategories are defined, one invariant-mass histogram is created for every combination of at least two objects (electrons, muons, photons, $V_h$, Z, top, HM, and any of the four leading-\pT\ jets) within that subcategory. The invariant-mass is calculated as the magnitude of the sum of the four-momenta of the selected objects. For each of these histograms, a second histogram is created, referred to as \textit{MassMET}, that is calculated similarly but includes the missing transverse energy  four-momentum, with the \MET\ longitudinal momentum component (\( p_z \)) and mass both set to zero. The limit on the number of jets included in the combinations is set arbitrarily to prevent an excessive number of combinations. Additionally, if the selection criteria require at least one \( b \)-tagged jet, the combinations are also constructed by considering the flavor of the jets (light-flavor or \( b \)-tagged).

Finally, to address the discontinuity introduced by the \( Z \)-boson candidate selection, the mass spectra of opposite-charge, same-flavor dileptons are divided into two separate histograms: one covering the mass range from 10\,GeV to 76\,GeV, and another for masses greater than 106\,GeV.


BumpNet is designed to find mass bumps on smoothly falling distributions. Therefore, only the bins following the histogram's maximum are retained, effectively dropping the bins before the peak. 

The histograms are defined with varying bin sizes adjusted according to an approximate mass resolution.
This approach intends to make the signal bumps appear similar in units of bins to BumpNet, even though they have different widths in units of mass (GeV). Although achieving perfect uniformity is unavoidably imperfect, this method enhances the network's sensitivity to narrow signals.
For a bin starting at a mass \( m \), the bin width is set to half the resolution of the combined mass (\( \sigma(m) \)), where \( \sigma(m) \) is the quadratic sum of the resolutions of the objects entering the combination, estimated at a transverse momentum of \( m/2 \):
\[
\text{width}(m) = \frac{1}{2} \sqrt{ \sum_{i \in \{ e,\, \mu,\, \gamma,\, V_h,\, \dots,\, \text{MET} \} } \sigma_i^2\left( p_{\mathrm{T}} = \frac{m}{2} \right) }
\]

\noindent
Here, the \( \sigma_i \) are the mass resolutions of the individual objects, extracted from DELPHES for leptons and from Ref.~\cite{ATLAS:2019oxp} for jets. The \( V_h \) candidates are considered to consist of two jets, while top quark and HM candidates consist of three jets. The resolution on \MET\ is computed as the quadratic sum of the resolutions of all the objects in the event. To perform optimally, BumpNet requires histograms with a relatively large number of bins. Only histograms containing at least 100 events and with greater than 30 bins are retained for further analysis.


This comprehensive categorization process results in a total of 8,104 histograms in channel 2b and 31,664 histograms in channel 3. The smaller number of histograms in channel 2b is due to its requirement of exactly two leptons, which limits the possible object combinations and leads to fewer histograms passing the criterion of having at least 100 events and at least 30 bins.

\paragraph{Smoothing procedure}
Each histogram obtained through the procedure described earlier is fitted with an analytical function to create a smooth "mother" histogram, which captures the underlying distribution in the limit of infinite statistics. In this work, a 4th-order log-polynomial function, \( \ln(y) = \sum_{n=0}^4 P_n x^n \), has been used for this purpose. The minimization is performed using the \texttt{curve\_fit} routine from the SciPy library \cite{2020SciPy-NMeth}. 

\begin{figure}[htbp]
    \centering
    \begin{subfigure}[b]{0.45\textwidth}
        \centering
        \includegraphics[width=\textwidth]{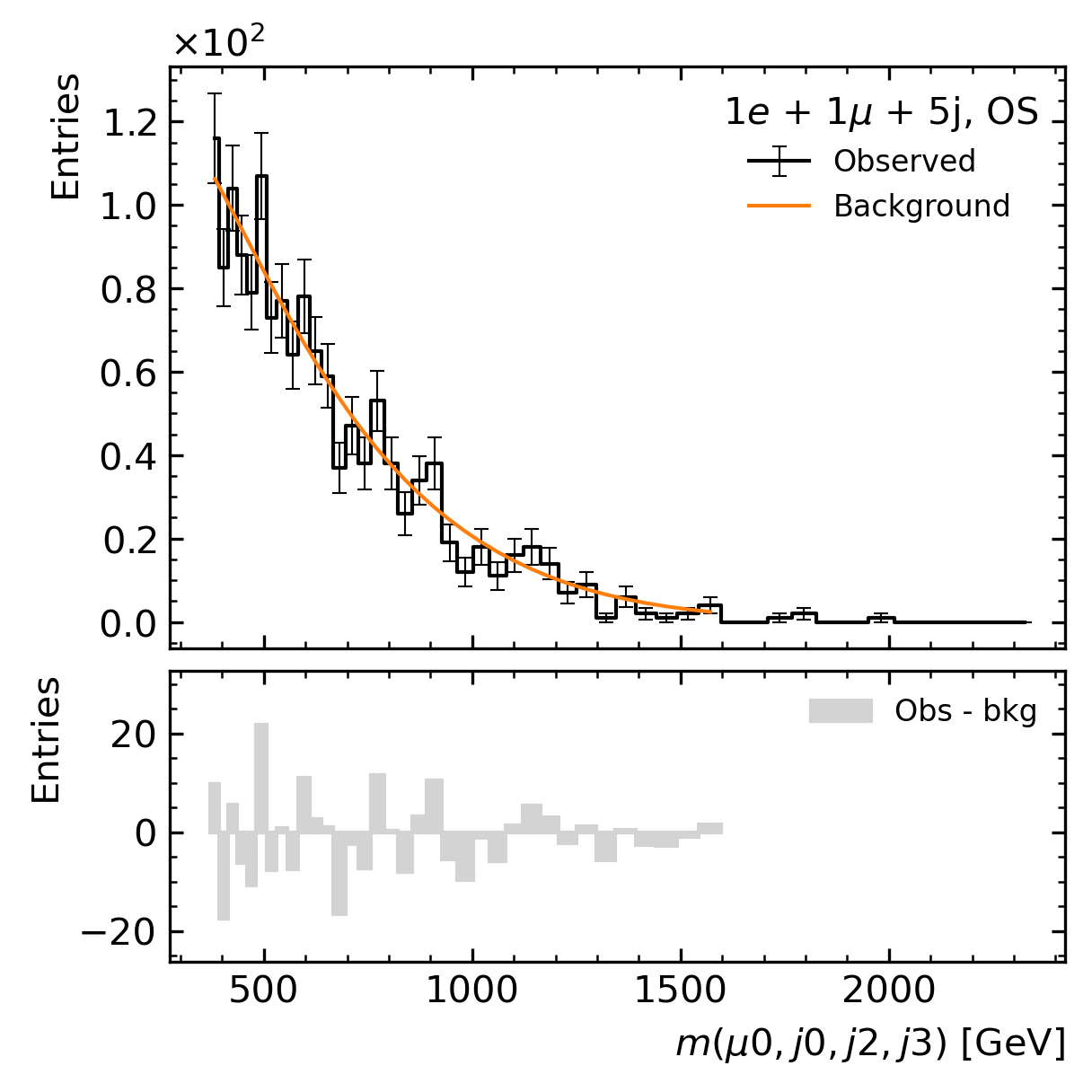}
    \end{subfigure}
    \hfill
    \begin{subfigure}[b]{0.45\textwidth}
        \centering
        \includegraphics[width=\textwidth]{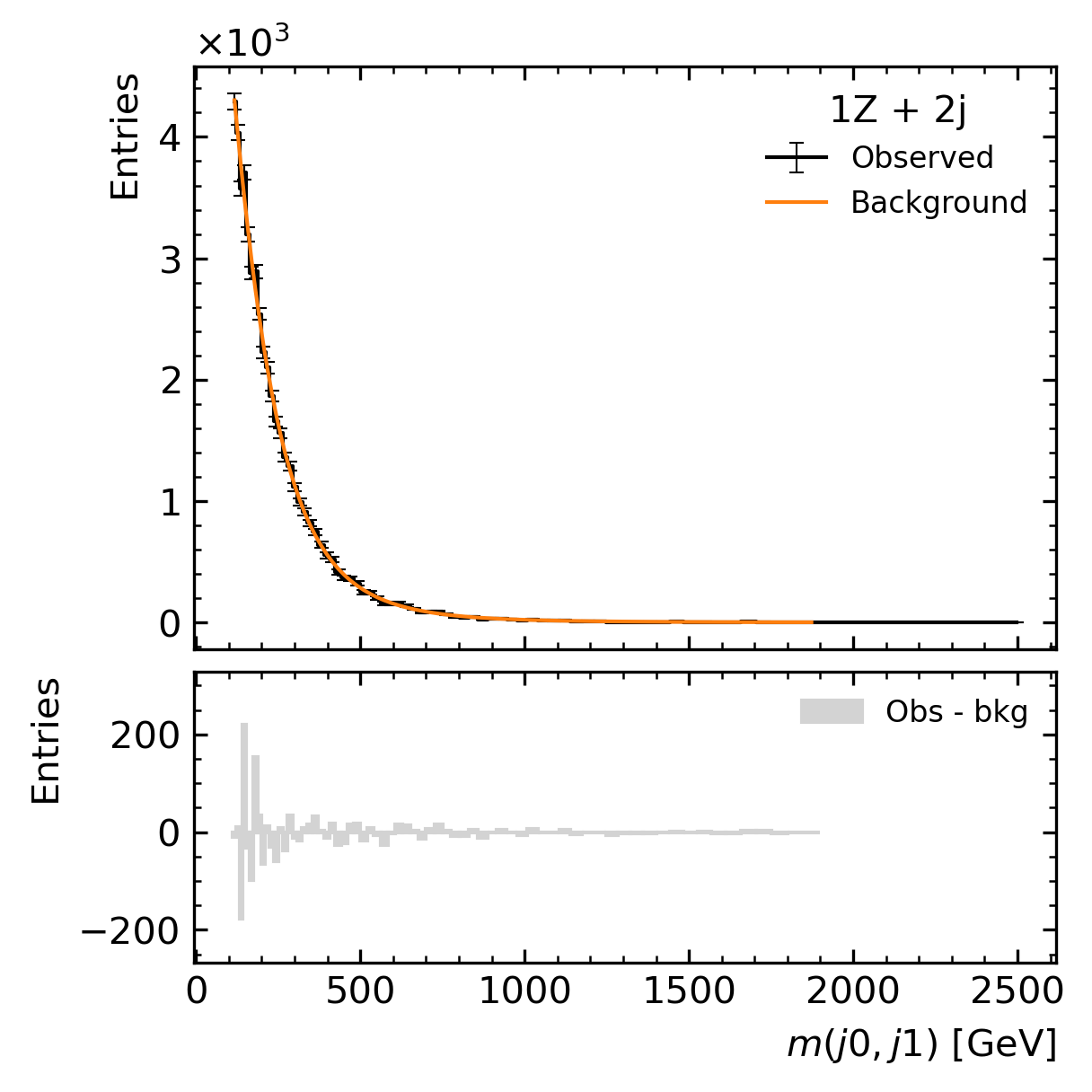}
    \end{subfigure}
    \vskip\baselineskip
    \begin{subfigure}[b]{0.45\textwidth}
        \centering
        \includegraphics[width=\textwidth]{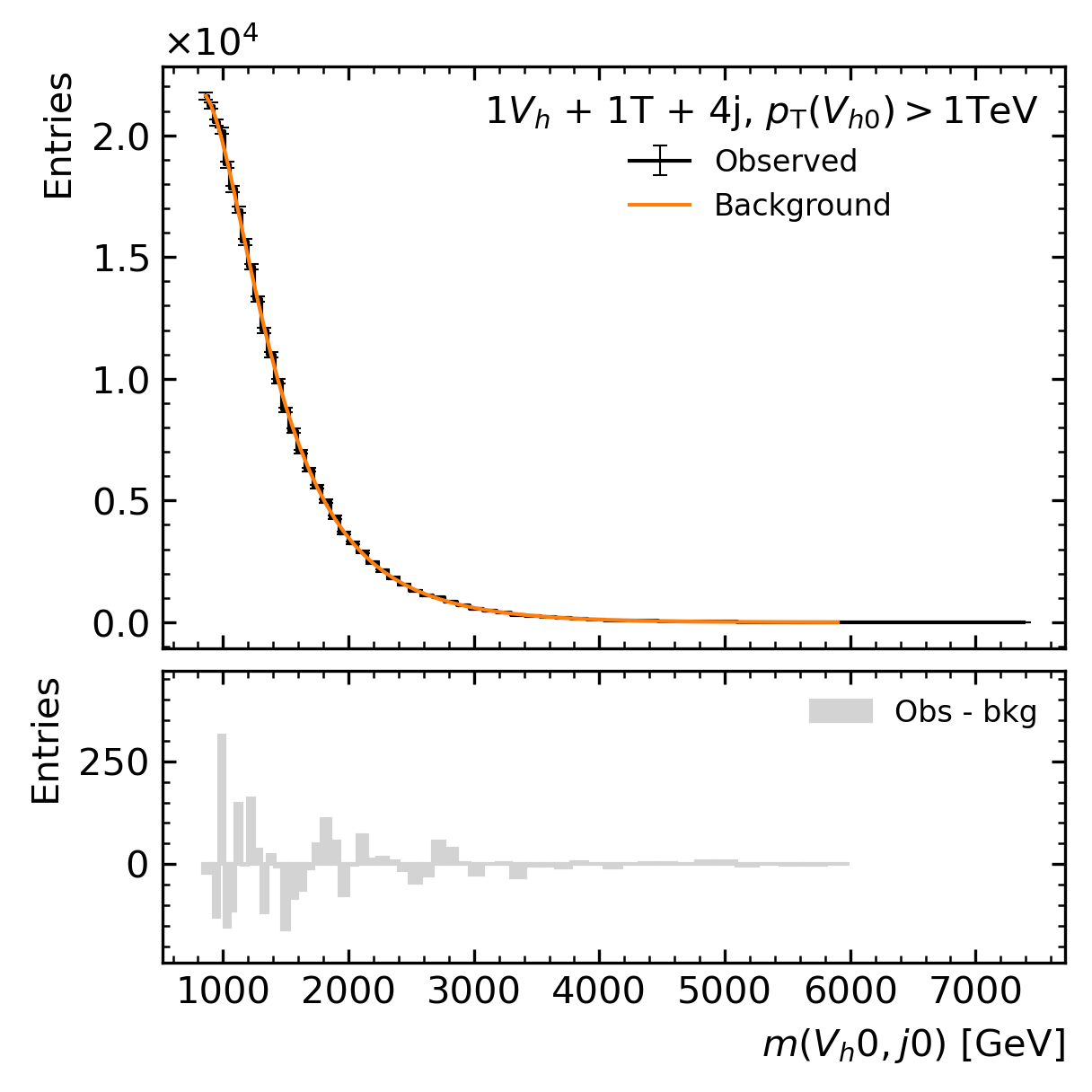}
    \end{subfigure}
    \hfill
    \begin{subfigure}[b]{0.45\textwidth}
        \centering
        \includegraphics[width=\textwidth]{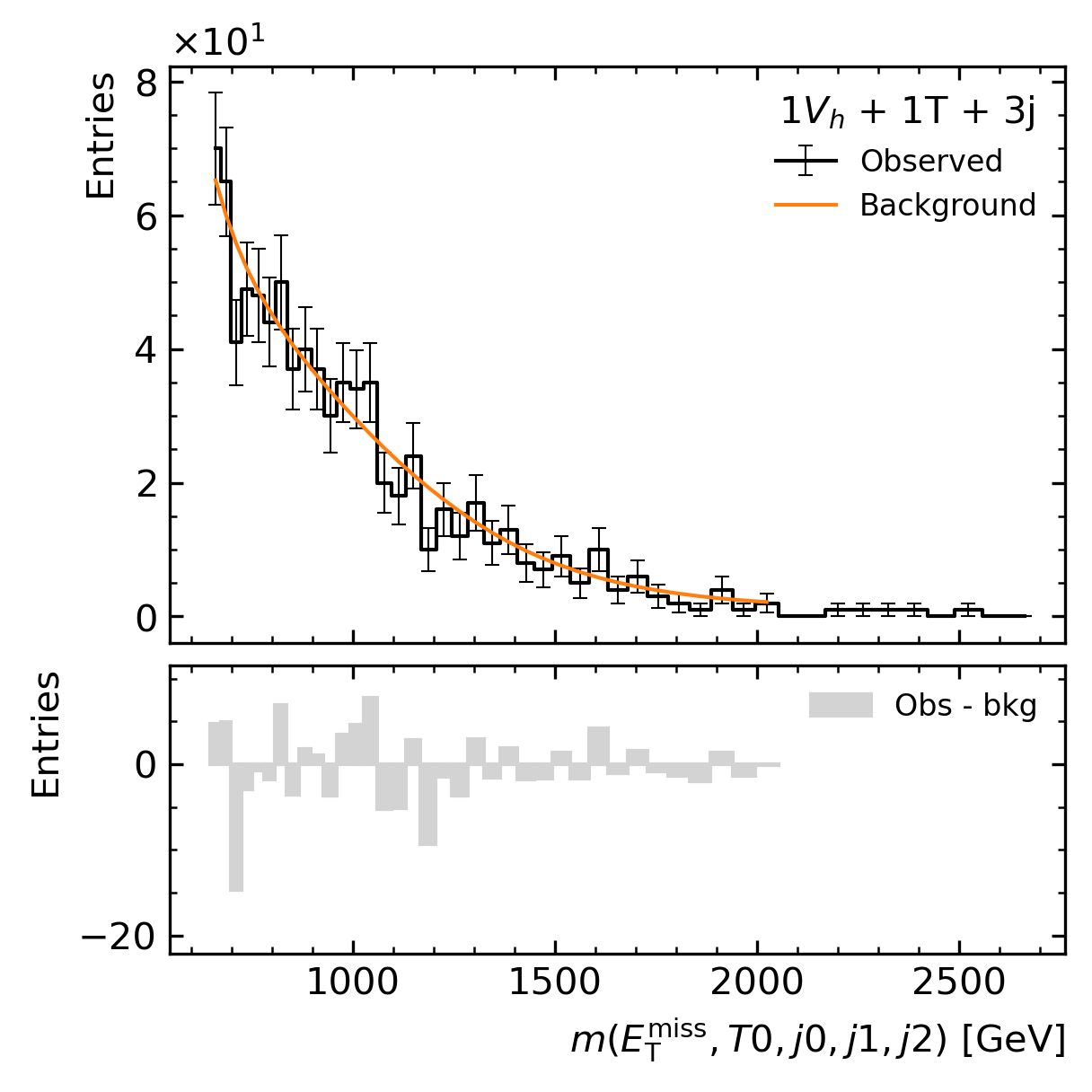}
    \end{subfigure}
    \caption{Example background-only mass histograms generated from the \ac{DM} background samples. The top and bottom plots are from channel 2b and 3 of the \ac{DM} samples, respectively. The fitted smoothed curve is overlaid in orange on the raw histogram. The event category is indicated in the top-right corner of each histogram, and the objects involved in the invariant-mass calculation are labeled on the \( x \)-axis. The number "0" at the end of an object label denotes the leading object in the event, with "1" indicating the sub-leading object, and so on. The bottom panel displays the residuals between the observed and expected (smoothed) number of events.}
    \label{fig:smoothing}
\end{figure}

Figure~\ref{fig:smoothing} illustrates several examples of the "raw" background mass histograms obtained from the \ac{DM} samples using the histogram production described previously. The smoothed distribution, shown in orange, is overlaid on top of the raw histogram in the top panel, while the residual between the two is displayed in the bottom panel. 
Although this simple smoothing procedure works well for the wide majority of the \ac{DM} histograms, future work may explore the use of a broader set of functions or more sophisticated techniques for improved performance.

Background shapes obtained from the pre-determined analytical functions described in Section~\ref{sec:functions} are also added to the dataset, which help BumpNet generalizes over unseen background shapes. Both of these types of background shapes are subsequently used to inject synthetic gaussian signal shapes, as described in Section~\ref{sec:Znet}, producing the training and validation datasets. 

\begin{figure}[htbp]
    \centering
    \begin{subfigure}[b]{0.45\textwidth}
        \centering
        \includegraphics[width=\textwidth]{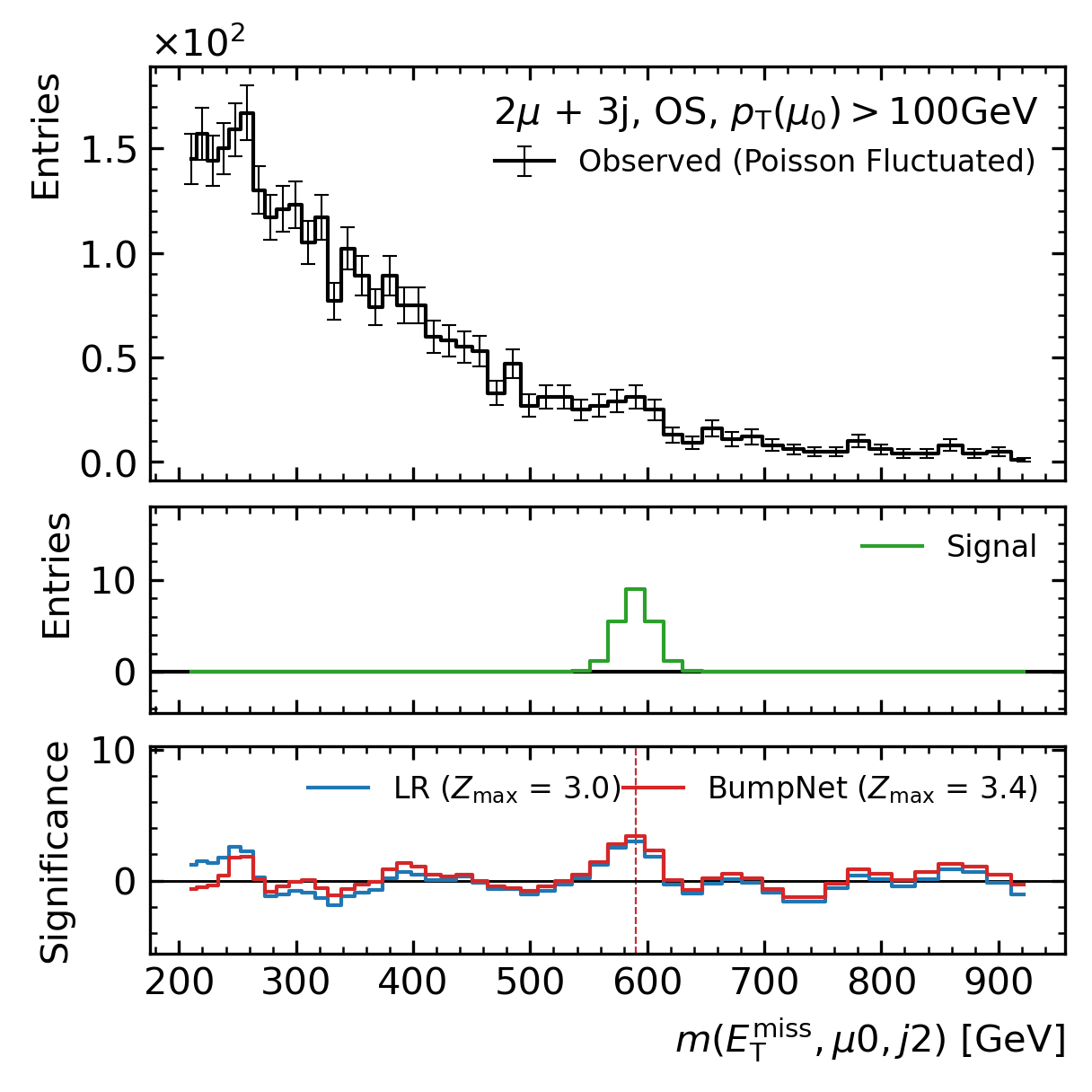}
    \end{subfigure}
    \hfill
    \begin{subfigure}[b]{0.45\textwidth}
        \centering
        \includegraphics[width=\textwidth]{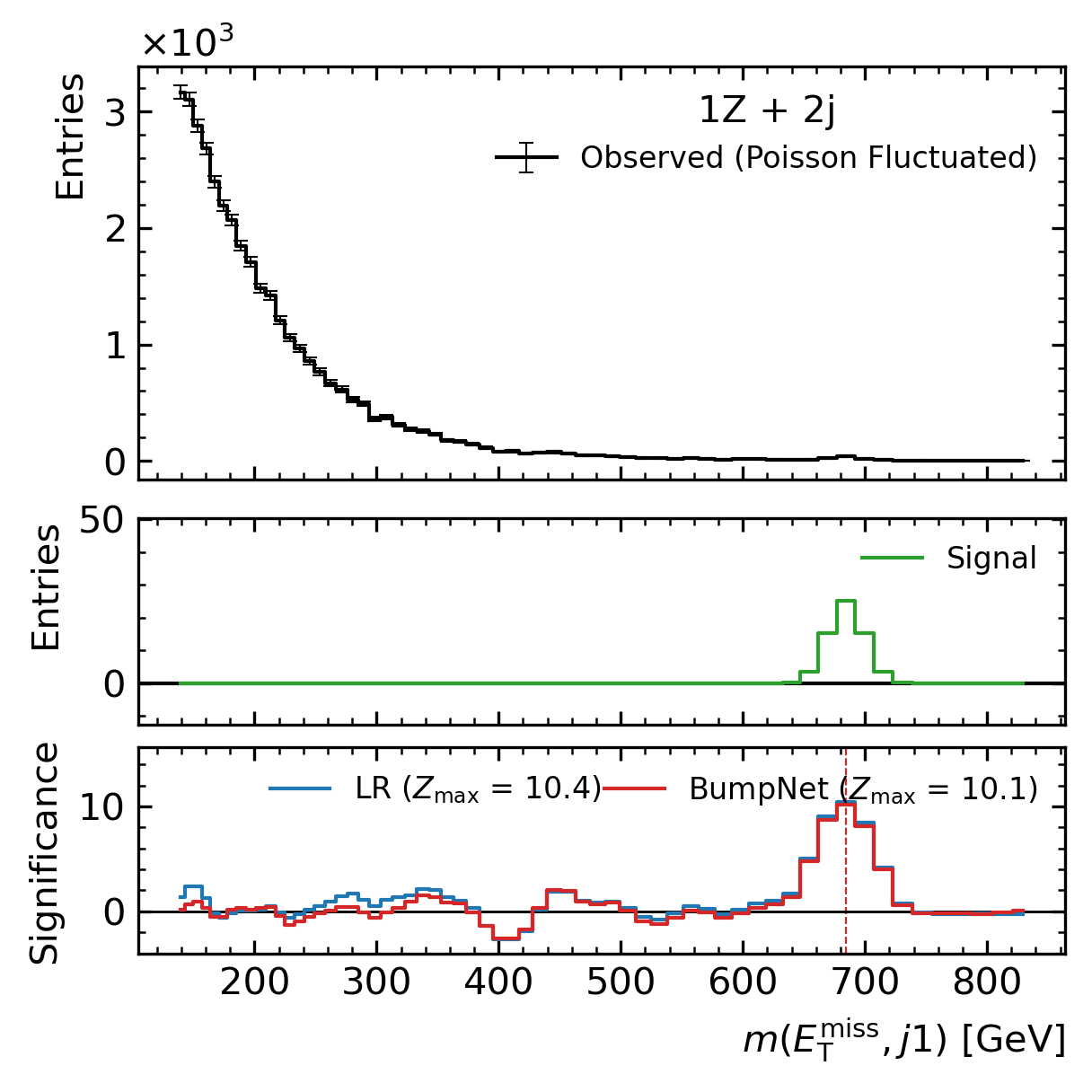}
    \end{subfigure}
    \vskip\baselineskip
    \begin{subfigure}[b]{0.45\textwidth}
        \centering
        \includegraphics[width=\textwidth]{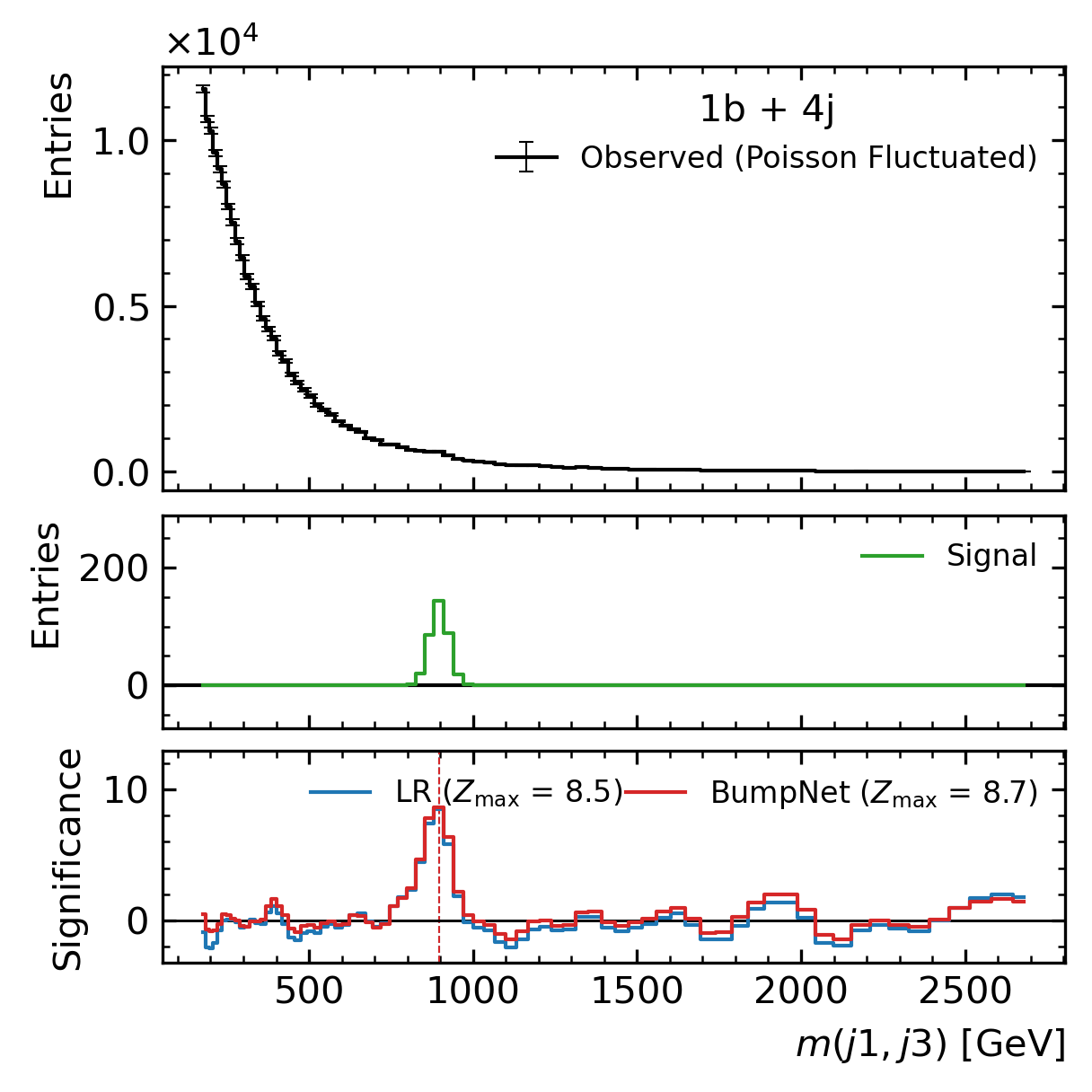}
    \end{subfigure}
    \hfill
    \begin{subfigure}[b]{0.45\textwidth}
        \centering
        \includegraphics[width=\textwidth]{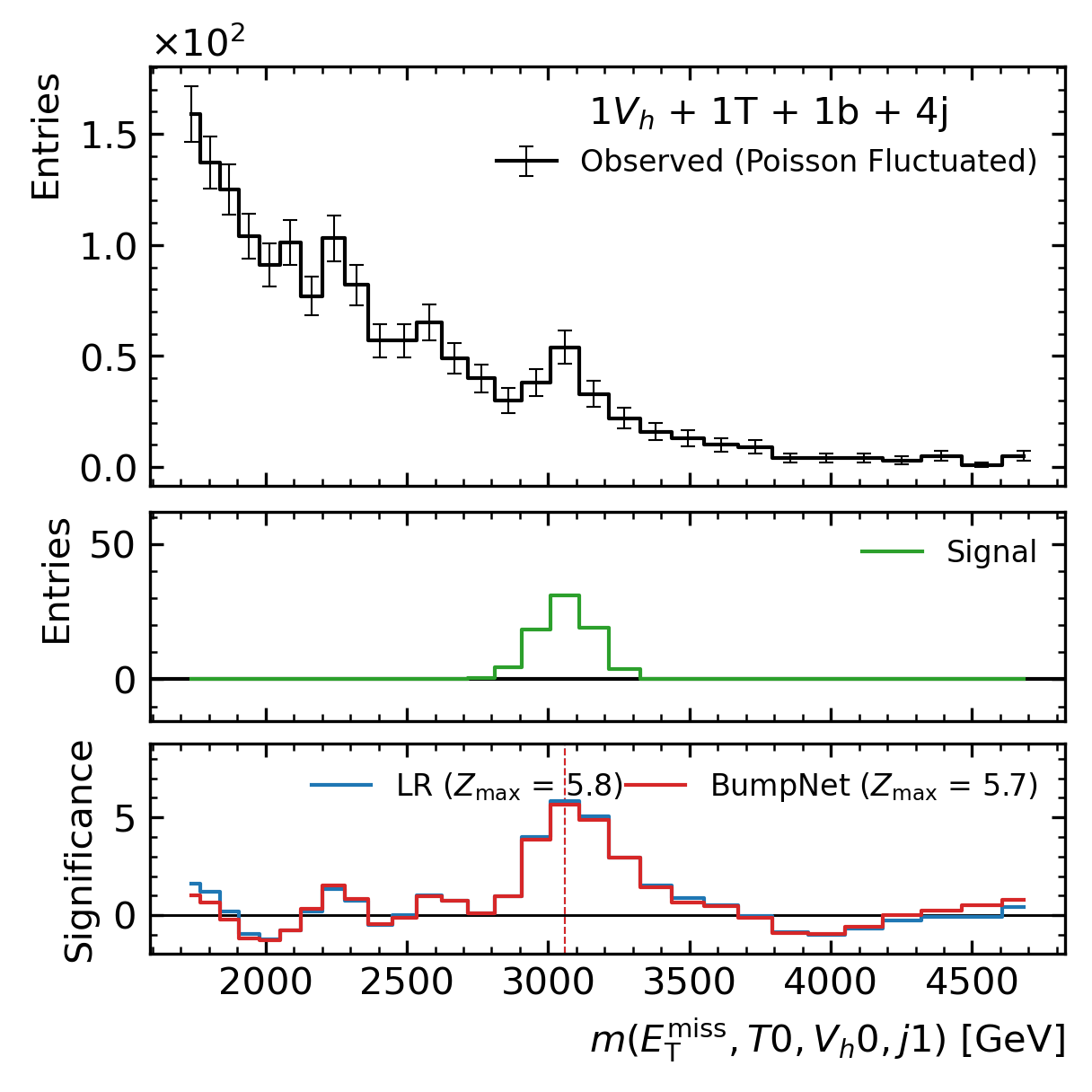}
    \end{subfigure}
    \caption{Example mass histograms generated for the training and validation datasets of BumpNet, derived from the \ac{DM} background samples. The top and bottom plots correspond to channels 2b and 3 of the \ac{DM} samples, respectively. The event category is indicated in the top-right corner of each histogram, and the objects involved in the invariant-mass calculation are labeled on the \( x \)-axis. A number at the end of an object label specifies its ranking in the event: "0" denotes the leading object, "1" indicates the sub-leading object, and so on. The middle panel shows the number of injected signal events, while the bottom panel compares the true significance (blue) to the significance predicted by BumpNet after it was trained (red).}
    \label{fig:exampleDMMass}
\end{figure}

A few example mass histograms of the validation datasets  are shown in Figure~\ref{fig:exampleDMMass}. The background shapes of these examples have been extracted from \ac{DM} samples. The middle panel shows the number of signal events injected. The bottom panel shows in blue the true significance computed from the known signal and background shapes that have been used to generate these histograms using the aformentioned procedure, and the significance predicted by BumpNet is shown in red. Although these exact histograms have never been seen by BumpNet during the training procedure, a very good agreement is observed between the two predicted and the true significances. The performance of BumpNet is characterized in more detail in the subsequent sections.


\section{Performance over injected gaussian signals}
\label{sec:performanceGaussian}

This section presents the performance of BumpNet on nominal histograms—histograms generated similarly to those used in training—as well as on unseen background shapes, in most cases with injected Gaussian signals as in the nominal case. The following section will examine BumpNet’s performance on histograms from real data and on fully simulated \ac{BSM} signals injected into the \ac{DM} background samples.

\subsection{Performance over nominal histograms}
\label{sec:perfDMsamples}

The performance of BumpNet is first evaluated using histograms generated in the same manner as described in Section~\ref{sec:datasets}, but which were not used during training. A total of 500,000 function-based histograms and 1,000,000 \ac{DM}-based histograms, each with injected Gaussian signals, were used for this evaluation.

\begin{figure}[htbp]
\centering
    \begin{subfigure}[b]{0.45\textwidth}
        \centering\includegraphics[width=\textwidth]{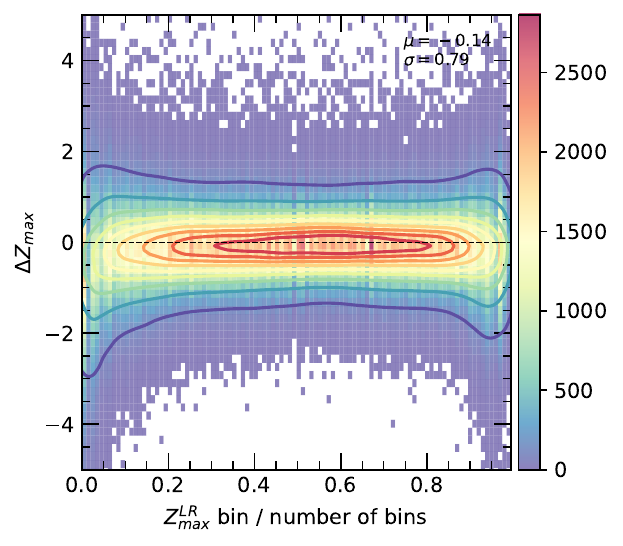} 
        \caption{{\DZ~as a function of the position of \ZLR~along the histogram including all bins.}
        \label{fig:DZvsRelativeBin}}
    \end{subfigure}
    \hfill
    \begin{subfigure}[b]{0.45\textwidth}
        \centering\includegraphics[width=\textwidth]{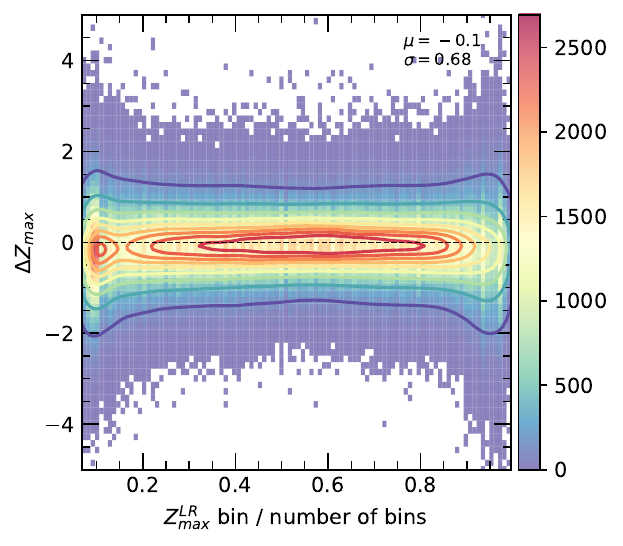} 
        \caption{{\DZ~as a function of the position of \ZLR~along the histogram excluding the first 10\% of the bins.}
        \label{fig:DZvsRelativeBinCut}}
    \end{subfigure}

    \caption{\DZ~as a function of the position of \ZLR~along the histogram including (\ref{fig:DZvsRelativeBin}) and excluding (\ref{fig:DZvsRelativeBinCut}) the first 10\% of the bins.}
        \label{fig:effectOfFirstBins}
\end{figure}

The accuracy of BumpNet's predicted significance (\( z_\mathrm{pred} \)) is assessed by comparing it to the true significance (\( z_\mathrm{LR} \)), which is calculated using the likelihood ratio based on the known background shape of the "mother" histogram, as well as the known strength and position of the injected Gaussian signal. The difference \DZ, defined as the difference between the maximum values of \( z_\mathrm{pred} \) (\Zpred) and \( z_\mathrm{LR} \) (\ZLR) within a given histogram, is shown in Figure~\ref{fig:effectOfFirstBins} as a function of the relative position of \ZLR in a histogram. The results are shown for cases where the first 10\% of the bins are included (left) and excluded (right) from consideration for determining a mass bump. The comparison is performed on histograms generated from both analytical functions and \ac{DM} background shapes. 
As shown in the figure, excluding the first 10\% of the bins reduces the occurrence of significant deviations between \Zpred\ and \ZLR. This is likely because BumpNet struggles to distinguish actual bumps from subtle variations in the background slope near the beginning of the histogram. 
As a result, all subsequent analyses exclude the first 10\% of the histogram range, focusing only on bumps in the remaining 90\%. A slight worsening of the performance is also observed at the high end of histograms which results from smoothing imperfections in that region, and which will be fixed in future iterations of BumpNet. 


\begin{figure}[htbp]
\centering
    \begin{subfigure}[b]{0.45\textwidth}
        \centering
        \includegraphics[width=\textwidth]{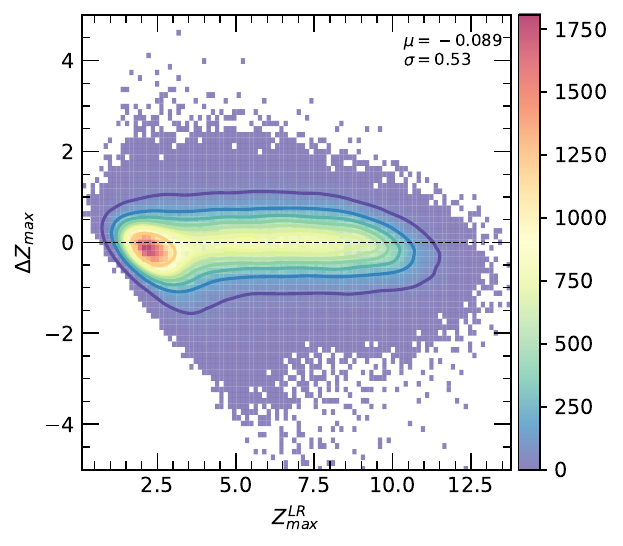}
        \caption{Function background shapes}
        \label{fig:DiffZ_DM_fcn}
    \end{subfigure}
    \hfill
    \begin{subfigure}[b]{0.45\textwidth}
        \centering
        \includegraphics[width=\textwidth]{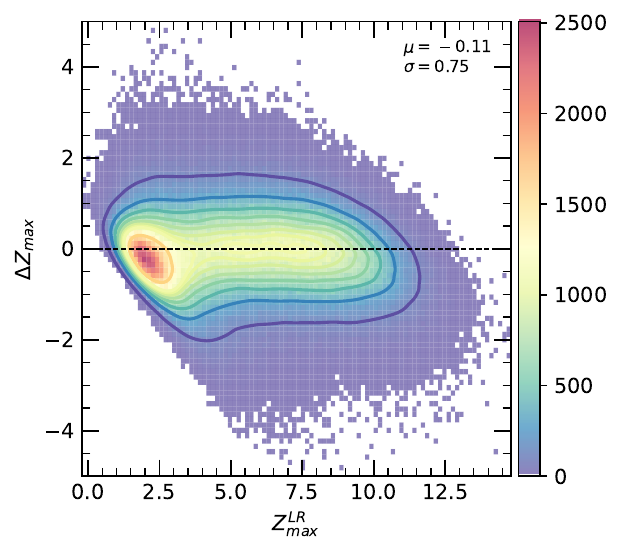}
        \caption{\ac{DM} background shapes}
        \label{fig:DiffZ_DM_dm}
    \end{subfigure}
    \caption{Distributions of \DZ~as a function of \ZLR~over testing data with backgrounds from (\subref{fig:DiffZ_DM_fcn}) analytical functions and  (\subref{fig:DiffZ_DM_dm}) \ac{DM} fits.} 
\label{fig:DiffZ_comp}
\end{figure}

The accuracy \DZ~is shown in Figure~\ref{fig:DiffZ_comp} as a function of maximal true signal strength \ZLR~in each histogram. The central value of \DZ~is close to zero, indicating that the BumpNet significance is unbiased. The spread of \DZ~is relatively small, with standard deviations of 0.53\,$\sigma$ and 0.75\,$\sigma$ for function-based and \ac{DM}-based background shapes, respectively. The slightly larger spread observed for \ac{DM} histograms is likely due to the more complex background structures in that sample, where the left-hand side of the histogram is sometimes shaped by \ac{DM} kinematic selections, such as \HT\ $>$ 600~GeV for channel 3.

\begin{figure}[htbp]
\centering
    \begin{subfigure}[b]{0.45\textwidth}
        \centering
        \includegraphics[width=\textwidth]{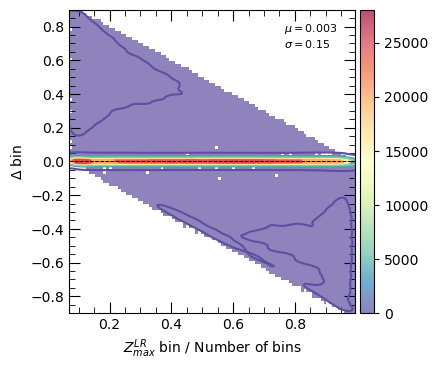}
        \caption{Full \ZLR~range}
        \label{fig:ZPos-a}
    \end{subfigure}
    \hfill
    \begin{subfigure}[b]{0.45\textwidth}
        \centering
\includegraphics[width=\textwidth]{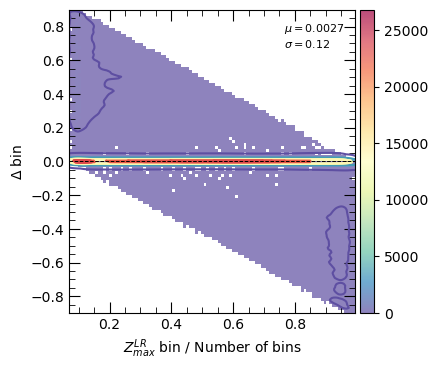}
        \caption{\ZLR$\geq 2$}
        \label{fig:ZPos-b}
    \end{subfigure}
\caption{The difference between the position of the signal predicted by BumpNet (in terms of bin number) and the position of the true signal as a function of the position of the true signal when BumpNet is applied to a dataset combining analytical functions and \ac{DM} samples. The difference is shown for (\subref{fig:ZPos-a}) the full \ZLR~range  and (\subref{fig:ZPos-b}) for \ZLR$\geq 2$.}
\label{fig:ZPos}
\end{figure}

Figure~\ref{fig:ZPos} shows the difference between the injected and predicted signal position (in terms of bin number) as a function of the true position of the injected signal for the combined dataset. BumpNet predicts precisely bump positions, with the small number of prediction inaccuracy observed in Figure~\ref{fig:ZPos-a} are associated with small injected \ZLR~as indicated in Figure~\ref{fig:ZPos-b} when the difference is shown only for distributions with \ZLR~$\geq 2$. For a modest injected signal strength with \ZLR~$\leq 2$, there is a high probability that a background fluctuation elsewhere in a histogram results in a stronger signal than \ZLR and is picked up by BumpNet.

%
\begin{figure}[htbp]
\centering
    \includegraphics[width=0.45\textwidth]{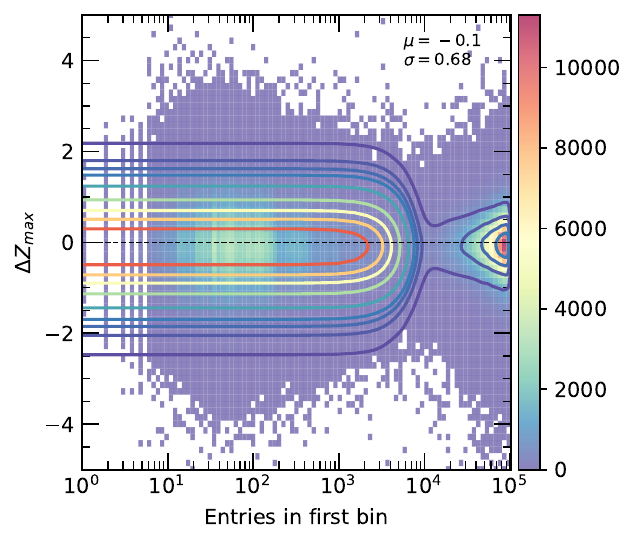}
    \caption{\DZ~as a function of the number of entries in the first bin.}
\label{fig:DZvsEntriesFirstBin}
\end{figure}

Figure \ref{fig:DZvsEntriesFirstBin} shows the accuracy as a function of the number of events in the first bin of the histogram, providing a measure of BumpNet’s performance relative to the statistical content of each histogram. The results indicate that BumpNet’s performance is independent of this histogram characteristic\footnote{The apparent cut-off around $10^4$ events arises from differences in the populations of DM histograms and function histograms, with the latter typically having higher statistics.}.

\begin{figure}[htbp]
\centering
    \begin{subfigure}[b]{0.45\textwidth}
        \centering\includegraphics[width=\textwidth]{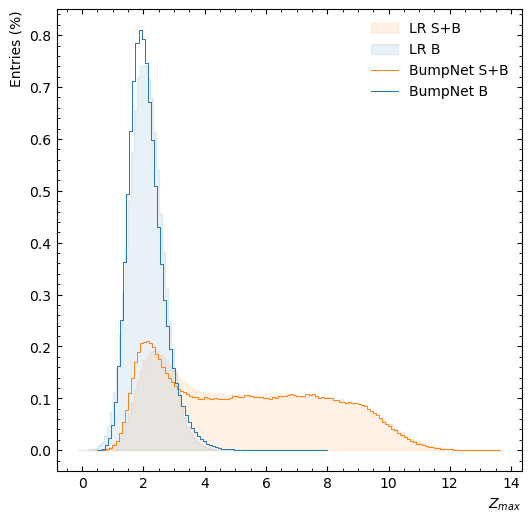} 
        \caption{Function background shapes}
        \label{fig:ZMax_dist_DM_fcn}
    \end{subfigure}
    \hfill
    \begin{subfigure}[b]{0.45\textwidth}
        \centering
\includegraphics[width=\textwidth]{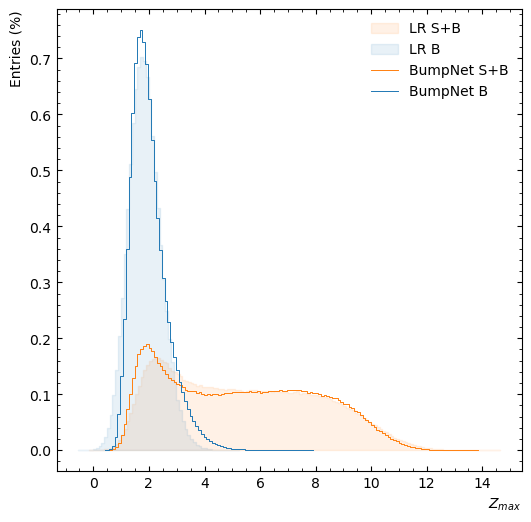}
        \caption{\ac{DM} background shapes}
       \label{fig:ZMax_dist_DM_dm}
    \end{subfigure}
    \begin{subfigure}[b]{0.45\textwidth}
        \centering\includegraphics[width=\textwidth]{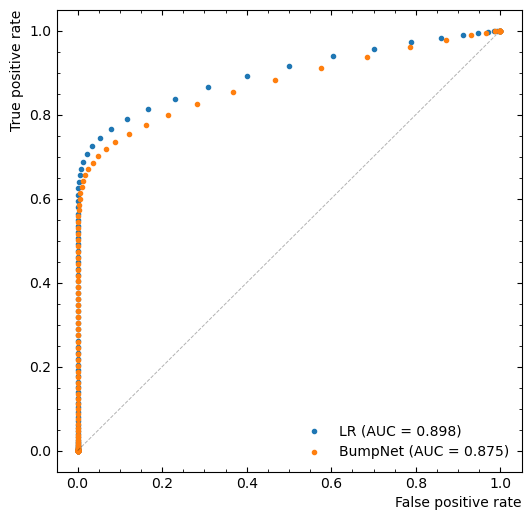}
        \caption{Function background shapes}
        \label{fig:ROC_DM_fcn}
    \end{subfigure}
    \hfill
    \begin{subfigure}[b]{0.45\textwidth}
        \centering
\includegraphics[width=\textwidth]{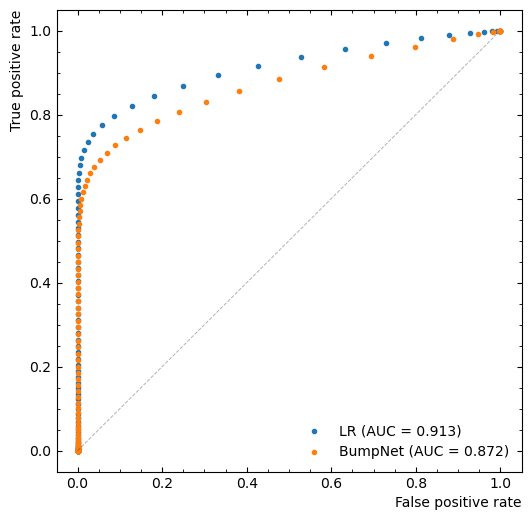}
        \caption{\ac{DM} background shapes}
       \label{fig:ROC_DM_dm}
    \end{subfigure}
\caption{Top plots: Distributions of the predicted (dotted lines) and true (full lines) maximum significances in a given histogram when a signal is present (orange lines) or not (blue lines), when BumpNet is applied to Dark Machines data for the cases where the background training data is generated from (\subref{fig:ZMax_dist_DM_fcn}) analytical functions and  (\subref{fig:ZMax_dist_DM_dm}) Dark Machines samples. Bottom plots: ROC curves of BumpNet applied dataset generated from (\subref{fig:ROC_DM_fcn}) analytical functions and (\subref{fig:ROC_DM_dm}) Dark Machines samples.} \label{fig:ZMax_dist_DM}
\end{figure}

Figure~\ref{fig:ZMax_dist_DM} shows, at the top, the distributions of the predicted (dotted lines) and true (solid lines) maximum significances in each histogram, both with signal present (orange) and with background-only samples (blue). The "false discovery" rate can be extracted from these background-only distributions, revealing that BumpNet identifies a signal with a significance greater than 5\,$\sigma$ in  0.048\% and 0.129\% of background-only histograms for analytical functions and the Dark Machines samples, respectively. These distributions are further used to produce Receiver Operating Characteristic (ROC) curves, displaying the true positive rate versus the false positive rate, as shown in Figures~\ref{fig:ROC_DM_fcn} and \ref{fig:ROC_DM_dm}. For reference, the ideal ROC curve, based on the LR significance, is also included. BumpNet’s performance approaches that of the ideal LR test when both the signal and background shapes are well-defined.

\subsection{Performance of BumpNet over unseen shapes}
\label{sec:perfDistDMsamples}

The results in the previous section were obtained using histograms that were not seen during training, but with background shapes derived from the same "mother" histograms as the training data. To evaluate BumpNet's ability to generalize to entirely new background shapes, it is essential to test its performance on truly unseen data. This is particularly relevant because a possible strategy for applying BumpNet to real LHC data involves training it on background shapes extracted from fully-simulated datasets. As with any LHC analysis, such simulated data will inherently contain experimental and theoretical uncertainties, which may cause the simulated backgrounds to differ systematically from real data.

\subsubsection{Evaluation on Unseen Dark Machines Distributions}

\begin{figure}[htbp]
\centering
    \begin{subfigure}[b]{0.45\textwidth}
        \centering
        \includegraphics[width=\textwidth]{fig/BumpNet_on__chan3-chan2b/DeltaZ_vs_Zlr_signal.pdf}
        \caption{DM shapes included in the training data set.}
        \label{fig:SeenDMShapes}
    \end{subfigure}
    \hfill
    \begin{subfigure}[b]{0.45\textwidth}
        \centering
        \includegraphics[width=\textwidth]{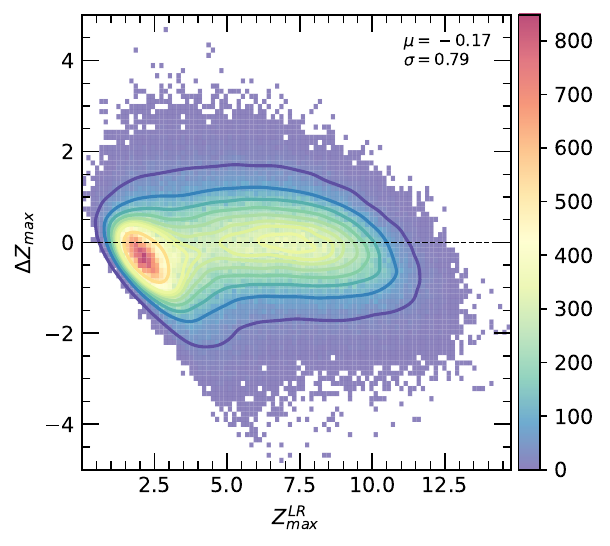}
        \caption{DM shapes excluded from the training data set.}
        \label{fig:UnSeenDMShapes}
    \end{subfigure}
    \caption{Distributions of \DZ~as a function of \ZLR~over testing data with backgrounds derived from \ac{DM} fits to \ac{DM} distributions. The backgrounds include those introduced during training in \protect{\subref{fig:SeenDMShapes}} and those not introduced during training in \protect{\subref{fig:UnSeenDMShapes}}.}
\label{fig:UnSeenVsSeenDMShapes}
\end{figure}

During the training process, 25\% of the \ac{DM} background histograms were randomly excluded from the training dataset. These excluded distributions were reserved for testing purposes and used to evaluate BumpNet's performance on unseen data. 
The results, shown in Figure~\ref{fig:UnSeenVsSeenDMShapes}, demonstrate that BumpNet performs equally well on both training distributions (Figure~\ref{fig:SeenDMShapes}) and the excluded distributions (Figure~\ref{fig:UnSeenDMShapes}). This consistency confirms BumpNet's capability to generalize its predictions effectively, even when applied to data outside the training set.

\subsubsection{Systematic distortions of background shapes}

In this section, systematic differences between the training and application datasets are emulated. Four new application sets are created by applying the following mathematical transformations to the nominal \ac{DM} background shapes. These transformations are intentionally exaggerated to represent uncertainties larger than those typically encountered in nowaday's mature LHC experiments:


\begin{IEEEeqnarray*}{lrCl}
  \text{Standard bias:} &\quad
  y'_{i} &=& 1.5 \frac{N_{\text{bins}} - x_{i}}{N_{\text{bins}}}y_{i},\\[0.5em]
  \text{Reverse bias:} &\quad
  y'_{i} &=& 0.5 \frac{N_{\text{bins}} - x_{i}}{N_{\text{bins}}}y_{i}, \\[0.5em] 
  \text{Inferior bias:} &\quad
  y'_{i} &=& 0.5 y_{i},\\[0.5em]
  \text{Superior bias:} &\quad
  y'_{i} &=& 1.5 y_{i},
\end{IEEEeqnarray*}
where \(x_{i}\) and \(y_{i}\) are, respectively, the bin number and the number of entries in bin \(i\), and \(N_{\text{bins}}\) is the total number of bins in the histogram. Each of these distortions are applied separately to create four distinct application sets that are systematically biased with respect to the nominal histograms that have been used to train BumpNet.

\begin{figure}[t]\centering
    \begin{subfigure}{0.235\linewidth}\centering
        \includegraphics[width=0.99\linewidth]{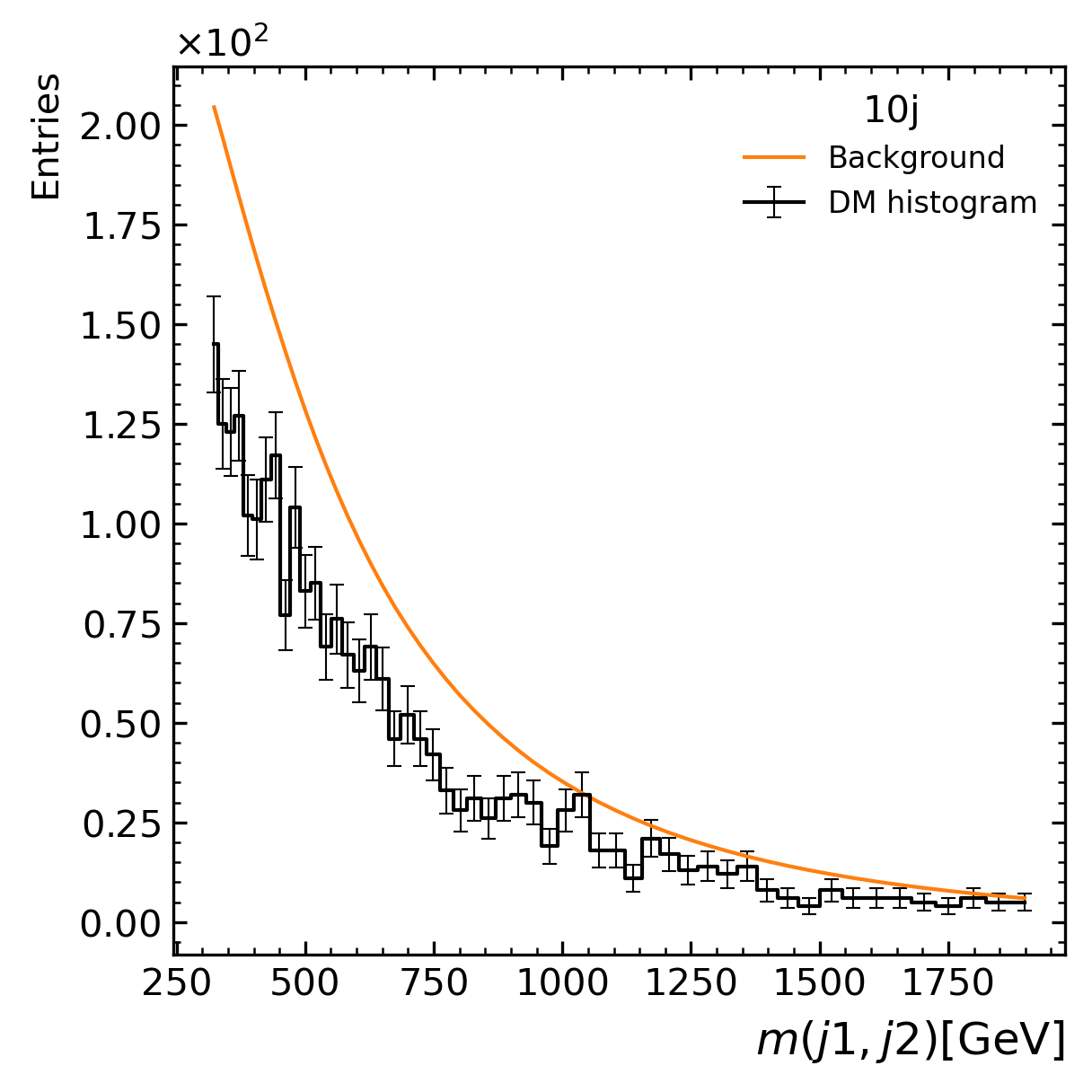}
        \includegraphics[width=\linewidth]{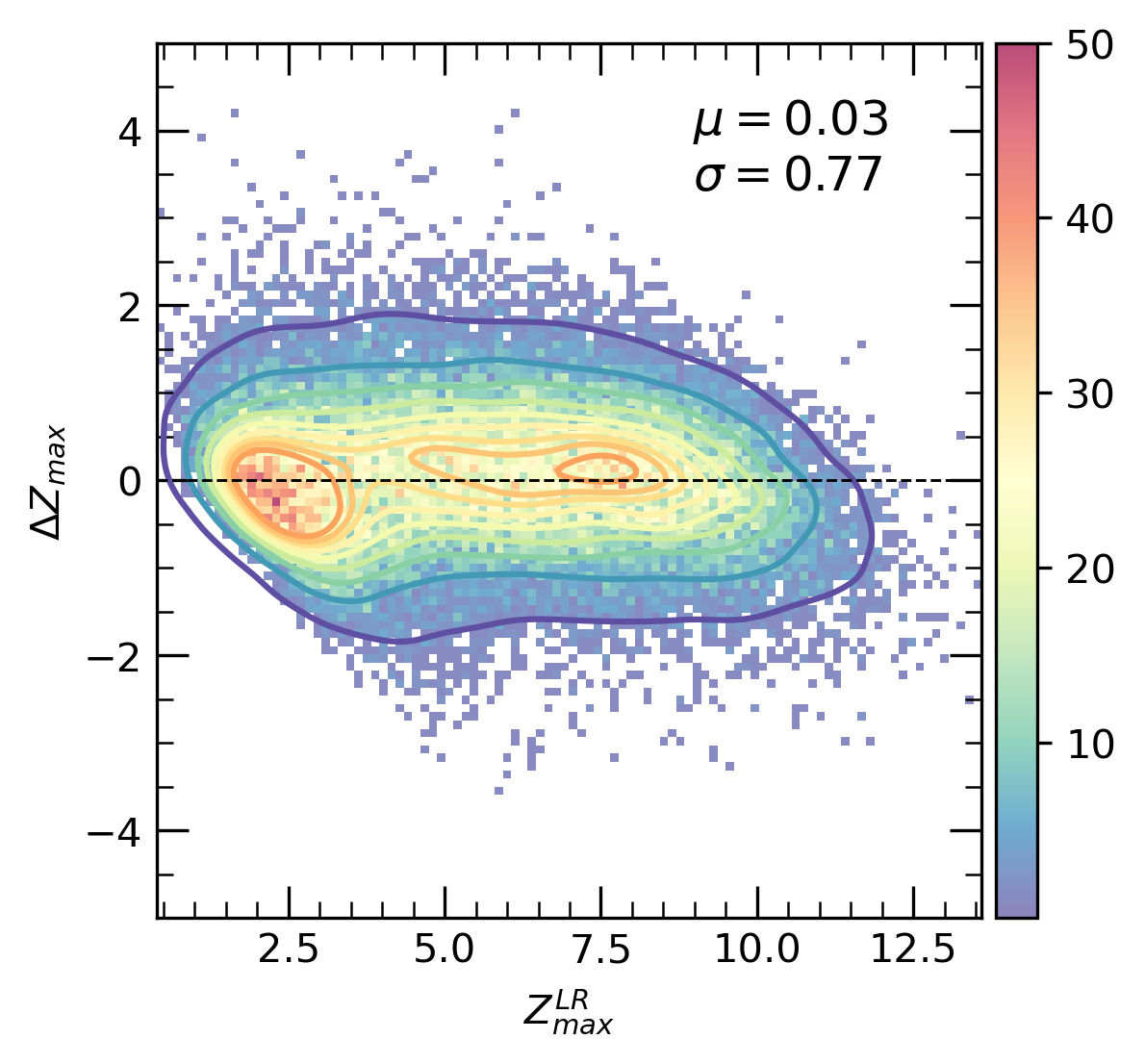}
        \caption{Superior bias}\label{}
    \end{subfigure}
    \begin{subfigure}{0.24\linewidth}\centering
        \includegraphics[width=0.95\linewidth]{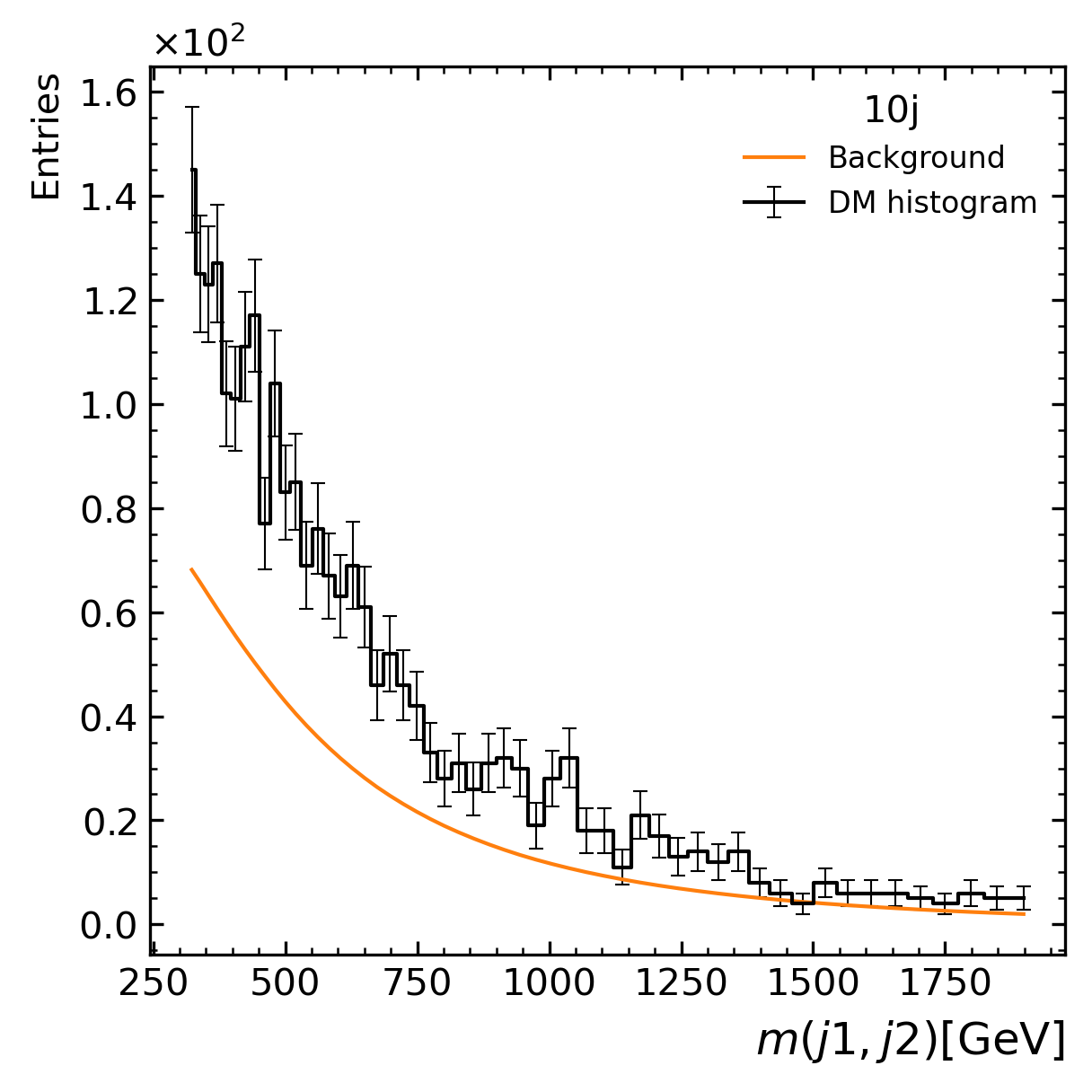}
        \includegraphics[width=\linewidth]{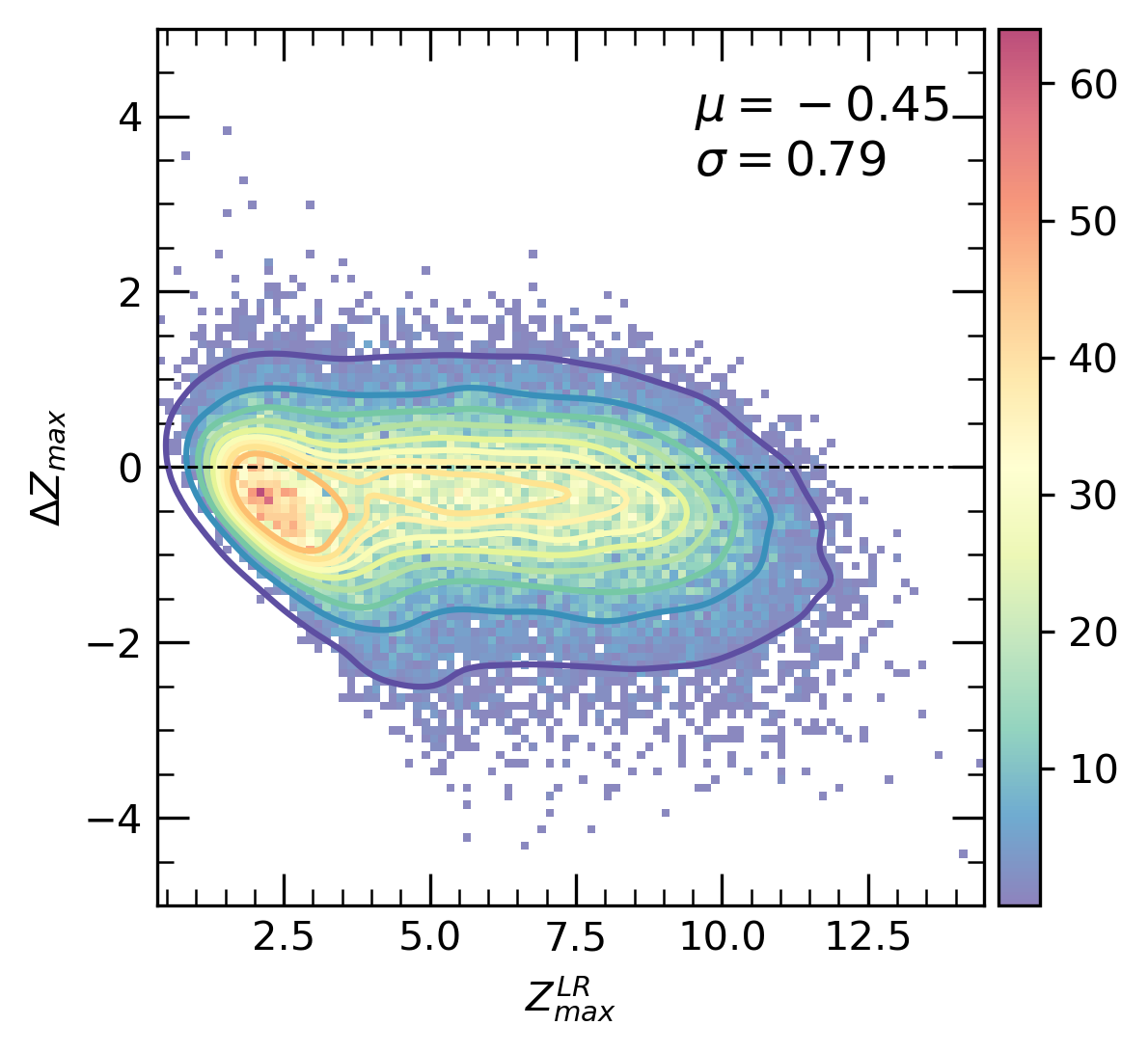}
        \caption{Inferior bias}\label{}
    \end{subfigure}
    \begin{subfigure}{0.24\linewidth}\centering
        \includegraphics[width=0.95\linewidth]{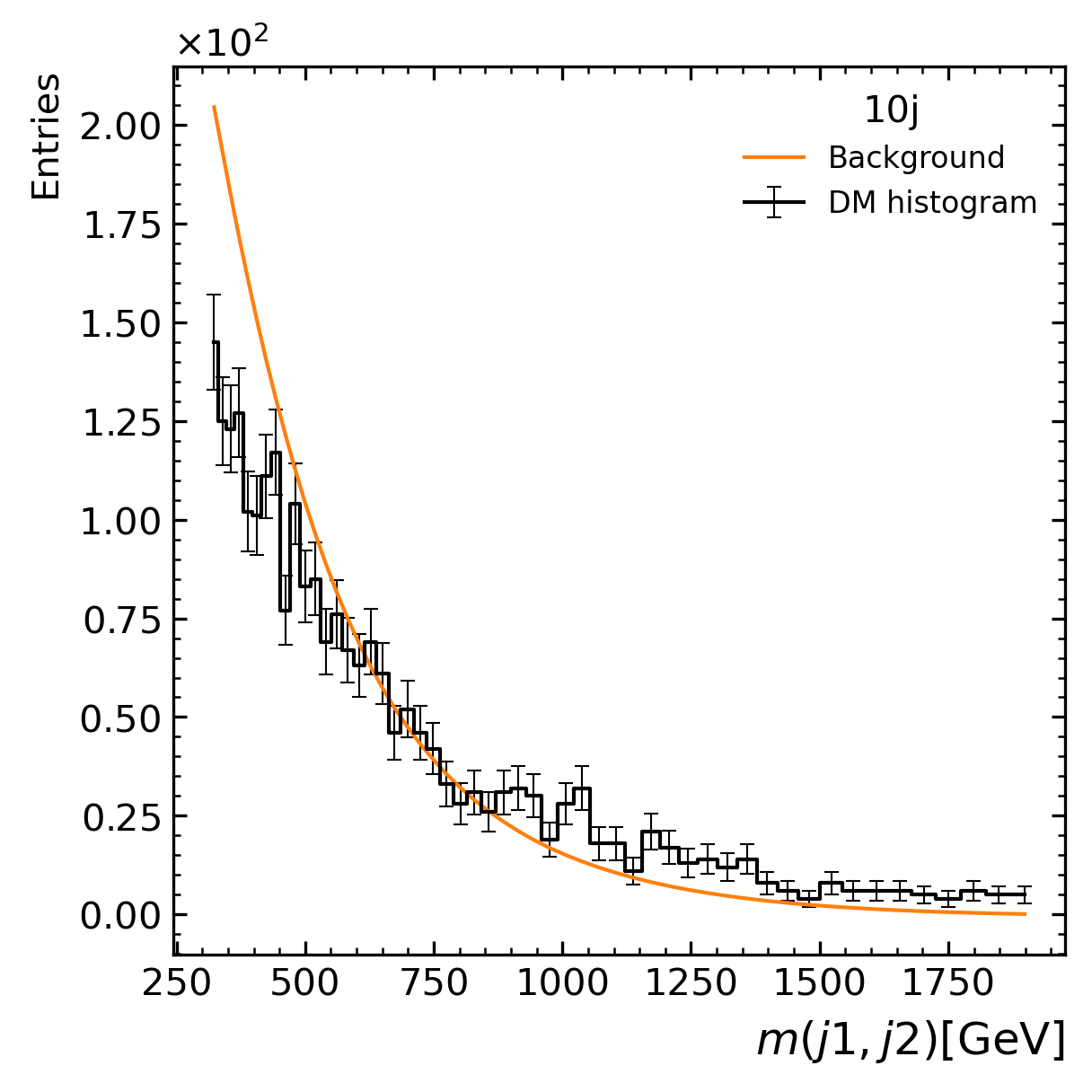}
        \includegraphics[width=\linewidth]{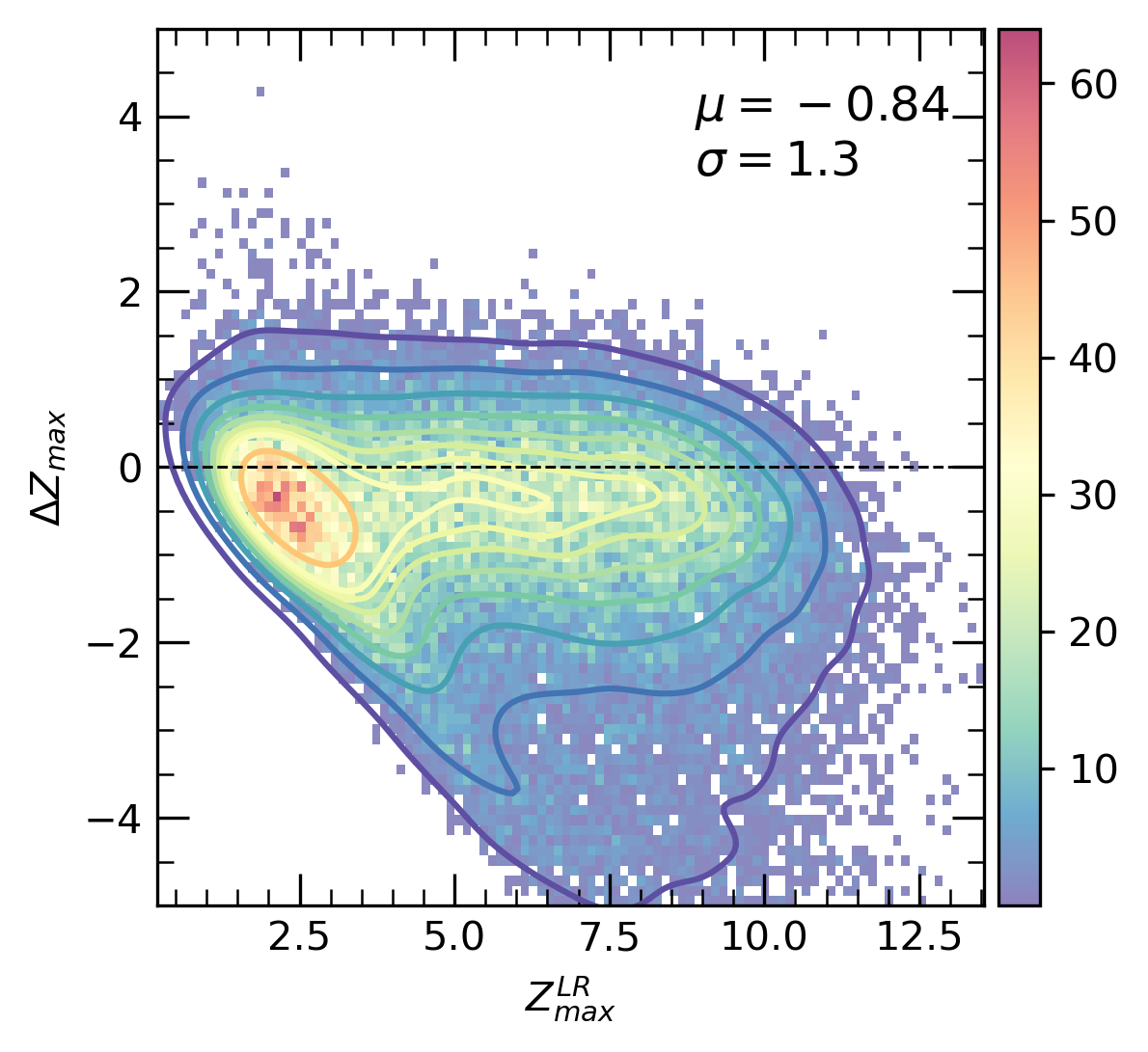}
        \caption{Standard bias}\label{fig:bias_standard}
    \end{subfigure}
    \begin{subfigure}{0.24\linewidth}\centering
        \includegraphics[width=0.95\linewidth]{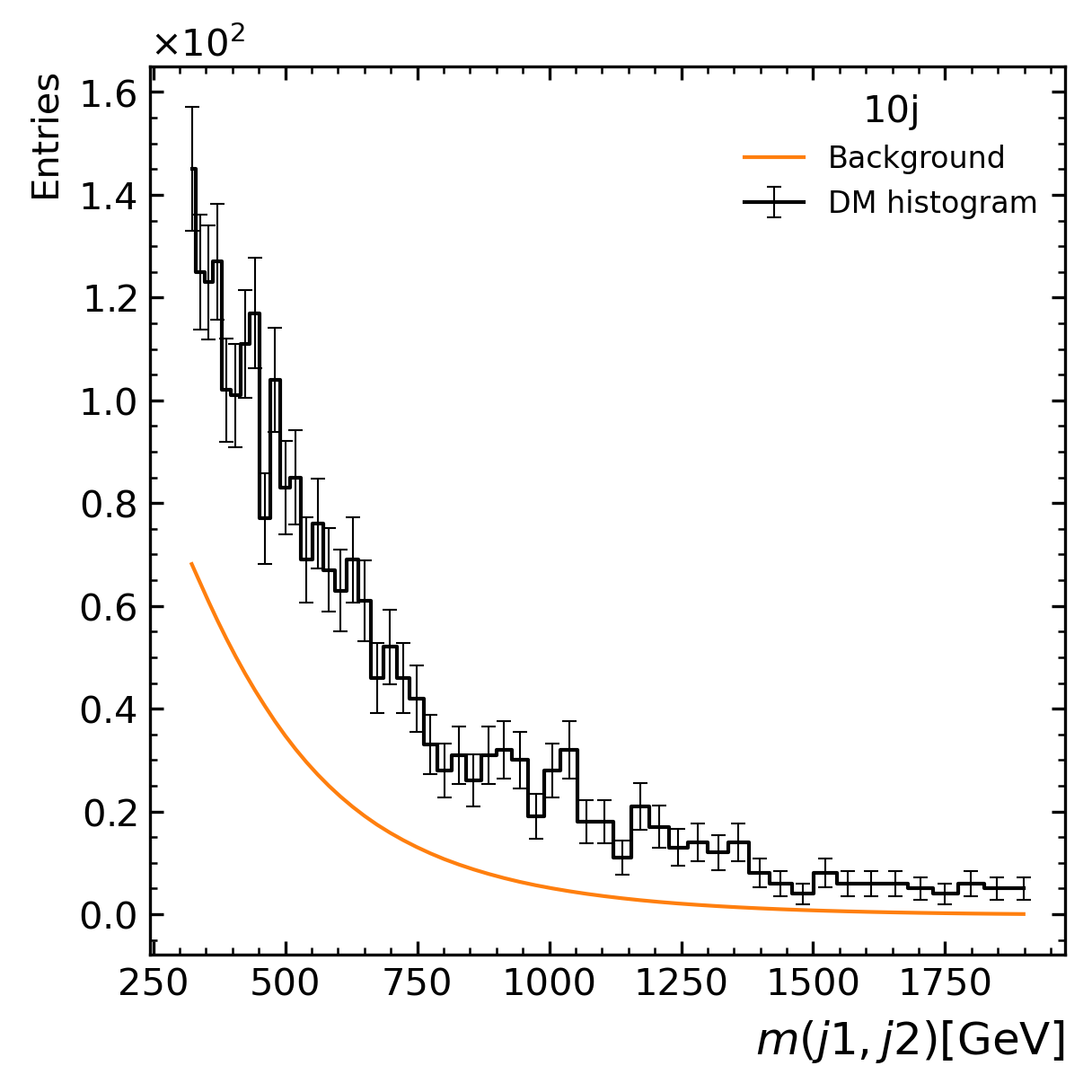}
        \includegraphics[width=\linewidth]{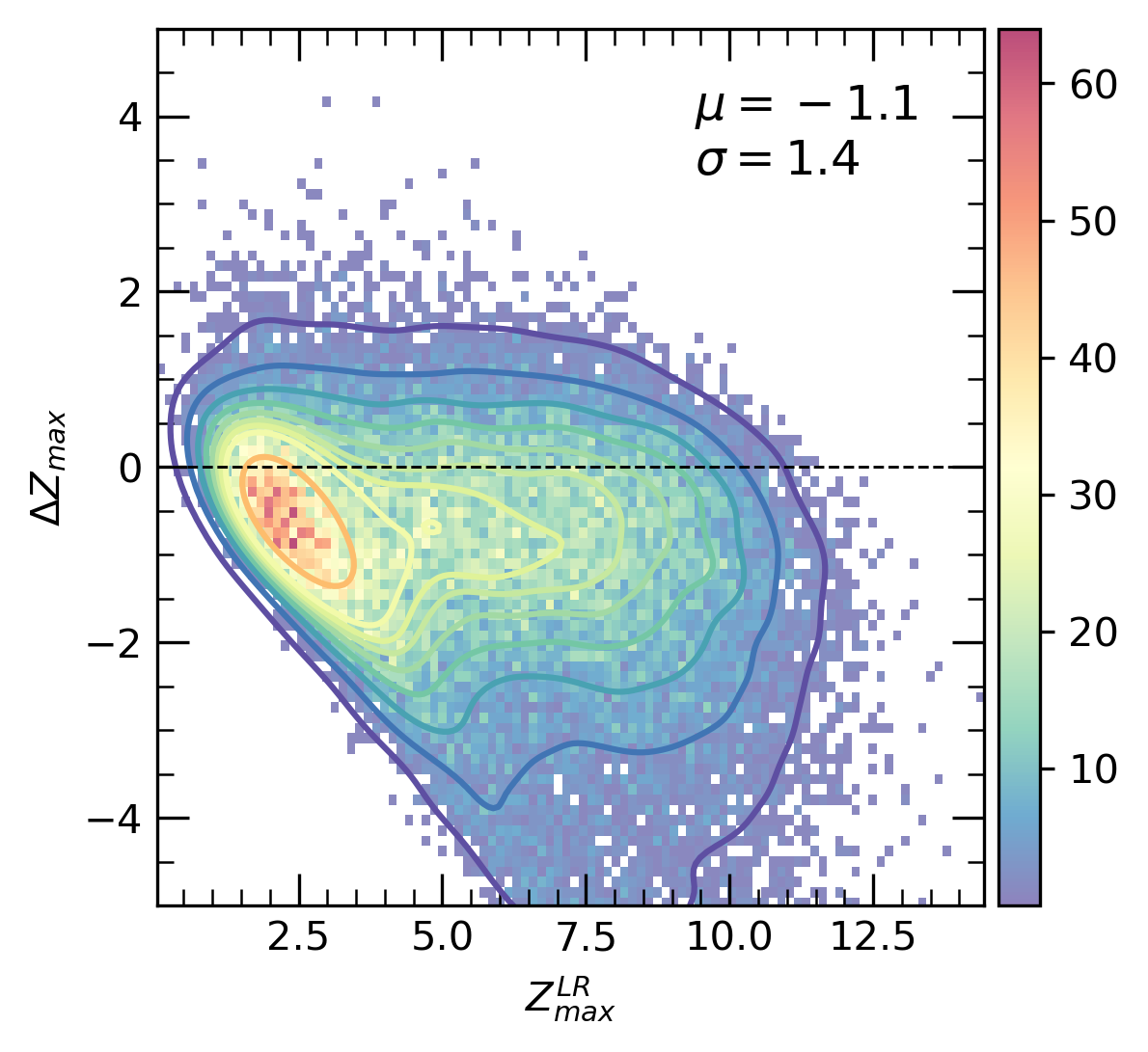}
        \caption{Reverse bias}\label{fig:bias_reverse}
    \end{subfigure}
   
  \caption{The top panels show the effect of four transformations on an example DM histogram with a selection requiring exactly 10 jets, and for which the mass is computed from the 2nd and 3rd leading jet four-vectors. The bottom panels show the distributions of \DZ~as a function of the target signal strength \ZLR\ over the four systematically-biased application sets.
  }
  \label{fig:systematic_bias}
\end{figure}
\begin{figure}[htbp]
\centering
    \begin{tabular}{cc}
\subfloat[]{\includegraphics[width=.45\textwidth]{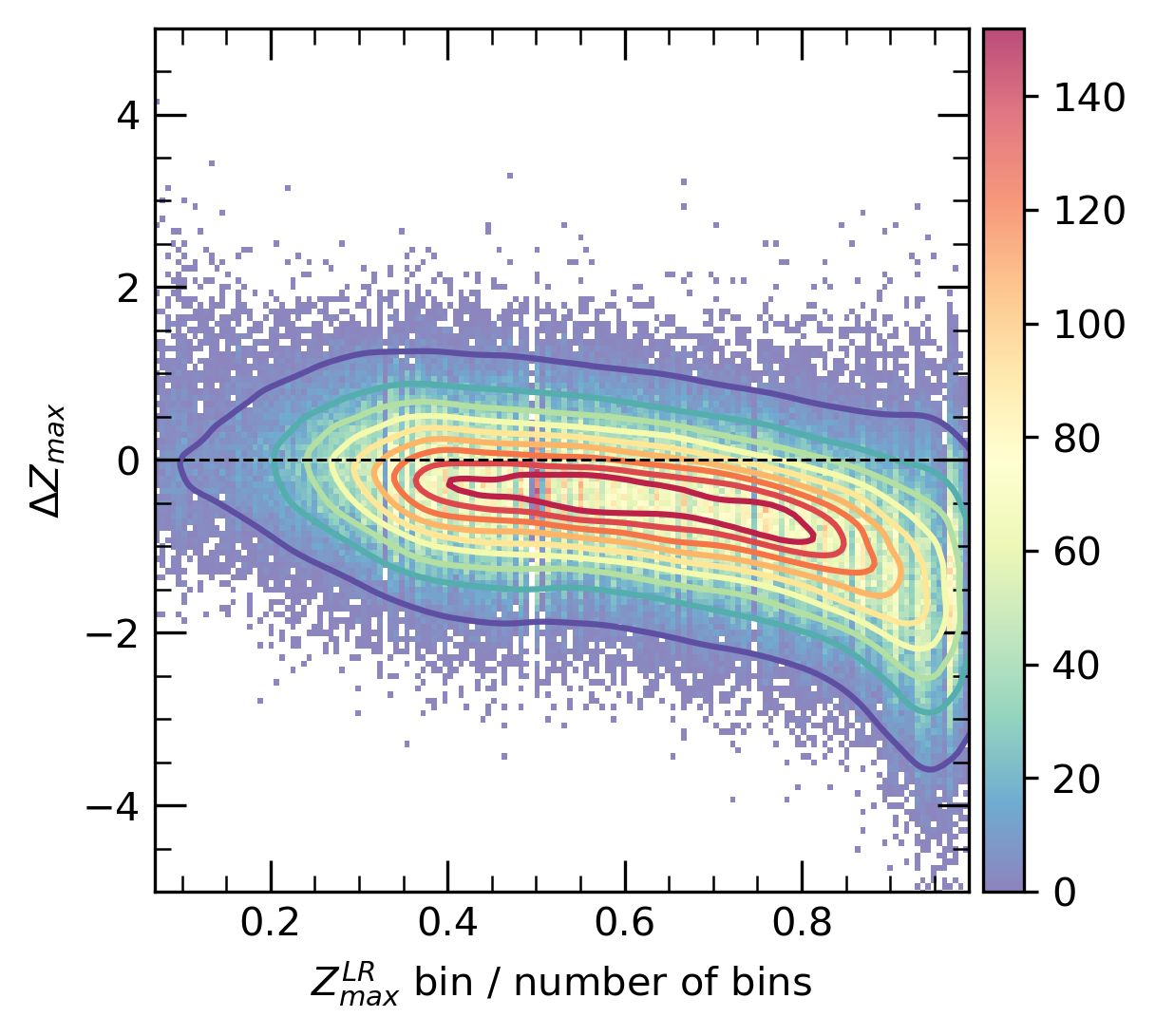}\label{fig:reverse-bins}}
 \subfloat[]{\includegraphics[width=.45\textwidth]{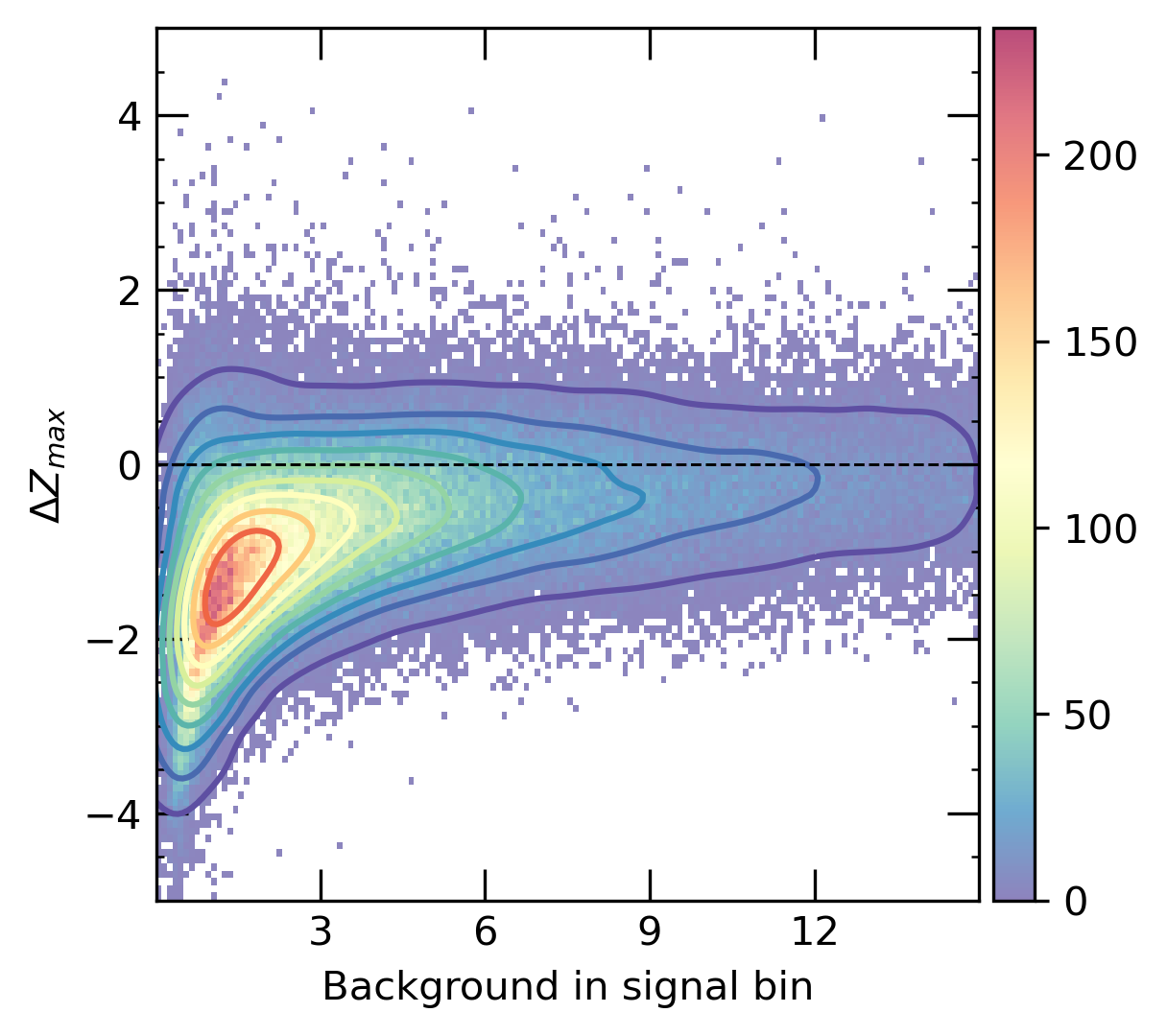}\label{fig:bkg_Sbin}}
   \end{tabular}
\caption{Distributions of \DZ\ for the reverse transformation as a function of (a) the true position of the injected signal, and (b) and number of background events in the bin with injected signal zoomed-in for low number of background events.} \label{fig:revers_bias_vs_bins}
\end{figure}

    The top panels of Figure~\ref{fig:systematic_bias} depicts the effect of each of the four transformations on an example DM histogram with a selection requiring exactly 10 jets, and for which the mass is computed from the 2nd and 3rd leading jet four-vectors. 
The bottom panel of Figure \ref{fig:systematic_bias} shows the performance of BumpNet, i.e. \DZ~as a function of \ZLR, on the four systematically biased application sets. The performance over the inferior and superior transformations is similar to the one over the baseline \ac{DM}-based dataset shown in Figure~\ref{fig:DiffZ_DM_dm}, highlighting BumpNet's ability to generalize over these new background shapes. 
 However, for the other two transformations, standard and reverse, BumpNet tends to underestimate the true LR test significance. This behavior is further explored in Figure~\ref{fig:revers_bias_vs_bins}, which illustrates the distributions of \DZ\ as a function of the relative position in the histogram (\subref{fig:reverse-bins}) and the number of background events under the signal peak (\subref{fig:bkg_Sbin}).
The biases are predominantly located in the high-mass regions of the histograms, which correspond to areas with sparse background statistics. Notably, both the standard and reverse transformations exhibit a pronounced suppression of events in these regions, as shown in the top panels of Figures~\ref{fig:bias_standard} and \ref{fig:bias_reverse}. This bias is hypothesized to arise from two key factors: BumpNet's training examples predominantly featured more background events in the high-mass regions, and the inherent constraint of Poisson distributions, where negative fluctuations are not possible. 
Together, these factors suggest that BumpNet can be thought of as overpredicting the background in these regions, which then leads to an underestimation of the significance.
Further work will be conducted to address and correct this bias, ensuring that BumpNet achieves consistent performance even in regions with sparse background statistics.

Another systematic evaluation of BumpNet's performance was conducted using a test dataset generated by stretching \ac{DM} distributions, as described in Section~\ref{sec:functions}. Despite the significant variability introduced by the new background shapes, no degradation in performance relative to the reported results was observed, demonstrating BumpNet's robustness in handling diverse data scenarios.

Overall, apart from the specific case of systematic biases that significantly suppress the background shapes in the high-mass region, BumpNet demonstrates excellent generalization to previously unseen background shapes. It is important to note that this bias is localized to regions with very sparse background statistics. When the number of background events under the signal peak exceeds a small threshold, BumpNet's performance consistently returns to its expected high standard, further underscoring its adaptability and reliability across a wide range of conditions.

\subsubsection{Sensitivity to varying signal widths}
\label{sec:perDiffWidth}

\begin{figure}[htbp]
\centering
    \begin{tabular}{cc}
\subfloat[]{\includegraphics[width=.3\textwidth]{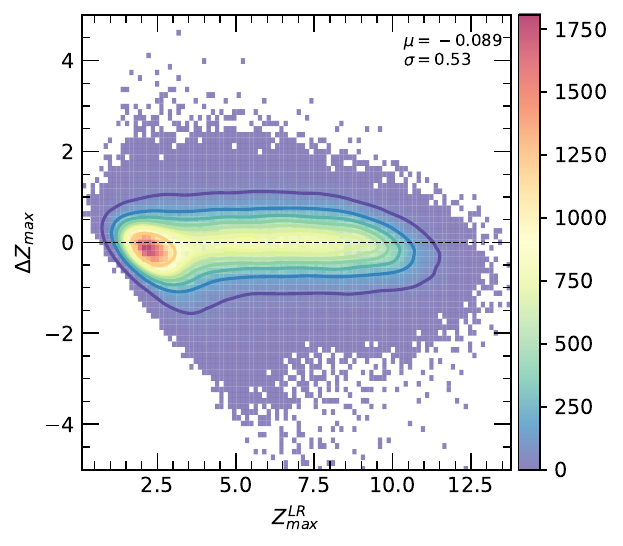}\label{fig:func_width1}}
\subfloat[]{\includegraphics[width=.3\textwidth]{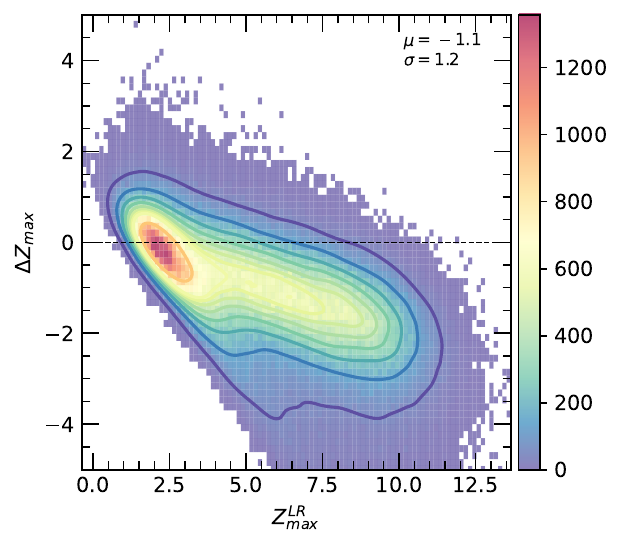}\label{fig:func_width2}}
\subfloat[]{\includegraphics[width=.3\textwidth]{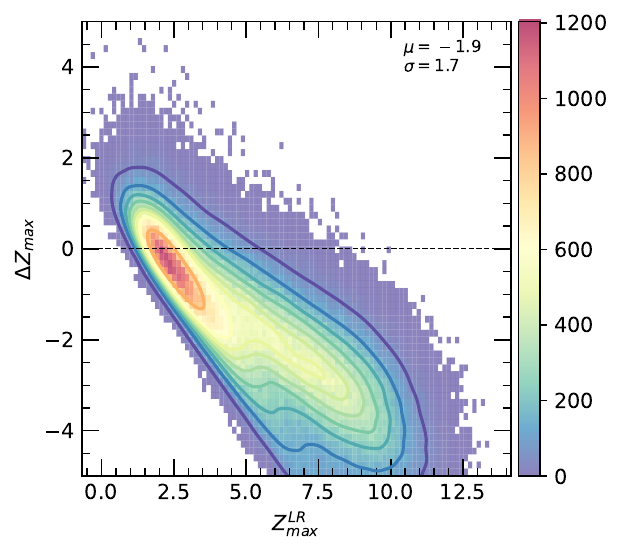}\label{fig:func_width3}}
    \end{tabular}
\caption{The difference between BumpNet prediction and the LR significance for dataset generated from functions with (\subref{fig:func_width1}) signal width on 1 bin, (\subref{fig:func_width2}) 2 bins and (\subref{fig:func_width3}) 3 bins.} \label{fig:funcDiffWidth}
\end{figure}

BumpNet was trained with injected Gaussian signals of 1-bin width, based on the assumption that it is optimized to detect narrow resonances and that the bin widths are calibrated to reflect the experimental resolution. However, due to factors like the dependency of resolution on object \( p_{\mathrm{T}} \) and \( \eta \) — which cannot be fully captured within a single mass bin — this calibration process is inherently imperfect. It is also useful to evaluate BumpNet’s sensitivity to broader resonances. To this end, a set of 1.5 million histograms was generated from smoothly falling functions, with Gaussian signals added at widths of 1, 2, and 3 bins. The difference between BumpNet's prediction and the LR significance is shown as a function of injected significance for each of the three signal widths in Figure~\ref{fig:funcDiffWidth}. 

As expected, since BumpNet was trained on 1-bin-wide signals, its performance degrades with increasing signal width. The predicted significance, primarily based on the height of the peak’s central region, tends to be lower than the LR significance, which also incorporates the peak’s tails. This result underscores the importance of the re-binning procedure discussed in Section~\ref{sec:DMsamples}, which is intended to mitigate detector resolution effects and ensure consistent signal widths. To address this sub-optimal performance on broader signals, future work will explore incorporating signals of varying widths into the training data.

It is also worth noting that, although BumpNet’s significance predictions  are less accurate and somewhat biased for broader signals, its primary purpose is to identify the presence of a bump rather than to determine its exact significance. As shown will be shown in Section~\ref{sec:resBSM}, BumpNet remains effective at identifying realistic \ac{BSM} signals injected into the \ac{DM} data, even when those signals are, on average, broader than one bin.

\subsubsection{Background shapes from ATLAS dilepton resonance search}
\label{sec:perDilep}

The ATLAS dilepton resonance search \cite{ATLAS:2019erb} used the following function to model the background shape:

\begin{equation}
\label{eq:HEP}
f_{\ell \ell}(m_{\ell\ell}) = f_{BW, Z}(m_{\ell \ell}) \cdot (1 - x^c)^b \cdot x^{\sum_{i=0}^{3} p_i \log(x)^i},
\end{equation}
where \( x = m_{\ell \ell}/\sqrt{s} \), and the parameters \( b \) and \( p_i \) (for \( i = 0, \ldots, 3 \)) are free in the fit to data, with independent values for the di-electron and di-muon channels. The parameter \( c \) is set to 1 for the di-electron channel and \( 1/3 \) for the di-muon channel. The function \( f_{BW, Z} (m_{\ell \ell}) \) represents a non-relativistic Breit–Wigner distribution centered at \( m_Z = 91.1876 \) GeV. As discussed in Ref.~\cite{ATLAS:2019erb}, this function was fitted to mass histograms with 1 GeV bin widths; hence, it was convolved with the resolution function extracted from Figure 15 (auxiliary material) of Ref.~\cite{ATLAS:2019erb} for histogram production.

\begin{figure}[htbp]
\centering
    \begin{tabular}{cc}
\subfloat[]{\includegraphics[width=.45\textwidth]{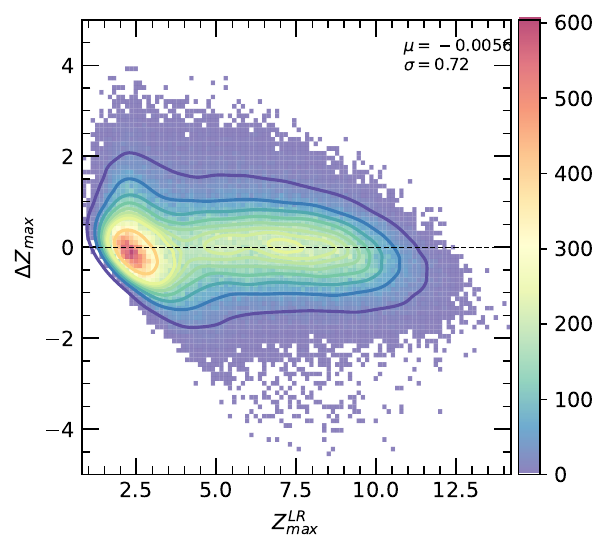}\label{fig:di-electron}}
\subfloat[]{\includegraphics[width=.45\textwidth]{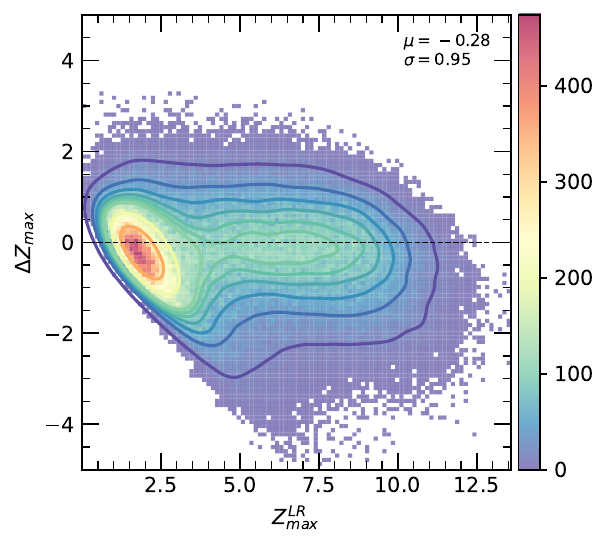}\label{fig:di-muon}}
   \end{tabular}
\caption{The difference between BumpNet prediction and the likelihood-ratio significance for dataset generated from the (\subref{fig:di-electron}) di-electron and (\subref{fig:di-muon}) di-muon backgrounds. The functional forms are taken from Ref.~\cite{ATLAS:2019erb}.} \label{fig:DiffZ_dilep}
\end{figure}

This background model was used to generate a BumpNet test dataset. Histograms of 98 and 28 bins were produced for the di-electron and di-muon distributions, respectively, ensuring a minimum of 10 entries per bin. 
The smaller number of bins in the muon channel is due to the poorer momentum resolution for muons compared to electrons at high-\pT, leading to significantly wider bin widths. 
Gaussian signals with a 1-bin width and statistical significance in the range of 1–10\,$\sigma$ were injected into the fluctuated distributions following the procedure detailed in Section~\ref{sec:methodology}. 
Figure~\ref{fig:DiffZ_dilep} displays the difference between the BumpNet prediction and the likelihood-ratio significance as a function of the injected significance for the di-electron (left) and di-muon (right) distributions. The results for the di-electron distributions are excellent, showing no bias and a variance comparable to that observed when tested on the original set of functions, highlighting BumpNet's flexibility in accurately predicting significance even with background functions it was not trained on. However, the results for the di-muon distributions are slightly worse due to their smaller histogram size (28 bins), which falls just below the 30-bin lower limit used in training. Furthermore smaller histograms provide less granularity, making it inherently harder for BumpNet to resolve subtle features like bumps.

\section{Performance over data and data-like signals}
\label{sec:performanceDataLike}

While BumpNet is shown to predict well the significance of Gaussian shaped signals, it is necessary to evaluate its performance also on realistic data. This is done by
 by applying BumpNet to the $H \rightarrow \gamma\gamma$ and high-mass di-electron and di-muon real data as well as exploiting simulated \ac{BSM} signals within the \ac{DM} framework.

\subsection{HEP data}
\label{sec:resHEPdata}

\begin{figure}[htbp]
\centering
    \includegraphics[width=.67\textwidth]{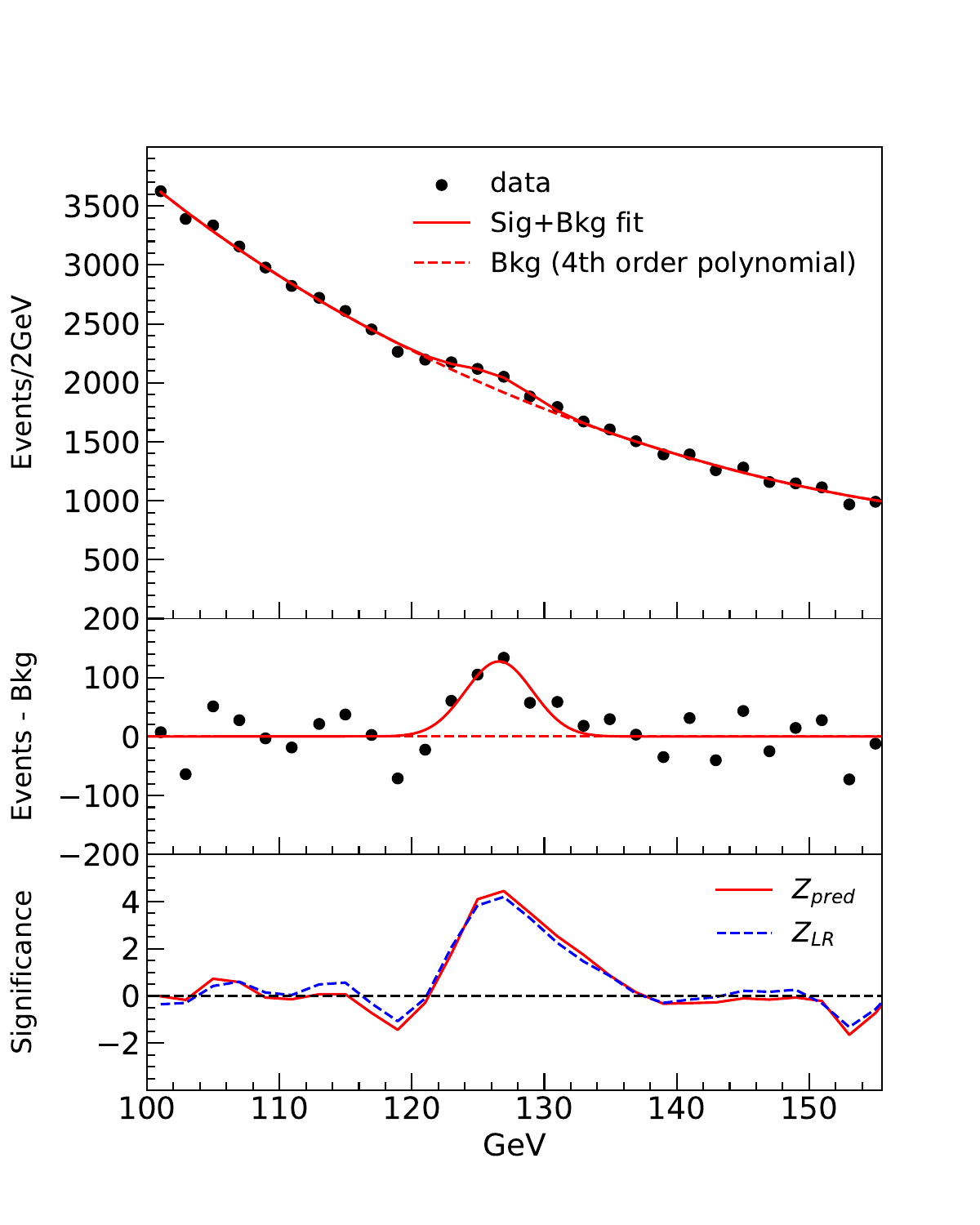}
\caption{BumpNet's significance prediction (bottom) over the ATLAS $H\rightarrow \gamma\gamma$ data (top). Data points are extracted from Figure 4 of Ref.~\cite{ATLAS:2012yve}.} 
\label{fig:higgsSearch}
\end{figure}

The data points from the \( H \rightarrow \gamma \gamma \) histogram were extracted from Figure 4 in the ATLAS Higgs discovery paper \cite{ATLAS:2012yve}. As described in the paper, the background was modeled using a 4th-order polynomial fit. These data points are shown in the top panel of Figure~\ref{fig:higgsSearch}. A Gaussian fit to the background-subtracted data is displayed in the middle panel, and the BumpNet prediction is compared with the bin-by-bin significance calculated using the LR test statistic in the bottom panel.

The Higgs signal is clearly visible, with BumpNet predicting a maximum significance of 4.5$\sigma$, consistent with the likelihood-ratio significance of 4.2$\sigma$, at the correct mass value.\footnote{Note that this likelihood-ratio significance does not correspond to the overall significance reported in Ref.~\cite{ATLAS:2012yve}, which is derived from a more comprehensive analysis including multiple signal regions. The significance quoted here is based solely on this single histogram.}

\begin{figure}[htbp]
\centering
    \begin{tabular}{cc}
\subfloat[]{\includegraphics[width=.49\textwidth]{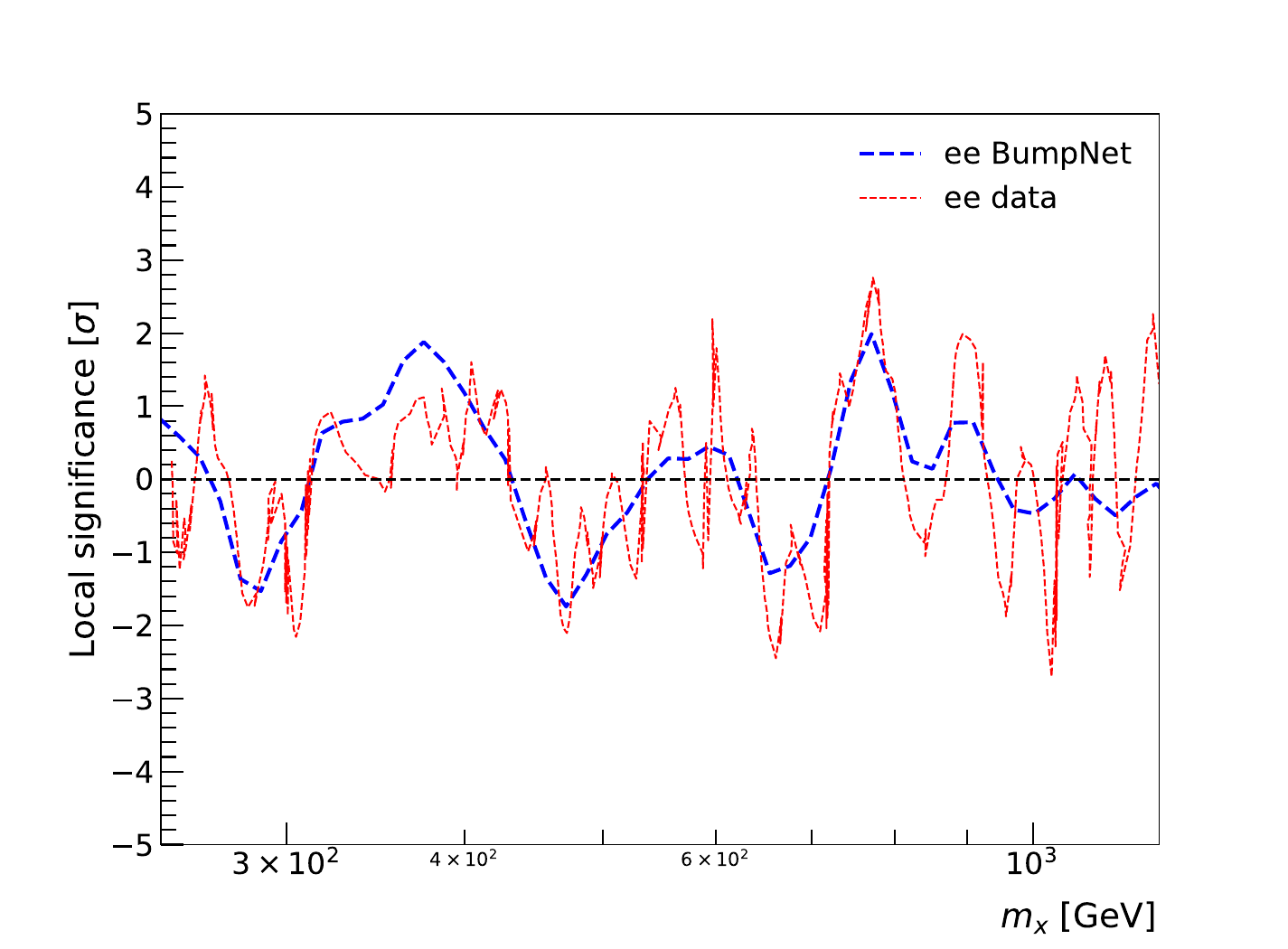}\label{fig:hep_ee}}
\subfloat[]{\includegraphics[width=.49\textwidth]{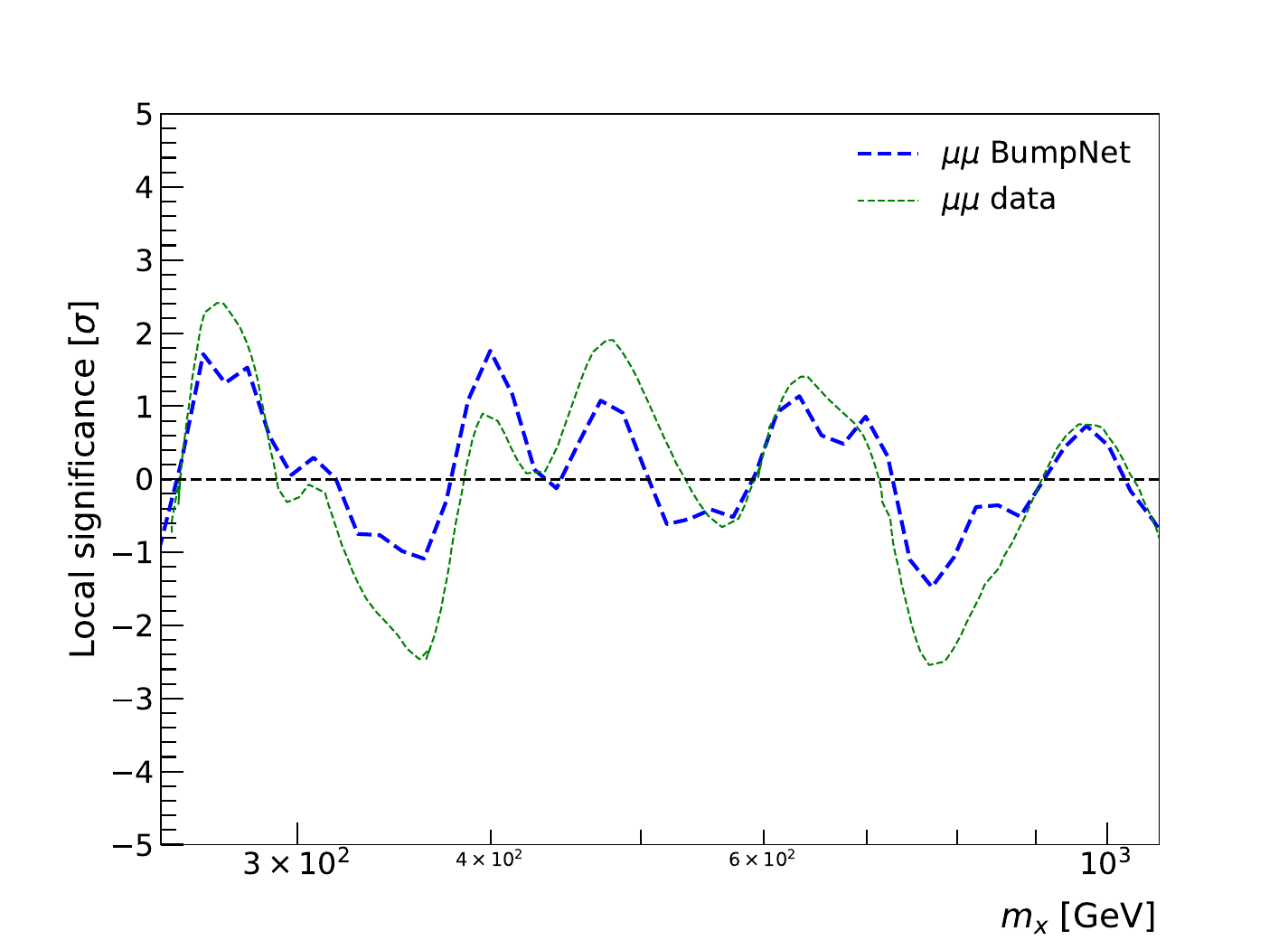}\label{fig:hep_mm}}
    \end{tabular}
\caption{Comparison of BumpNet’s predicted significance (dashed line) with the published significance (solid line) in the search for di-electron (left) and di-muon (right) resonances. The data points were extracted from Figure 2 of \cite{ATLAS:2019erb} and convolved with the di-electron and di-muon invariant-mass resolutions provided in Figure 15 of the same publication.} 
\label{fig:hep_ll}
\end{figure}

Figure~\ref{fig:hep_ll} compares the significance of the di-electron (left) and di-muon (right) invariant-mass distributions as reported in the ATLAS dilepton resonance search (dashed line) \cite{ATLAS:2019erb} with those predicted by BumpNet (solid line). Results are shown up to approximately 1 TeV, beyond which the event count drops below 10, leading to a slight degradation in BumpNet's performance, as discussed in Section~\ref{sec:perfDistDMsamples}.

BumpNet’s predictions closely follow the reported significance values, with any observed deviations well within BumpNet’s known variance.

\subsection{Dark Machines BSM signals}
\label{sec:resBSM}

In this section, we evaluate BumpNet in a realistic data analysis scenario by injecting simulated \ac{BSM} signals into the SM \ac{DM} samples presented in Section~\ref{sec:DMsamples}. This approach enables to assess BumpNet's performance on more realistic signal shapes, which may deviate from perfect Gaussian distributions due to factors such as combinatorial effects. We also discuss methods to address the unavoidable LEE that arises in realistic analyses employing BumpNet to scan a large number of mass histograms. Despite BumpNet's relatively low false-positive rate, some false-positive signals are expected to occur in such extensive analyses.

A set of BSM signals has been selected to provide a variety of final states and mass values. They are listed in Table~\ref{table:bsm_dm}, and include:
\begin{itemize}
\item Pair production of scalar leptoquarks with a mass of 600~GeV, each decaying to a $b$-quark and an electron or a muon \cite{Dorsner:2018ynv}. The three possible final states ($bebe$, $b\mu b\mu$, $beb\mu$ are tested.
    \item A low mass $Z$' of 50~GeV that decays to $\mu\mu$. The $Z$' can be emitted either from a $W$ or a $Z$ boson, producing two samples tested independently. Additional muon(s) are produced in the event from the decay of the $W$ ($W\to\mu\nu$) or $Z$ ($Z\to\mu\mu$) bosons. This sample was generated by the Dark Machines group \cite{Aarrestad:2021oeb}.
    \item Pair production R-parity violating (RPV) supersymmetric stop. Each stop has a mass of 1~TeV and decays to $be$ or $b\mu$. This sample was generated by the Dark Machines group \cite{Aarrestad:2021oeb}.
    \item $W'^\pm$ with a mass of 1.5~TeV that decays to $W^\pm Z$. Two samples are treated, one where $WZ \to qq\nu\nu$ and the other where $WZ \to \ell\nu qq$. 
\end{itemize} 
\begin{table}[ht]
\centering
\caption{BSM samples used to test the performance of BumpNet with simulated Gaussian signals injected into SM \ac{DM} samples.}
\label{table:bsm_dm}
\begin{tabular}{c|c|c|c}
\hline
\textbf{BSM Particle}      & \textbf{Mass (GeV)} & \textbf{Decay Channel} & \textbf{DM Channel} \\ \hline
Leptoquark (pair) & 600        & $b \mu b \mu$       & 2b         \\ \hline
Leptoquark (pair) & 600        & $b e b e$       & 2b         \\ \hline
Leptoquark (pair) & 600        & $b e b \mu$       & 2b         \\ \hline
$Z$' (3-muon events)               & 50         & $\mu \mu$    & 2b         \\ \hline
$Z$'  (4-muon events)              & 50         & $\mu \mu$    & 2b         \\ \hline
Stop (pair)       & 1000       & $b\ell$      & 3          \\ \hline
$W$'                & 1500       & $qq \nu \nu$    & 3          \\ \hline
$W$'                & 1500       & $\ell \nu qq$  & 3         \\ \hline
\end{tabular}
\end{table}
The leptoquarks and $W'$ samples have been generated privately using the  same configurations as the \ac{DM} project, including a fast detector simulation provided with DELPHES \cite{deFavereau:2013fsa}.
These BSM signals are  independently added on top of the \ac{DM} SM histograms, creating independent "data" samples of histograms to analyze, each corresponding to approximately 10~fb$^{-1}$ of LHC data. For the privately generated samples, the production cross-section of the BSM particles was selected such as providing a statistical significance of at least 5$\sigma$ for at least one mass histogram. As indicated in Table~\ref{table:bsm_dm}, the BSM samples have been added either to the \ac{DM} channels 2b or 3. 
 The stringent kinematic criteria of channel 3, particularly its $H_\mathrm{T}^\mathrm{jets} > 600 \mathrm{\ GeV}$ cut, resulted in severe sculpting of a small subset of 25 mass histograms, rendering them incompatible for processing by BumpNet; these have been excluded from the application sample. The resulting application sets are constituted of 8,104 and 31,642 histograms for channels 2b and 3, respectively.

\begin{figure}
     \centering
     \begin{subfigure}[b]{0.32\textwidth}
         \centering
         \includegraphics[width=\textwidth]{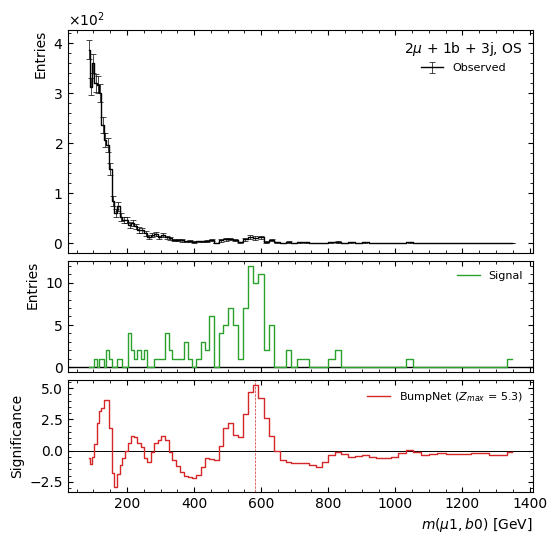}
         \caption{Two $LQ\to b\mu$}
         \label{fig:sub_bsm_examples_a}
     \end{subfigure}
     \hfill
     \begin{subfigure}[b]{0.32\textwidth}
         \centering
         \includegraphics[width=\textwidth]{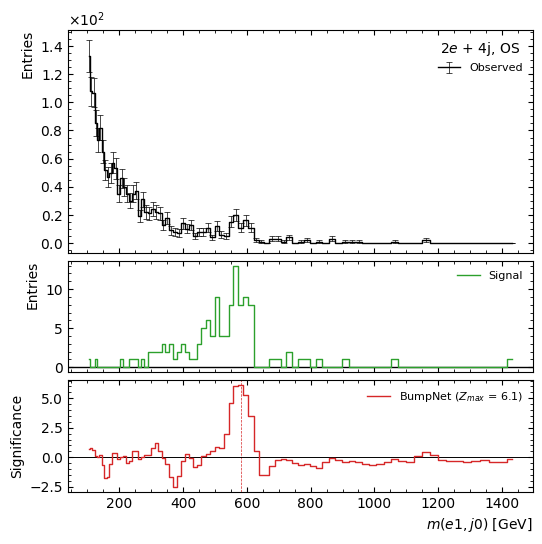}
         \caption{Two $LQ\to be$}
         \label{fig:sub_bsm_examples_b}
     \end{subfigure}
     \hfill
     \begin{subfigure}[b]{0.32\textwidth}
         \centering
         \includegraphics[width=\textwidth]{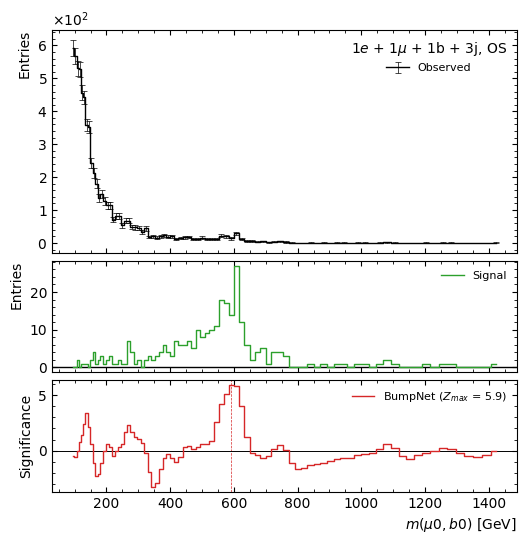}
         \caption{$LQ\to b\mu$, $LQ\to be$}
         \label{fig:sub_bsm_examples_c}
     \end{subfigure}
     \begin{subfigure}[b]{0.32\textwidth}
         \centering
         \includegraphics[width=\textwidth]{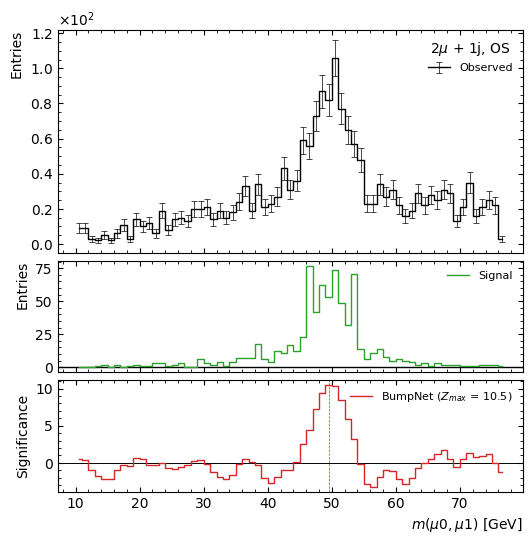}
         \caption{$Z'\to \mu\mu$}
         \label{fig:sub_bsm_examples_d}
     \end{subfigure}
     \hfill
     \begin{subfigure}[b]{0.32\textwidth}
         \centering
         \includegraphics[width=\textwidth]{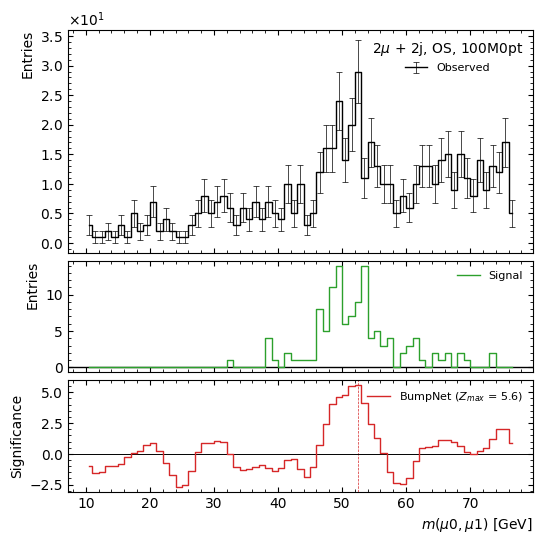}
         \caption{$Z'\to \mu\mu$}
         \label{fig:sub_bsm_examples_e}
     \end{subfigure}
     \hfill
     \begin{subfigure}[b]{0.32\textwidth}
         \centering
         \includegraphics[width=\textwidth]{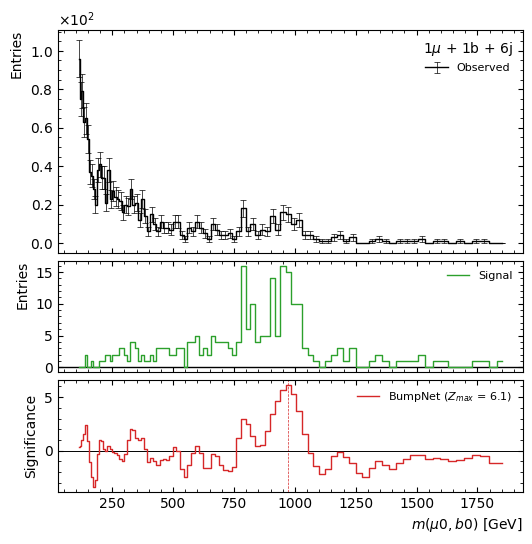}
         \caption{Two RPV Stop$ \to b\ell$}
         \label{fig:sub_bsm_examples_f}
     \end{subfigure}
     \begin{subfigure}[b]{0.32\textwidth}
         \centering
         \includegraphics[width=\textwidth]{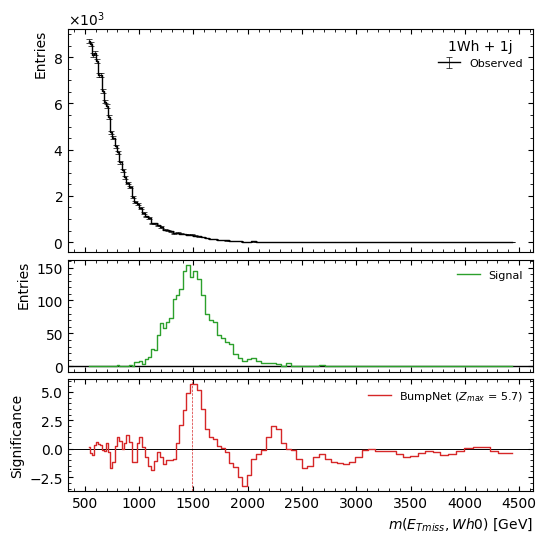}
         \caption{$W'\to qq\nu\nu$}
         \label{fig:sub_bsm_examples_g}
     \end{subfigure}
     \hfill
     \begin{subfigure}[b]{0.32\textwidth}
         \centering
         \includegraphics[width=\textwidth]{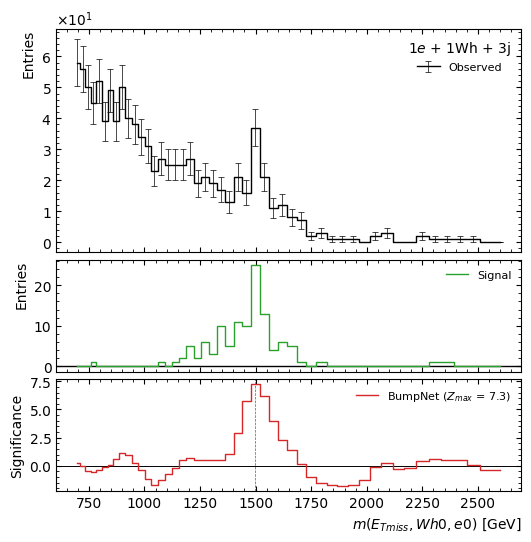}
         \caption{$W'\to \ell \nu qq$}
         \label{fig:sub_bsm_examples_h}
     \end{subfigure}
     \hfill
     \begin{subfigure}[b]{0.32\textwidth}
         \centering
         \includegraphics[width=\textwidth]{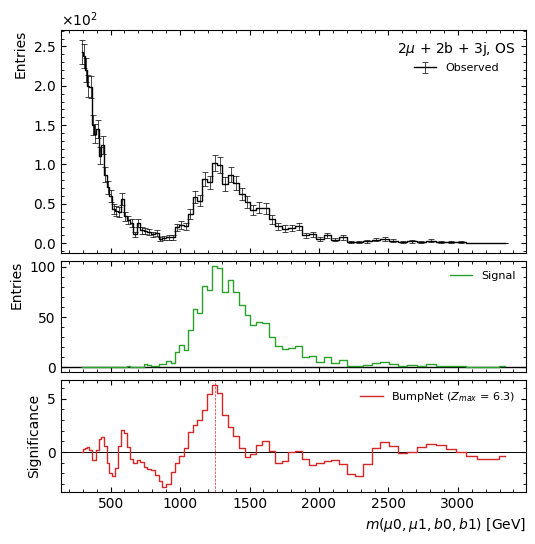}
         \caption{$LQ$ (pair) $\to b\mu b\mu$}
         \label{fig:sub_bsm_examples_i}
     \end{subfigure}
        \caption{Examples of histograms produced by BSM samples overlaid over SM background for \ac{DM} samples in which BumpNet predicted a significance $> 5\sigma$. The examples show events containing the objects shown above the histograms. For example, "$2\mu + 1b + 3j, OS$" from Figure (\subref{fig:sub_bsm_examples_a}) shows events containing 2 muons of opposite charges (OS), one $b$-jet and exactly 3 jets. Figures (\subref{fig:sub_bsm_examples_a}), (\subref{fig:sub_bsm_examples_b}) and (\subref{fig:sub_bsm_examples_c}) respectively come from the $b \mu b\mu$ , $b e b e$ and $b e b \mu$ LQs decay channels. Figures (\subref{fig:sub_bsm_examples_d}) and (\subref{fig:sub_bsm_examples_e}) respectively come from the 3$\mu$ and 4$\mu$ final state $Z'$, where in the name of the selection "100M0pt" means that the leading muon is required to have a $p_T$ above $100$ GeV. In Figure (\subref{fig:sub_bsm_examples_g}) and (\subref{fig:sub_bsm_examples_h}), "Wh" refers to hadronically decaying $W$ or $Z$ bosons that are reconstructed in a single large-radius jet. Figure~(\subref{fig:sub_bsm_examples_i}) shows the invariant-mass corresponding to 2 LQs. The chosen objects from the signature contributing to the shown invariant-mass are noted in the bottom right corner of each plot. The number of events purely due to the BSM particles is shown in green and the significances predicted by BumpNet are shown in red.}
        \label{fig:samples}
\end{figure}

Each of the \ac{BSM} signals listed in Table~\ref{table:bsm_dm} has at least one mass histogram with a predicted statistical significance of $5\sigma$ as determined by BumpNet. Figure~\ref{fig:samples} shows one example of such histograms per BSM model. In each of these histograms, the selections, the combinations of objects for the invariant-mass calculation, and the position of the signal peak are all consistent with the expected experimental signature of the corresponding \ac{BSM} signal\footnote{We note that all signal histograms for both 50~GeV $Z'$ models are found in events with two muons, despite the fact that these events are expected to contain either three or four muons. This is presumably because the  muons in such events are soft and likely to fail the $p_\mathrm{T}> 15$~GeV \ac{DM} selection.}. 
For some samples where the \ac{BSM} particles are produced in pairs and the signal strength is particularly prominent, a signal is even observed at twice the mass of the \ac{BSM} particle. Such an example is shown in Figure~\ref{fig:sub_bsm_examples_i}, where a signal is found at 1.2~TeV in the invariant-mass of all objects in the event originating from the decay of two 600 GeV leptoquarks. 

Despite this success in finding true positive signals, a challenge arises when using BumpNet to scan a large number of mass histograms. Even with its relatively low false-positive rate, the large LEE will unavoidably result in a non-zero number of histograms featuring false-positive signals. When BumpNet is applied to the 8,104 and 31,642 histograms of channels 2b and 3, respectively, in the absence of injected \ac{BSM} signals, 25 and 28 histograms exhibit a maximal statistical significance above $5,\sigma$, corresponding to a false-positive rate on the order of 0.1\%. To mitigate the LEE, a potential strategy for analyzing LHC data could involve initially unblinding only half of the dataset to identify histograms with maximal significance exceeding a defined threshold. These candidate excesses would then be validated by examining the remaining half of the data after unblinding.

\begin{figure}[htbp]
    \centering
    \begin{subfigure}[b]{0.45\textwidth}
        \centering
        \includegraphics[width=\textwidth]{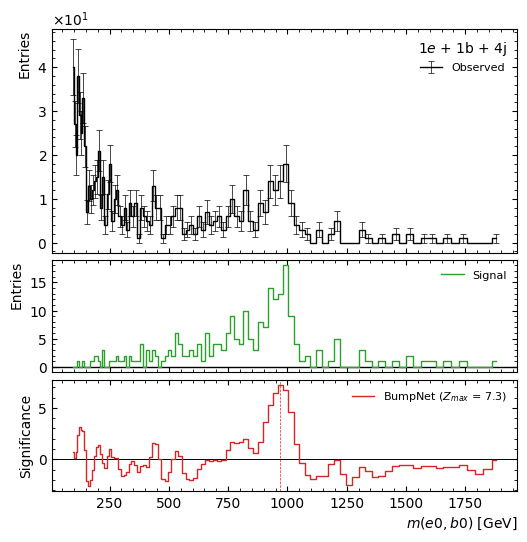}
        \caption{Stop $\to be$}
        \label{fig:e0b0}
    \end{subfigure}
    \hfill
    \begin{subfigure}[b]{0.45\textwidth}
        \centering
        \includegraphics[width=\textwidth]{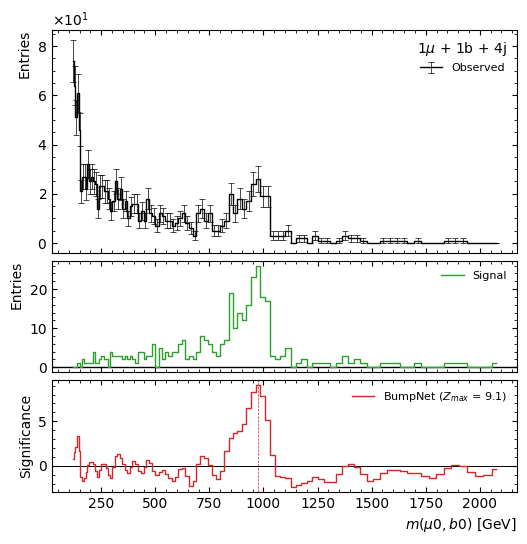}
        \caption{Stop $\to b\mu$}
        \label{fig:mu0b0}
    \end{subfigure}
    \vskip\baselineskip
    \begin{subfigure}[b]{0.45\textwidth}
        \centering
        \includegraphics[width=\textwidth]{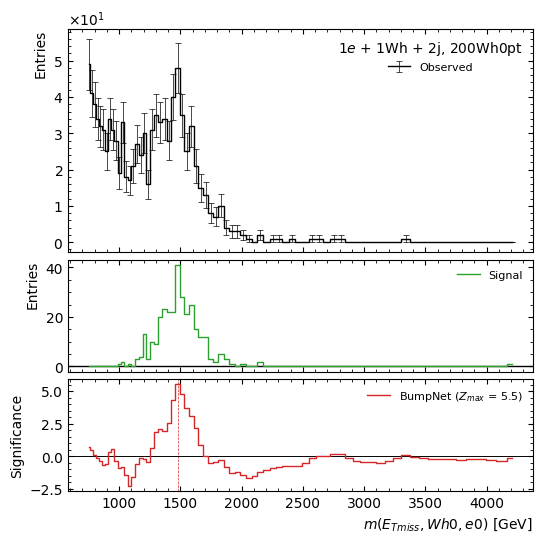}
        \caption{$W' \to \ell\nu qq$, exclusive 2-jet events}
        \label{fig:18684}
    \end{subfigure}
    \hfill
    \begin{subfigure}[b]{0.45\textwidth}
        \centering
        \includegraphics[width=\textwidth]{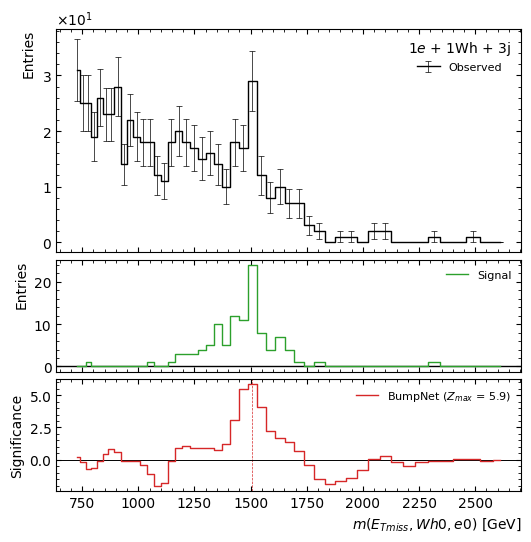}
        \caption{$W' \to \ell\nu qq$, exclusive 3-jet events}
        \label{fig:19064}
    \end{subfigure}
    \caption{Examples of physics correlations between histograms produced by BSM samples overlaid over SM background for \ac{DM} samples. Figures~(\subref{fig:e0b0}) and (\subref{fig:mu0b0}) show events containing four jets plus one $b$-jet plus either one electron (\subref{fig:e0b0}) or one muon (\subref{fig:mu0b0}) in \ac{DM} events containing RPV stop+ SM backgrounds. The invariant-mass plotted for these two figures is between the leading lepton and $b$-jet. Figures~(\subref{fig:18684}) and (\subref{fig:19064}) show events containing one electron, one boosted hadronic $W/Z$ boson (Wh) plus either two (\subref{fig:18684}) or three jets (\subref{fig:19064}) in \ac{DM} events containing $W'$ + SM backgrounds. For events with two jets, the boosted hadronic $W$ boson is required to have a $p_\mathrm{T}$ above 200~GeV. The invariant-mass plotted for these two figures is between the leading electron, boosted $W/Z$ boson and $E_\mathrm{T}^\mathrm{miss}$. The number of events purely due to the BSM sample are shown in green and the  significance predicted by BumpNet are shown in red.}
    \label{fig:DM_phys_corr}
\end{figure}

Another method involves exploiting physical correlations between histograms that feature a signal. Observing signals at the same mass value and for the same object combination in statistically uncorrelated histograms is highly unlikely to result from random false-positive fluctuations, suggesting the presence of a real signal. This is illustrated in Figure~\ref{fig:DM_phys_corr} with two examples. In the first example, Figures~\ref{fig:e0b0} and \ref{fig:mu0b0} show events containing four jets, one $b$-jet, and either one electron (\subref{fig:e0b0}) or one muon (\subref{fig:mu0b0}) in \ac{DM} events with RPV stop plus SM backgrounds. The invariant-mass plotted is between the leading lepton and the $b$-jet, corresponding to the expected products of the RPV stop decay. BumpNet detects clear signals at the same mass value in both plots. In the second example, Figures~\ref{fig:18684} and \ref{fig:19064} present events containing one electron, one boosted hadronic $W$ boson (denoted "Wh"), and either exclusively two (\subref{fig:18684}) or three jets (\subref{fig:19064}) in \ac{DM} events with $W'\to\ell\nu qq$ plus SM backgrounds. The invariant-mass plotted is between the leading electron, the boosted $W/Z$ boson, and $E_\mathrm{T}^\mathrm{miss}$, corresponding to the expected products of the $W'\to W Z$ decay, where $Z\to qq$ is reconstructed as a single large-radius jet. In both cases, the histograms are statistically uncorrelated, yet clear excesses are observed at the same mass value for the same decay products. Such coincidences are highly unlikely to arise from pure SM fluctuations and strengthen the case for a real signal.

A "Global Analysis Algorithm" (GAA) has been developed to detect such physical correlations. The GAA starts with a list of "seed" histograms that have a maximal significance predicted by BumpNet above a certain threshold, which is selected to be $5\sigma$ in this paper. It then iterates through all other histograms with a maximal significance exceeding another, potentially lower threshold, within the same mass region as the seed histogram (in this paper, $5\sigma$ is also used for that threshold). A tolerance of two histogram bins is used to define a "family" of histograms with an excess in the same mass region, as the bin width of the mass histograms roughly approximates the experimental mass resolution (see Section~\ref{sec:DMsamples}).
The GAA then examines all histograms in a family to determine whether there is a common combination of objects of the same type used in computing the invariant-mass (e.g., the invariant-mass of a lepton and a $b$-jet). For example, electrons and muons are considered objects of the same type, as are $b$-jets and generic jets. Histograms that do not share the same object combination for the mass calculation as the majority in the family are rejected.
Furthermore, if pairs of histograms within a family are statistically correlated—which can occur due to some selections being inclusive—only one of the correlated histograms is retained. Only histograms belonging to families that survive the GAA proceed to further analysis.

\begin{figure}[htbp]
    \centering
    \begin{subfigure}[b]{0.45\textwidth}
        \centering
        \includegraphics[width=\textwidth]{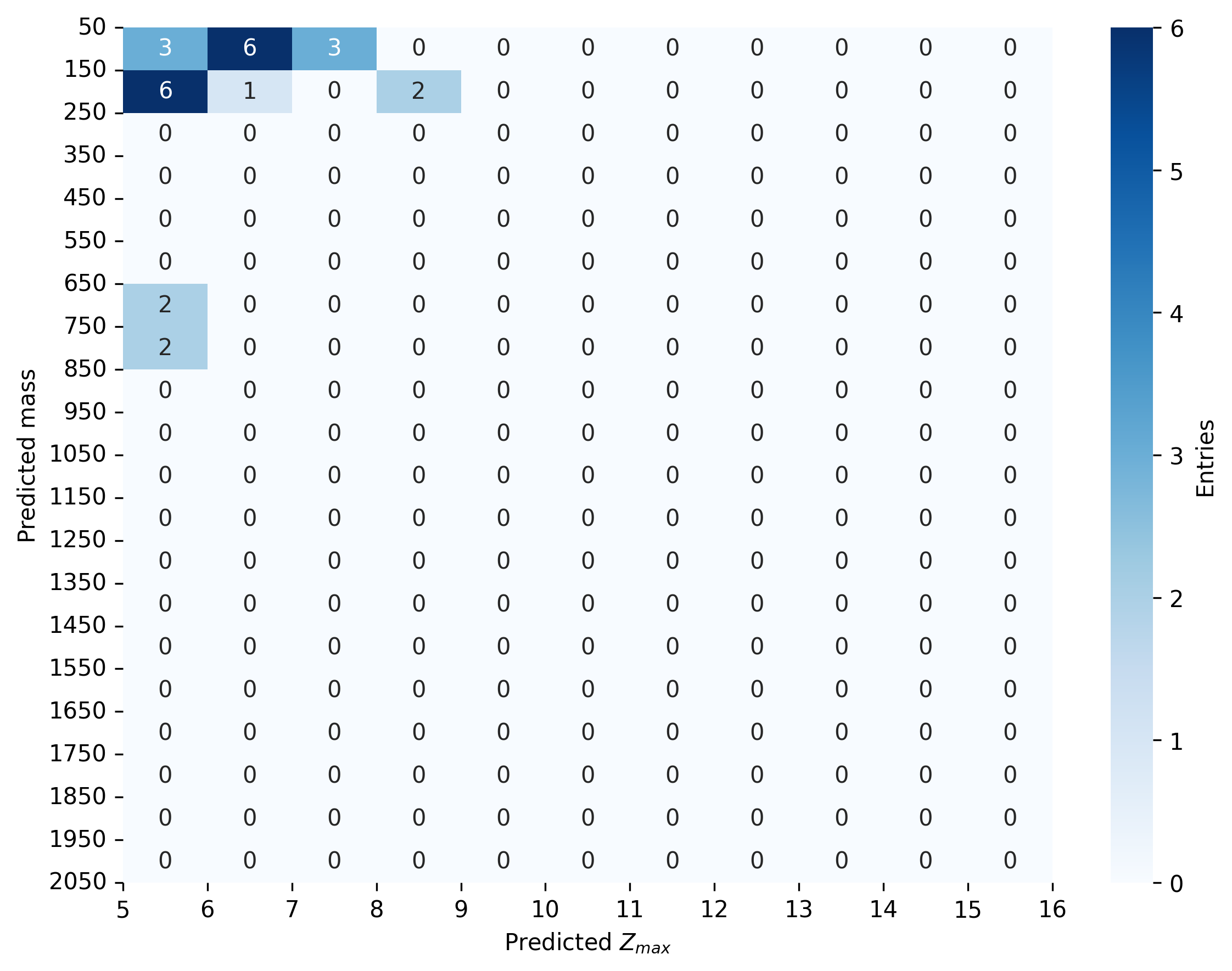}
        \caption{Channel 2b}
        \label{heatmap_ch2b}
    \end{subfigure}
    \hfill
    \begin{subfigure}[b]{0.45\textwidth}
        \centering
        \includegraphics[width=\textwidth]{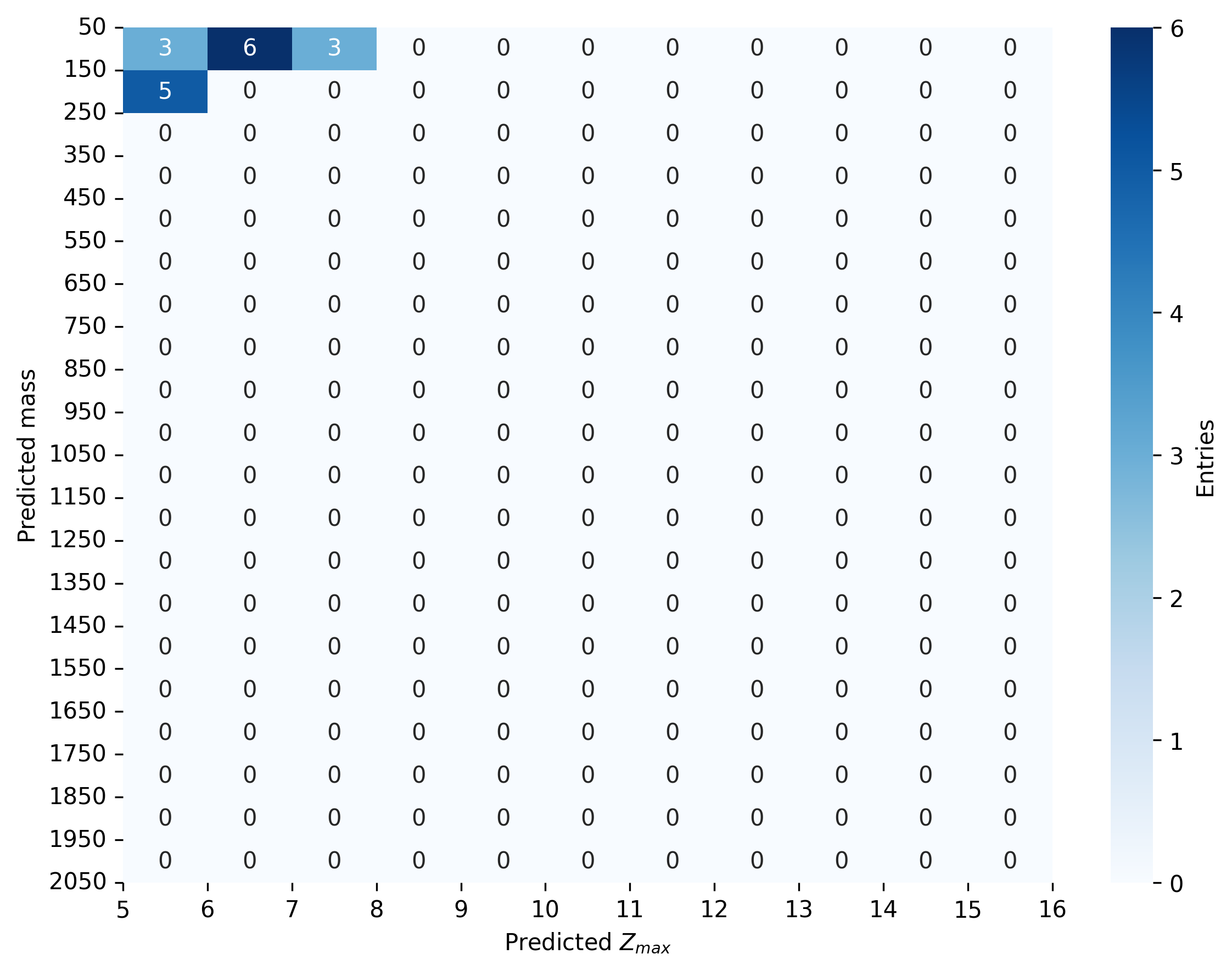}
        \caption{Channel 2b after GAA}
        \label{heatmap_ch2b_gaa}
    \end{subfigure}
    \vskip\baselineskip
    \begin{subfigure}[b]{0.45\textwidth}
        \centering
        \includegraphics[width=\textwidth]{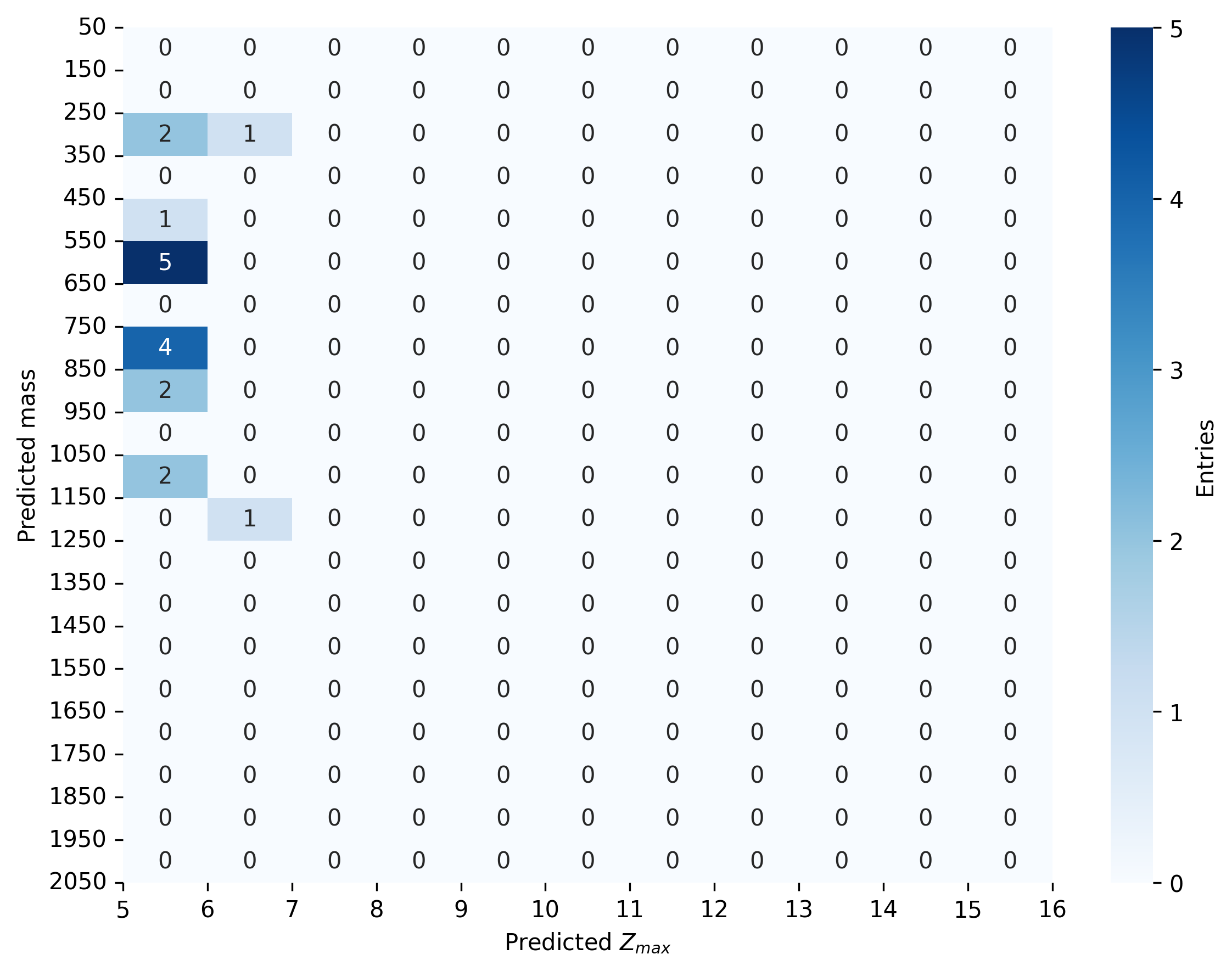}
        \caption{Channel 3}
        \label{heatmap_ch3}
    \end{subfigure}
    \hfill
    \begin{subfigure}[b]{0.45\textwidth}
        \centering
        \includegraphics[width=\textwidth]{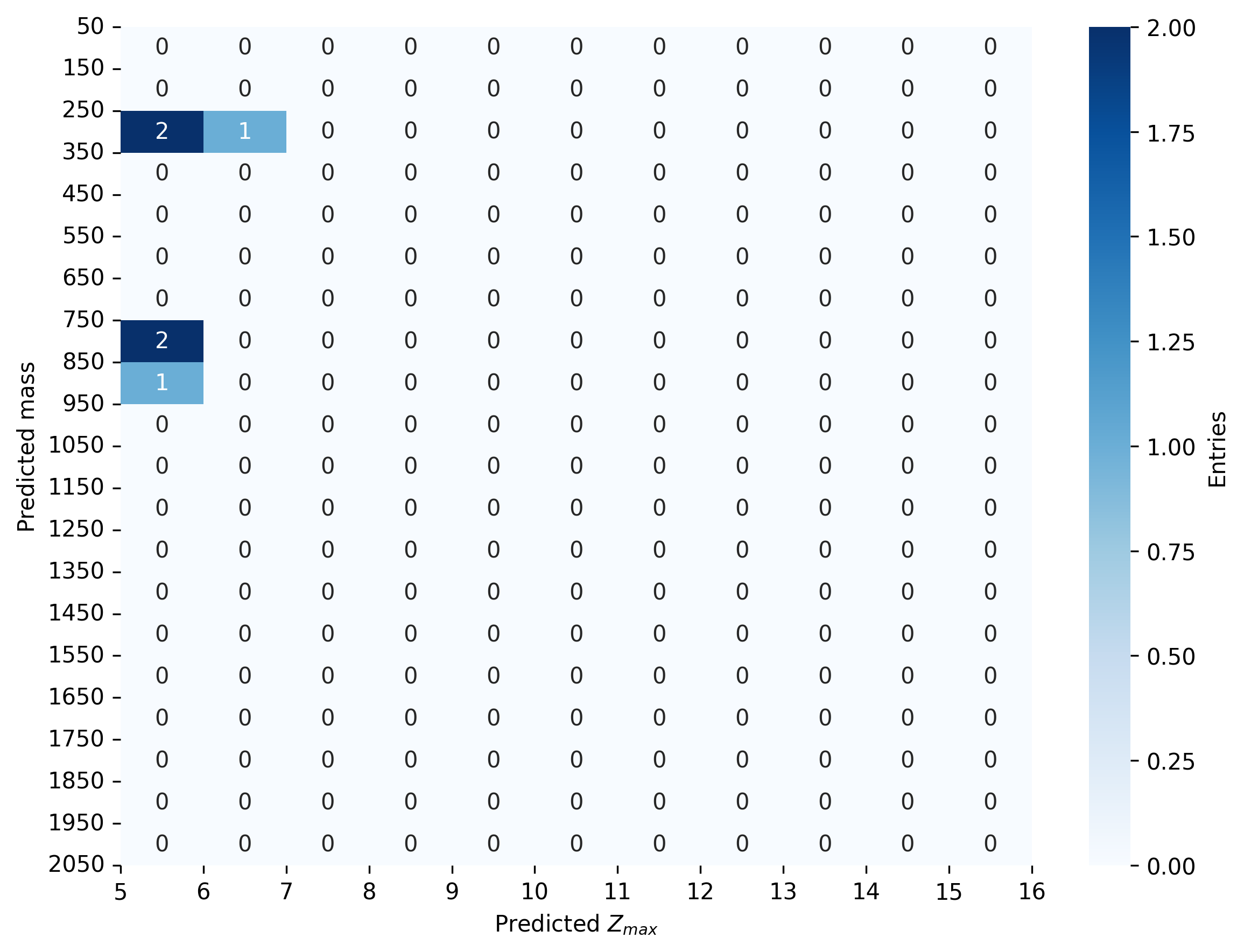}
        \caption{Channel 3 after GAA}
        \label{heatmap_ch3_gaa}
    \end{subfigure}
    \caption{Number of histograms as a function of BumpNet's maximal predicted significance and the mass position of the excess for channel 2b (top) and channel 3 (bottom) before (left) and after (right) the application of the GAA. No signals has been injected; the histograms contain only SM events and contain false-positive excesses.}
    \label{figDM_heatmaps}
\end{figure}

The GAA is first applied to the background-only \ac{DM} SM samples to verify how many false-positive signals survive the algorithm. The results are shown in Figure~\ref{figDM_heatmaps} for channels 2b (top plots) and 3 (bottom plots), illustrating the number of histograms as a function of BumpNet's maximal predicted significance and the mass position of the excess. The GAA reduces the number of false-positive signals from 25 to 18 for channel 2b and from 28 to 9 for channel 3, as observed by comparing the left (before GAA) and right plots (after GAA). The remaining histograms belong to 4 and 3 families in channels 2b and 3, respectively, and all feature excesses early in the mass histogram. Such false-positive signals could be eliminated in the future by fine-tuning the selection of the starting point for BumpNet's application on a histogram, which currently does not consider the first 10\% of bins. There is one exception in channel 2b, where three dimuon mass histograms show an excess around 70–75GeV, i.e., the low tail of the $Z$ boson mass. This could be eliminated by improving the definition of the $Z\to\ell\ell$ object (see Section~\ref{sec:DMsamples}).

\begin{figure}
     \centering
     \begin{subfigure}[b]{0.32\textwidth}
         \centering
         \includegraphics[width=\textwidth]{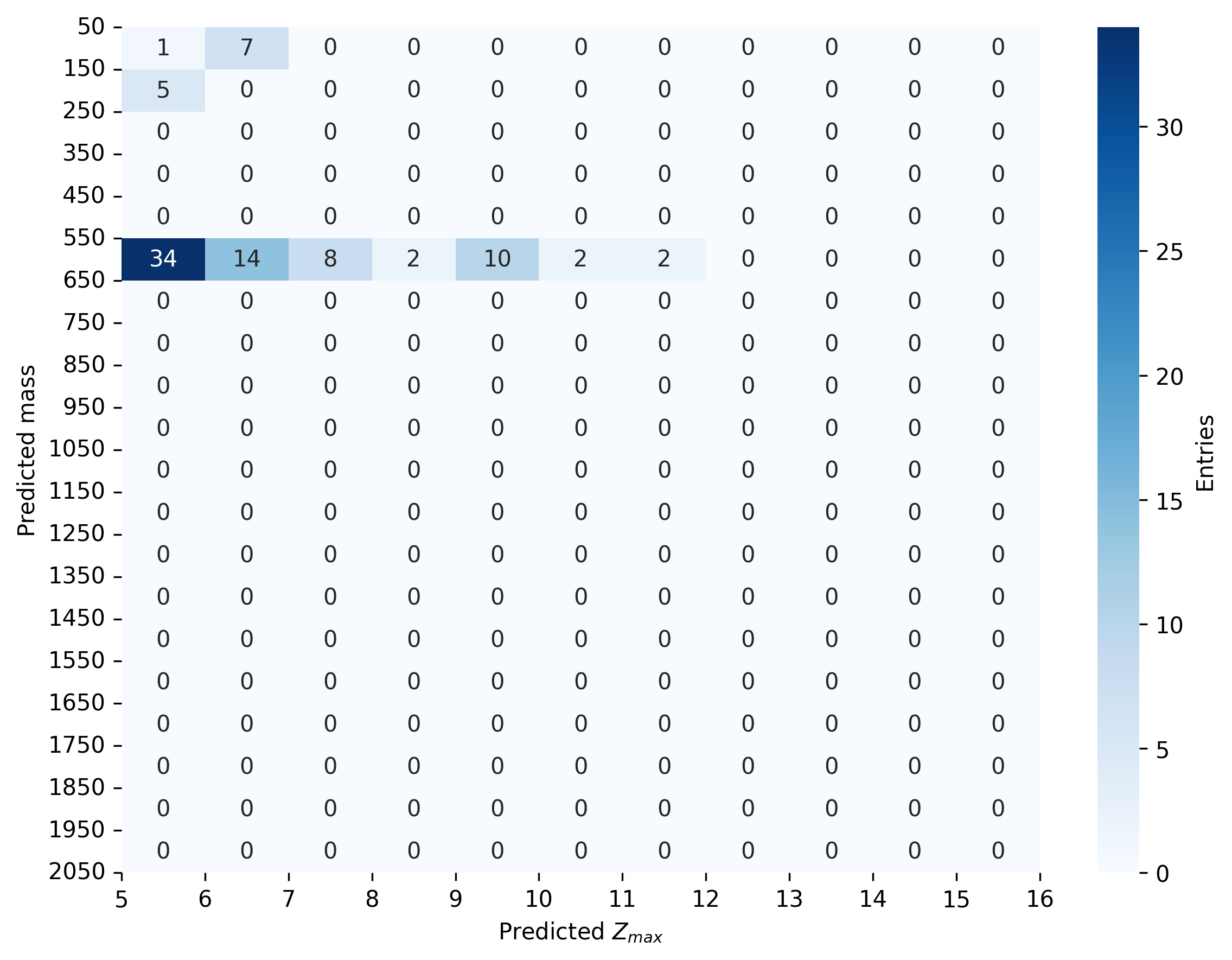}
         \caption{Two $LQ\to b\mu$}
         \label{fig:sub_bsm_heatmaps_a}
     \end{subfigure}
     \hfill
     \begin{subfigure}[b]{0.32\textwidth}
         \centering
         \includegraphics[width=\textwidth]{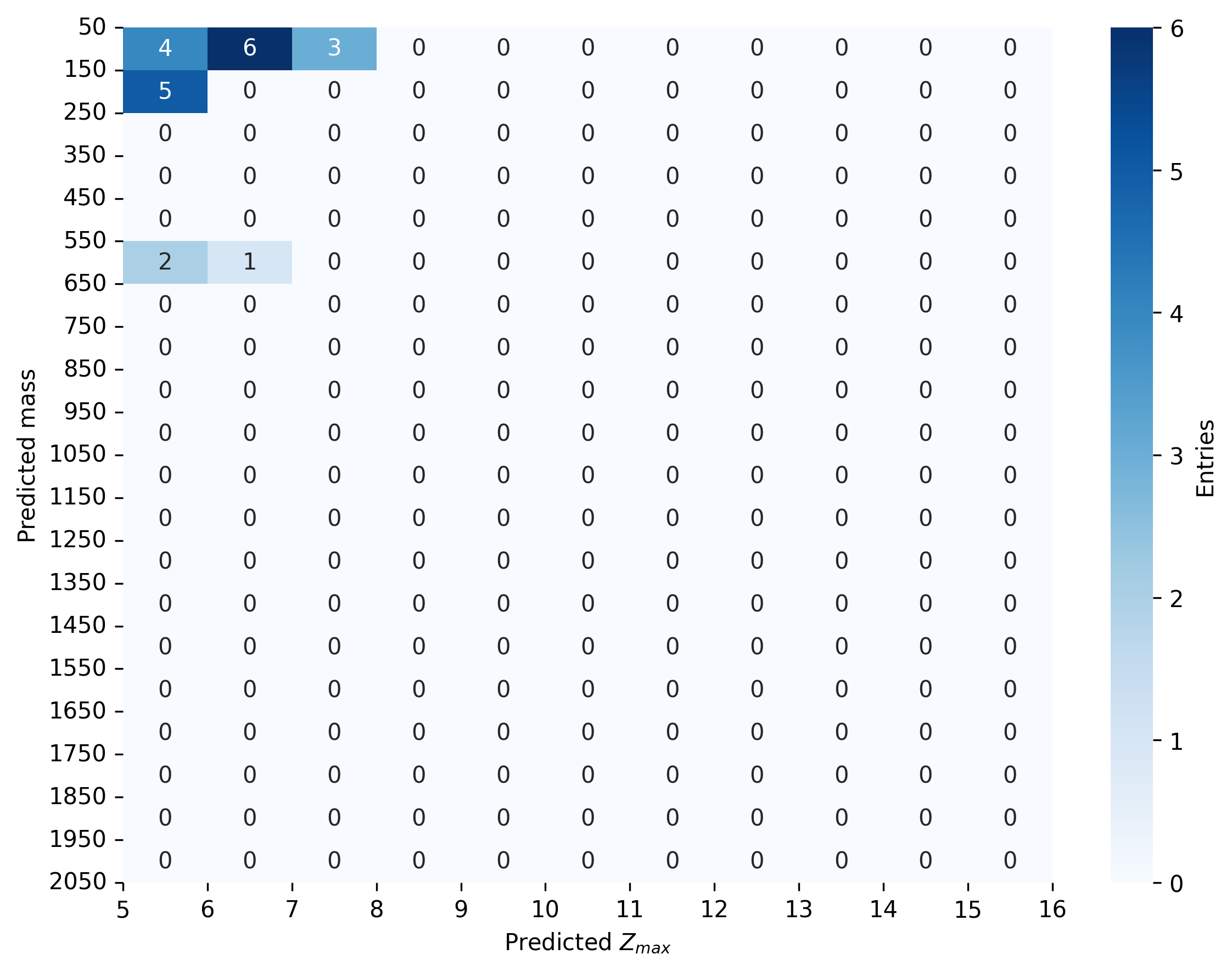}
         \caption{Two $LQ\to be$}
         \label{fig:sub_bsm_heatmaps_b}
     \end{subfigure}
     \hfill
     \begin{subfigure}[b]{0.32\textwidth}
         \centering
         \includegraphics[width=\textwidth]{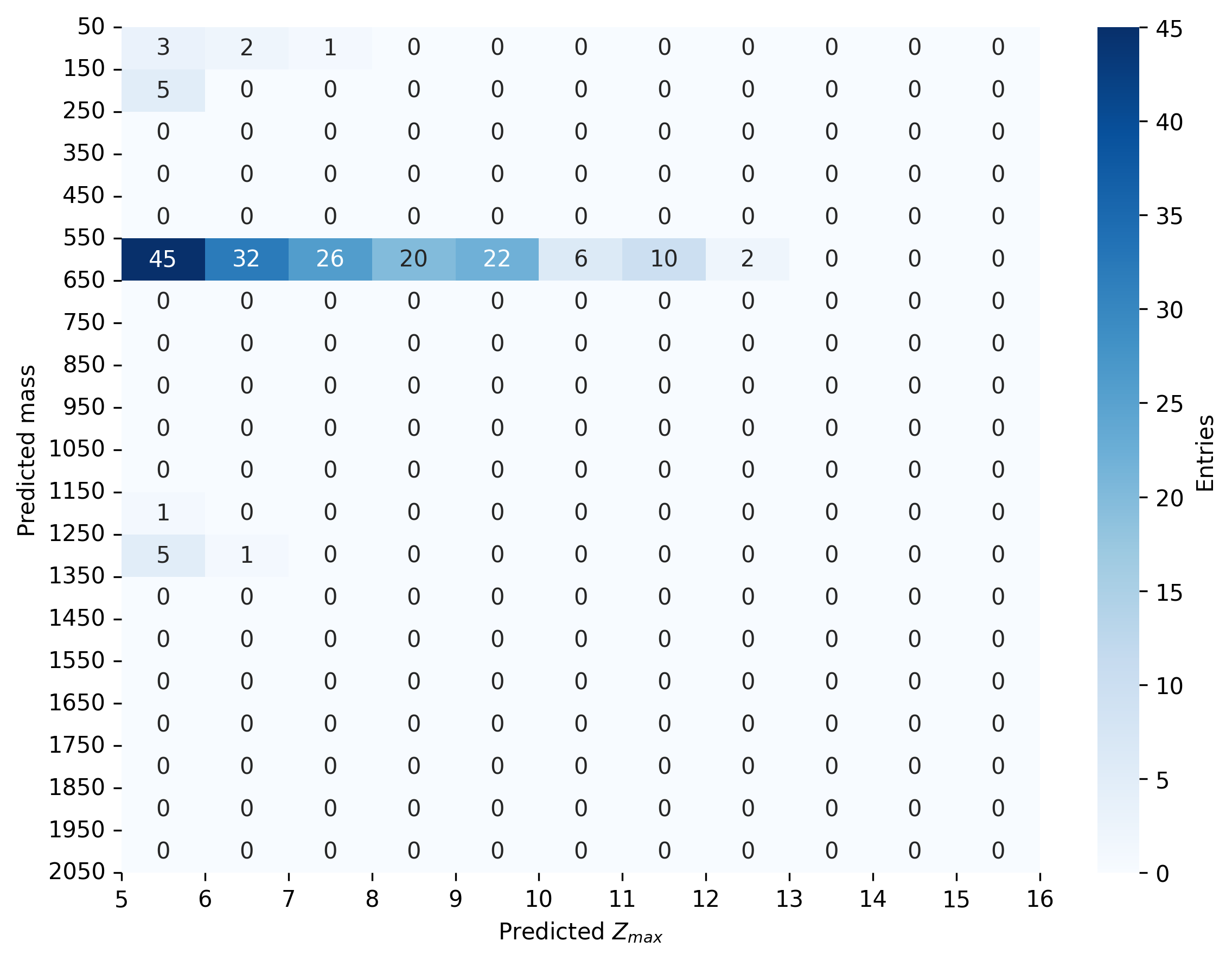}
         \caption{$LQ\to b\mu$, $LQ\to be$}
         \label{fig:sub_bsm_heatmaps_c}
     \end{subfigure}
     \begin{subfigure}[b]{0.32\textwidth}
         \centering
         \includegraphics[width=\textwidth]{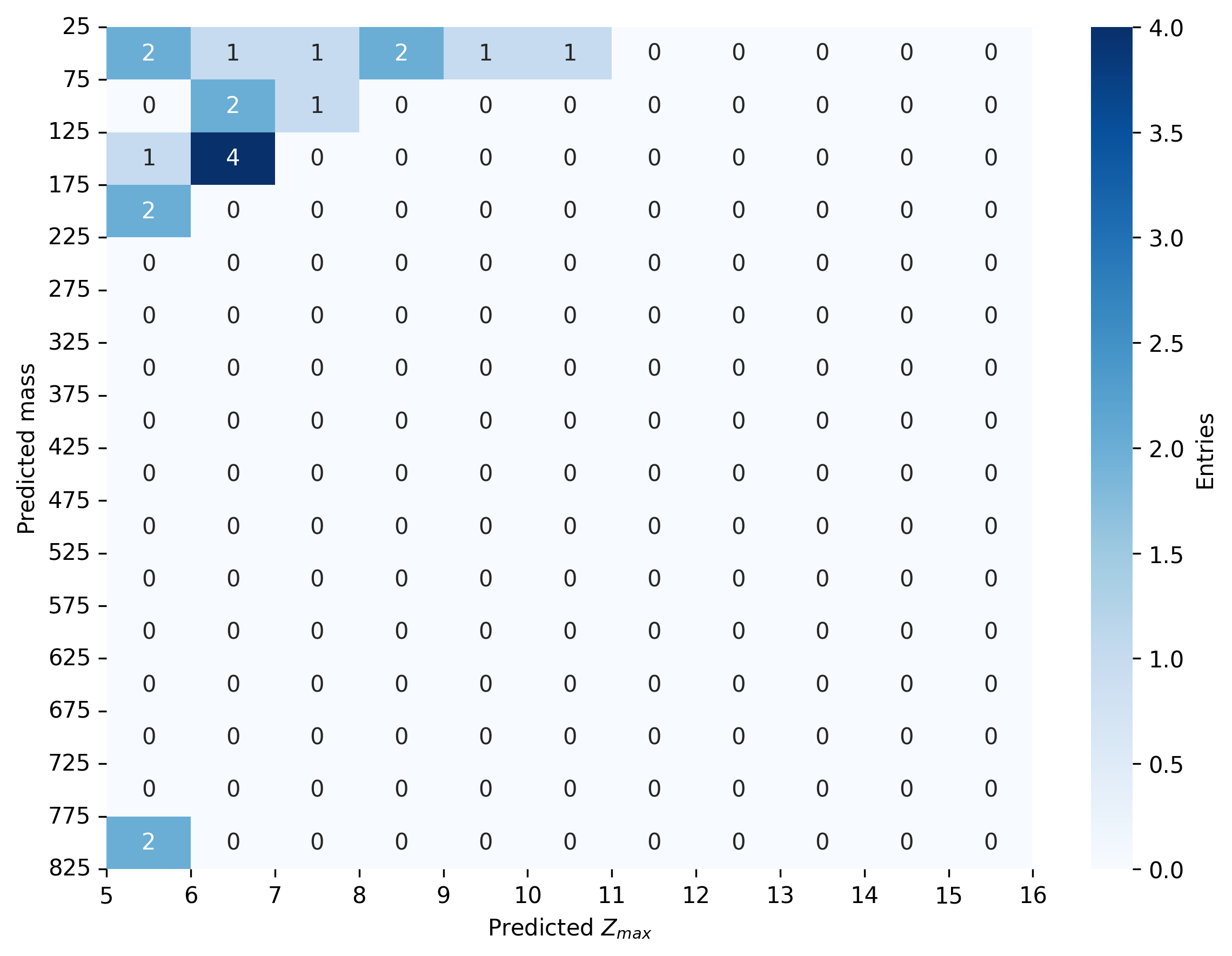}
         \caption{$Z'\to \mu\mu$}
         \label{fig:sub_bsm_heatmaps_d}
     \end{subfigure}
     \hfill
     \begin{subfigure}[b]{0.32\textwidth}
         \centering
         \includegraphics[width=\textwidth]{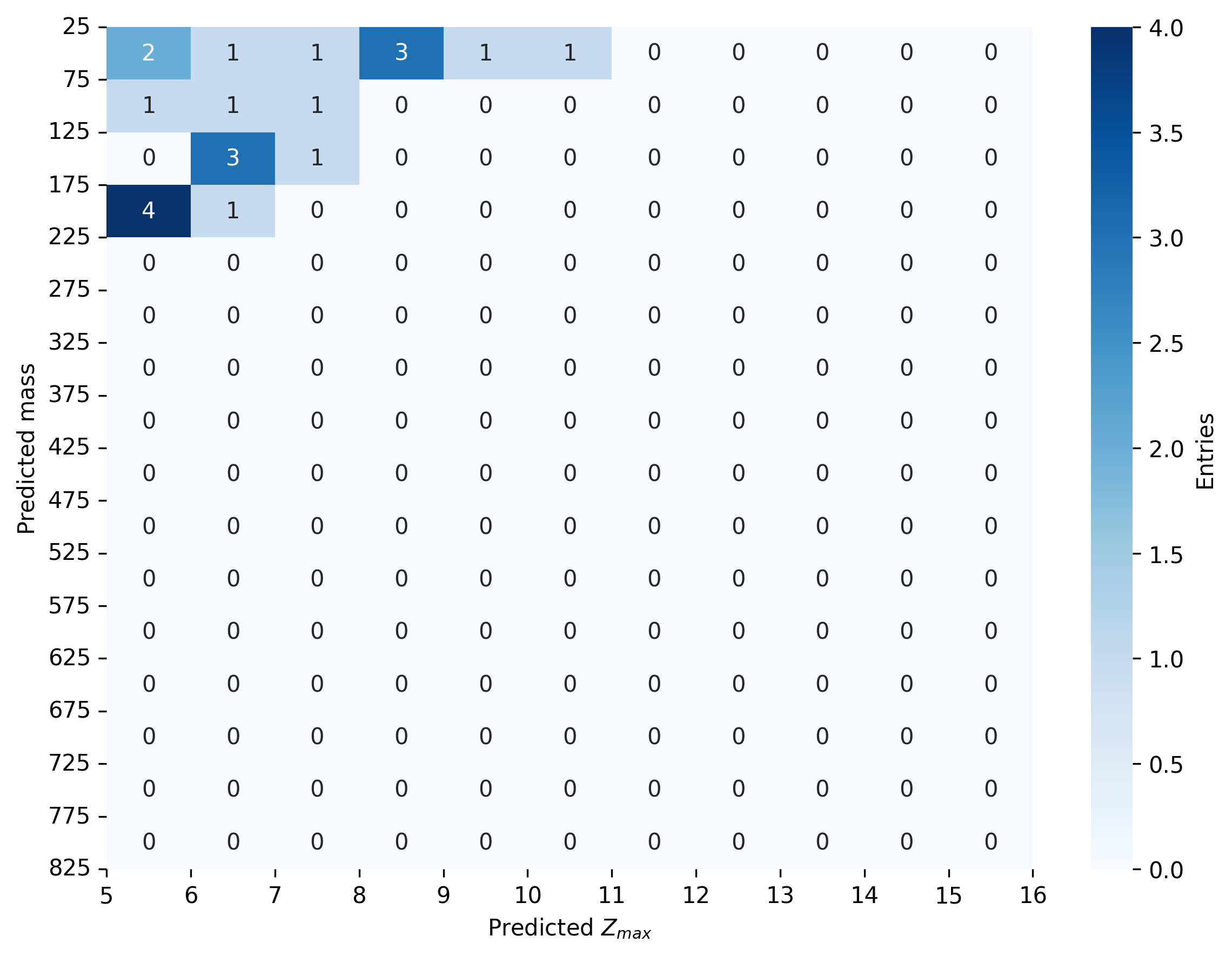}
         \caption{$Z'\to \mu\mu$}
         \label{fig:sub_bsm_heatmaps_e}
     \end{subfigure}
     \hfill
     \begin{subfigure}[b]{0.32\textwidth}
         \centering
         \includegraphics[width=\textwidth]{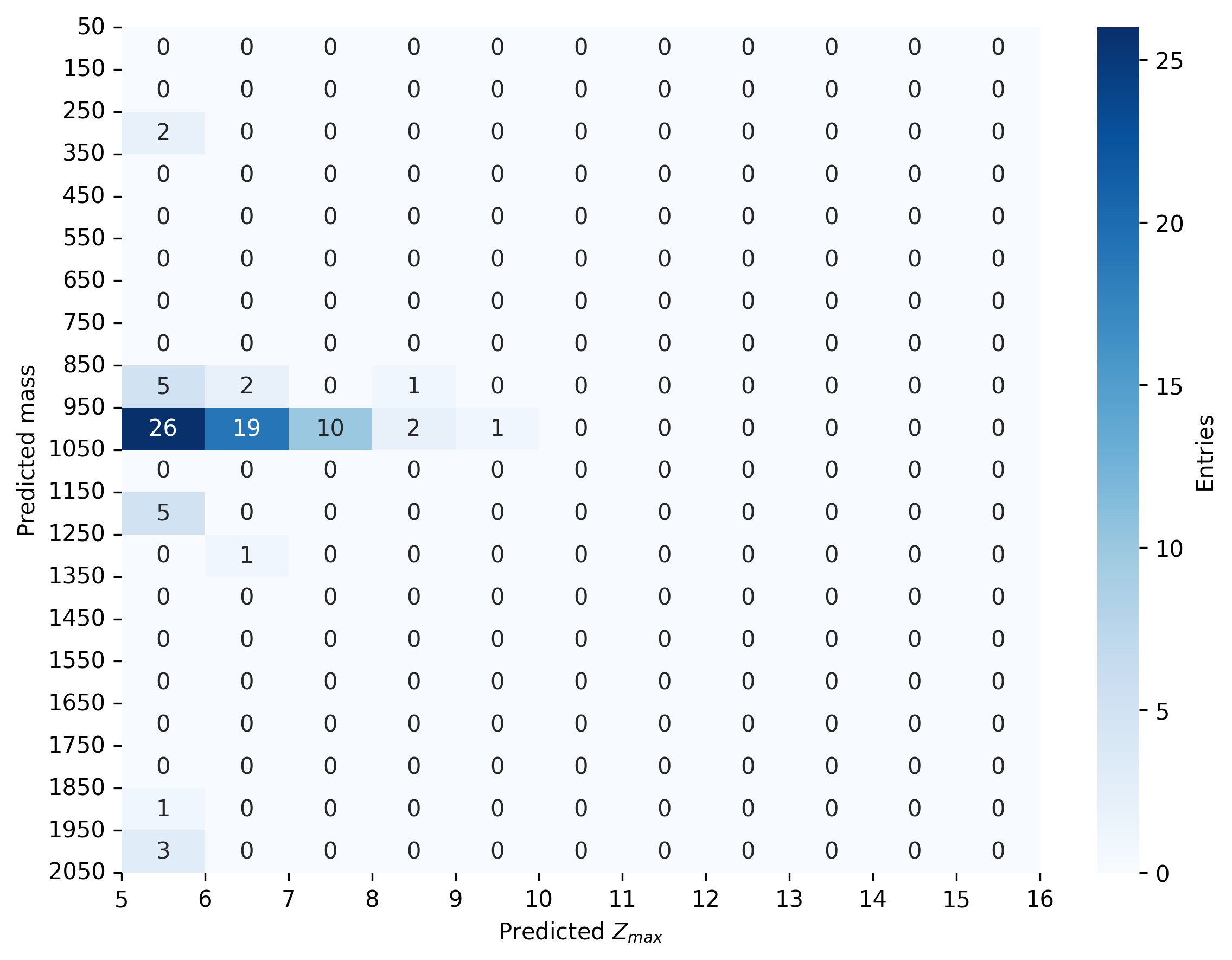}
         \caption{Two RPV Stop$ \to b\ell$}
         \label{fig:sub_bsm_heatmaps_f}
     \end{subfigure}
     \hfill
     \begin{subfigure}[b]{0.32\textwidth}
         \centering
         \includegraphics[width=\textwidth]{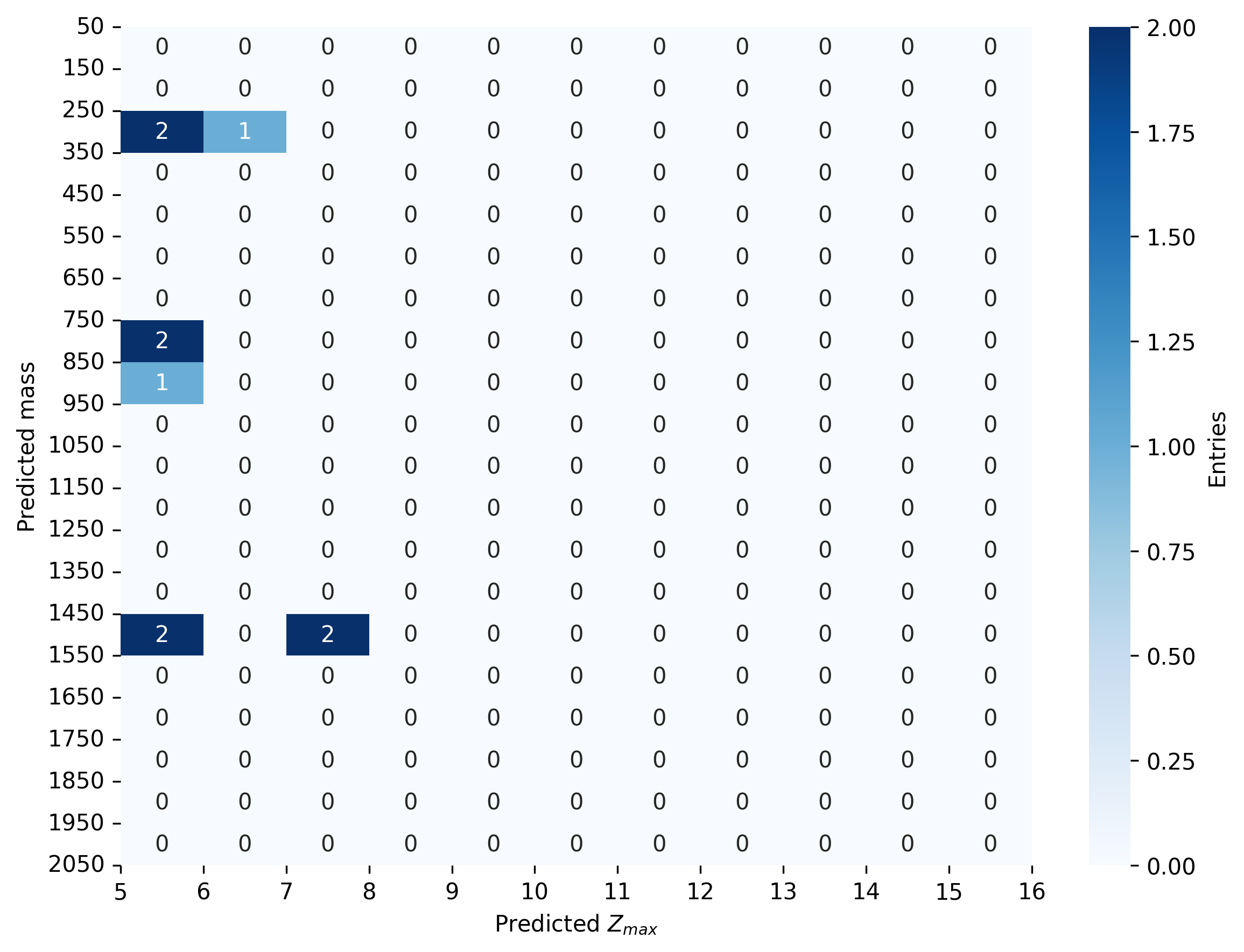}
         \caption{$W'\to \ell \nu qq$}
         \label{fig:sub_bsm_heatmaps_h}
     \end{subfigure}
     \hfill
        \caption{Number of histograms as a function of BumpNet's maximal predicted significance and the mass position of the excess for the DM samples after the application of the GAA. At least one group of histogram is visible at each sample's mass. The signal mass bumps are expected to be located at 600 GeV for figures (\subref{fig:sub_bsm_heatmaps_a})-(\subref{fig:sub_bsm_heatmaps_c}), 50 GeV for figures (\subref{fig:sub_bsm_heatmaps_d})-(\subref{fig:sub_bsm_heatmaps_e}), 1 TeV for figure (\subref{fig:sub_bsm_heatmaps_f}) and 1.5 TeV for figure (\subref{fig:sub_bsm_heatmaps_h}).}
        \label{fig:DM_samples_heatmap_after_GAA}
\end{figure}

The GAA is then applied to the various \ac{DM} \ac{BSM} + \ac{SM} samples. The results are shown in Figure~{\ref{fig:DM_samples_heatmap_after_GAA}}, which displays the number of histograms as a function of BumpNet's maximal predicted significance and the mass position of the excess. For all models except $W'\to qq\nu\nu$, the GAA successfully identifies families of histograms featuring true-positive signals at the expected mass and for the expected object combination. Figure~\ref{fig:DM_phys_corr} presents example histograms that have been identified by the GAA.  In the case of $W'\to qq\nu\nu$, only a single uncorrelated histogram exhibits a significant excess (Figure~\ref{fig:sub_bsm_examples_g}), which is insufficient to form a family—since the GAA requires at least two histograms to establish a correlation—and thus it does not survive the GAA. 

These results are promising, demonstrating that a true signal can be distinguished from false positives in LHC data if it is prominent enough to appear in multiple histograms. This indicates the potential of combining BumpNet with physics correlations to enhance signal detection in high-energy physics experiments.
To further improve the analysis, future work will focus on refining the histogram production process to reduce the number of false-positive signals that pass the GAA. Additionally, optimizing the parameters of the GAA, and potentially integrating it more closely with BumpNet could enhance its ability to discern true signals from background. Overall, this approach holds significant promise for advancing the search for new particles in LHC data.

\section{Conclusions}
\label{sec:conclusions}

We introduced BumpNet, a \ac{NN} designed to map invariant mass histograms into statistical inference distributions for signal detection in high-energy physics. BumpNet generalizes the Data-Directed Paradigm bump hunter by training on a diverse mixture of histograms generated from smoothly falling functions and Dark Machines samples that emulate realistic, data-like distributions. 

BumpNet's performance was benchmarked against that of an ideal analysis using the likelihood ratio test, assuming perfect knowledge of signal and background shapes. Its predictions demonstrated negligible to small biases and variance below 1$\sigma$ when tested on Gaussian-shaped signals 
added to backgrounds generated from both the smoothly falling functions used in training, \ac{DM} data, and the high-mass di-electron and di-muon background modeling employed by ATLAS in their resonance searches \cite{ATLAS:2019erb}. 

We validated BumpNet’s consistency with reported results by applying it to the \( H \rightarrow \gamma \gamma \) distribution from the ATLAS Higgs discovery paper \cite{ATLAS:2012yve}, as well as to high invariant mass di-electron and di-muon distributions, finding excellent agreement with ATLAS results \cite{ATLAS:2019erb}.

The application of BumpNet to \ac{BSM} signals injected into \ac{DM} backgrounds further underscores its potential to detect realistic particle resonances within complex data environments. BumpNet effectively identified significant signals across a range of \ac{BSM} models and mass values, accurately distinguishing them from \ac{SM} backgrounds. Additionally, combining BumpNet with the Global Analysis Algorithm (GAA) demonstrates enhanced signal detection capabilities while effectively managing the look-elsewhere effect in large datasets. These results validate BumpNet’s adaptability and robustness for challenging analysis scenarios, highlighting its promise for advancing signal detection in future high-energy physics applications.


\acknowledgments

We gratefully acknowledge the support of the Natural Sciences and Engineering Research Council of Canada (NSERC),  the Institut de valorisation des données IVADO,  the Canada First Research Excellence Fund, and the Israeli Science Foundation (ISF, Grant No. 2382/24).  We also extend our gratitude to the Krenter-Perinot Center for High-Energy Particle Physics, the Shimon and Golde Picker–Weizmann Annual Grant, and the Sir Charles Clore Prize for their support.
A special thanks is extended to Martin Kushner Schnur for his invaluable contribution to this research.



\bibliographystyle{apsrev4-2}
\bibliography{references}
\end{document}